\documentclass[12pt]{article}

\usepackage{amsmath}
\usepackage{amssymb}
\usepackage{graphicx}
\usepackage{comment}
\usepackage{xcolor}
\usepackage{threeparttable}
\usepackage{caption}
\usepackage{times}
\usepackage{bm}
\usepackage{natbib}
\usepackage{amsthm}

\usepackage{soul}
\setstcolor{red}

\newtheorem{proposition}{Proposition}

\graphicspath{{figs/}}

\usepackage{setspace}

\newcommand{\blind}{0}

\begin{document}
		
	\def\spacingset#1{\renewcommand{\baselinestretch}
		{#1}\small\normalsize} \spacingset{1}

	
	\if0\blind
	{
		\title{\bf On Construction and Estimation of Stationary Mixture Transition Distribution Models}

		\author{
		Xiaotian Zheng,
        Athanasios Kottas
        and
        Bruno Sans\'o
        \\
        Department of Statistics, University of California, Santa Cruz 
        }
		\maketitle
	} \fi
	
	\if1\blind
	{
		\bigskip
		\bigskip
		\bigskip
		\begin{center}
			{\LARGE\bf On Construction and Estimation of Stationary Mixture Transition Distribution Models}
		\end{center}
		\medskip
	} \fi
	
	\bigskip
	\begin{abstract}
		Mixture transition distribution time series models build high-order
		dependence through a weighted combination of first-order transition
		densities for each one of a specified number of lags. We present a
		framework to construct stationary mixture transition distribution
		models that extend beyond linear, Gaussian dynamics. We study
		conditions for first-order strict stationarity which allow for
		different constructions with either continuous or discrete families
		for the first-order transition densities given a pre-specified family
		for the marginal density, and with general forms for the resulting
		conditional expectations. Inference and prediction are developed 
		under the Bayesian framework with particular emphasis on flexible, 
		structured priors for the mixture weights.  Model properties are 
		investigated both analytically and through synthetic data examples.  
		Finally, Poisson and Lomax examples are illustrated through real data applications.
	\end{abstract}

	\noindent
	{\it Keywords: Bayesian inference; First-order strict stationarity; Markov chain Monte Carlo; Non-Gaussian time series.} 
	
	\vfill
	
	\newpage

	\spacingset{1.5} 

\doublespacing

\section{Introduction}

Mixture transition distribution (MTD) models describe a time series
$\{X_t: t\in\mathbb{N}\}$, where 
$X_t \in \mathcal{S} \subseteq \mathbb{R}$ for all $t$, by 
specifying the distribution of $X_t$
conditional on the past as 
\begin{equation}\label{MTD}
F(x_t\mid \bm{x}^{t-1}) = \sum_{l=1}^{L} w_l \, F_l(x_t\mid x_{t-l}),
\end{equation}
for $t > L$, based on initial values for
$(x_1,\dots, x_L)^\top$.
In Equation~(\ref{MTD}), $F(x_t\,|\,\bm{x}^{t-1})$ 
is the conditional cumulative distribution function (cdf) of $X_t$ 
given that $\bm{X}^{t-1} = \bm{x}^{t-1}$, and $F_l(x_t\,|\,x_{t-l})$ is the 
conditional cdf of $X_t$ with respect to the $l$th transition component 
given that $X_{t-l} = x_{t-l}$, where $\bm{X}^{t-1} = \{X_i: i\leq t-1\}$ and 
$\bm{x}^{t-1} = \{x_i: i\leq t-1\}$. The parameters $w_l\geq 0$, $l=1,\dots,L$, assign 
weights to the transition components, such that $\sum_{l=1}^Lw_l = 1$. 
On a finite state space this model provides a parsimonious 
approximation of high-order Markov chains 
\citep{raftery1985model, raftery1994estimation,
berchtold2001estimation}. On a more general space, the model structure 
can represent time series that depict non-Gaussian features such
as burst, outliers, and flat stretches \citep{le1996modeling}, or
change-points \citep{raftery1994change}. We refer to 
\cite{berchtold2002mixture} for a review.
An MTD model consists of $L$ first-order transition components. 
The mixture autoregressive model of \cite{wong2000mixture} is a
generalization that allows for each transition component to depend 
on a different number of lags;
\cite{lau2008bayesian} consider a Bayesian nonparametric 
prior for the transition component of such models.
There are several related extensions 
that consider mixtures of  autoregressive conditional heteroscedastic 
terms, including \cite{wong2001mixture}, \cite{berchtold2003mixture}, 
\cite{zhu2010mixture} and \cite{li2017mixture}. 
Other extensions 
include multivariate model settings
\citep{hassan2006modeling,fong2007mixture,
kalliovirta2016gaussian}, time-varying mixture weights
\citep{wong2001logistic,bartolucci2010note,bolano2016general}, 
non-linear transition dynamics \citep{GP-MTD},
and order/lag selection \citep{khalili2017regularization,heiner2019estimation}.
Applications of these models appear in many fields such as finance,
and the environmental and medical sciences; see,
for example, \cite{macdonald1997hidden, lanne2003modeling,
escarela2006flexible, cervone2014location}.

Stationarity for MTD models, and their extensions,
is generally difficult to attain due to the mixture model structure.
This limits the choices of parametric families for the transition
components for these models. Families considered in the literature include: 
Gaussian \citep{le1996modeling,wong2000mixture,kalliovirta2015gaussian};
Student-t \citep{wong2009student,meitz2021mixture};
Laplace \citep{nguyen2016laplace}; Weibull \citep{luo2009parameter};
and Poisson \citep{zhu2010mixture}. 
These models are typically parameterized in ways that 
result in conditional expectations that are linear functions of the lags. This particular parameterization facilitates the study of stationarity,
though only in a weak sense, at the cost of reducing model flexibility. 
Indeed, the conditional expectation of an MTD
model has the general form $\sum_{l=1}^L w_l \, \mu_{l}(x_{t-l})$, 
where $\mu_{l}(y)=$ $\int xdF_l(x\,|\,y)$, allowing for non-linear dependence 
of the mean, conditional on past observations. 

The primary goal of this article is to develop conditions for first-order 
strictly stationary MTD models, that is, 
stationary models with an invariant marginal distribution.
We show that a sufficient condition is to assume the same marginal
distribution for all the components of the mixture. It turns out that
this marginal distribution is also the invariant marginal
distribution of the time series. Under this condition, first-order
strict stationarity is achieved with respect to any particular 
parameterization. We thus obtain a rich class of distribution specifications 
for the model, facilitating the study of component distributions
that have not been explored in the literature,
and enhancing the modeler's ability to extend beyond high-order 
linear dependence in the conditional expectation. Although the focus 
of our methodology is on strictly stationary models, we also study
weak stationarity conditions for MTD models with linear conditional expectation.

MTD models are usually built by
specifying transition densities $f_{U_l\mid V_l}$  for
each component $l = 1, \dots, L$. These correspond to conditional
densities for random variable $U_l$ given random
variable $V_l$. This specification raises a question of existence
of a coherent bivariate density $f_{U_l,V_l}$. Our 
second goal is to provide a constructive approach to building MTD models 
that satisfy our strict stationarity condition under a coherent bivariate 
density $f_{U_l,V_l}$. We present two distinct approaches: 
the bivariate distribution method, which is based on specifying 
the bivariate distribution of the pair $(U_l,V_l),l = 1,\dots, L$; and 
the conditional distribution method, which consists 
of finding pairs of compatible conditional distributions 
$f_{U_l\mid V_l}$ and $f_{V_l\mid U_l}$ for all $(U_l,V_l)$.

Our final goal is to develop a Bayesian framework
for MTD model inference and prediction.
We assume that the order of dependence is unknown, but is bounded
above by a finite number $L$. We use an 
over-specified model with $L$ chosen conservatively, 
under the expectation that only a few of the 
lags contribute to the dynamics of the series. We consider two priors 
for the mixture weights, one based on a truncated stick-breaking process, 
and the other obtained by discretization of 
a c.d.f. which is assigned a nonparametric prior. 
While the former supports stochastically decreasing weights, the 
latter favors important, but not necessarily consecutive weights.

The rest of the article is organized as follows. In Section 2 we 
review the issues related to establishing stationarity conditions for 
MTD models. 
We then introduce the invariant condition that 
yields the class 
of first-order strictly stationary MTD 
models, and connect it to weak stationarity. Section 3 illustrates 
two methods to construct such models with many examples. 
In Section 4, we outline the Bayesian 
approach for model estimation and prediction, followed in Section 5 
by an illustration of the properties of two structured priors for mixture 
weights on synthetic data, and applications of the models on two real 
data sets of different nature. Finally, we conclude 
with a discussion in Section 6. Proofs and details of Markov chain Monte 
Carlo (MCMC) algorithms are provided in the Appendix and the Supplementary Material.

\section{First-order strict stationarity}

Consider the conditional density specification of the model
in Equation \eqref{MTD}:
\begin{align}\label{eq:MTD2}
f(x_t\mid \bm{x}^{t-1}) = \sum_{l=1}^L w_l \, f_l(x_t\mid x_{t-l}).
\end{align}
Under our modeling framework, each transition component is taken to 
correspond to the distribution for a random vector $(U_l, V_l)$, for 
$l=1,...,L$, where $f_l \equiv$ $f_{U_l\mid V_l}$ denotes the associated 
conditional density. 

Earlier work has studied necessary and sufficient conditions for 
constant first and second moments \citep{le1996modeling}. In general, 
such conditions are difficult to establish, especially 
for the second moment $ \int_{\mathcal{S}} x_{t}^{2} \, g_t(x_t) dx_t$, 
where $g_t(x_t) = \sum_{l=1}^L w_l \int_{\mathcal{S}}f_l(x_t\,|\, x_{t-l})
g_{t-l}(x_{t-l})dx_{t-l}$ is the marginal density of the process $\{X_t\}$.
This restricts the choices of parametric families for the component 
transition densities. In particular, those choices result in linear
conditional expectations. 
Even when conditions for time-independent first and second moments can be 
obtained, the resulting constrained parameter spaces complicate estimation.

The key result for our methodology is given in the following proposition, the 
proof of which can be found in the Appendix. The result provides the 
foundation for different constructions of first-order strictly stationary 
MTD models. Rather than imposing restrictions on 
the parameter space, the proposition formulates a substantially easier to 
implement condition on the marginals of the bivariate distributions that 
define the transition components. 

\begin{proposition} 
Consider a set of bivariate random vectors $(U_l,V_l)$ taking values in 
$\mathcal{S} \times \mathcal{S}$, $\mathcal{S} \subseteq \mathbb{R}$, with conditional 
densities $f_{U_l\mid V_l}$, $f_{V_l\mid U_l}$ and marginal densities 
$f_{U_l}$, $f_{V_l}$, for $l = 1,\dots, L$, and let $w_l \geq 0$, for
$l=1,...,L$, with $\sum_{l=1}^{L} w_l = 1$.
Consider a time series $\{X_t: t\in\mathbb{N}\}$, where
$X_t \in \mathcal{S}$, generated from
\begin{align}\label{eq:pp1}
f(x_t\mid \bm{x}^{t-1}) = \sum_{l=1}^{L} w_l \, f_{U_l\mid V_l}(x_t\mid x_{t-l}),
\,\,\,\,\,t > L,
\end{align}
and from 
$$
f(x_t\mid \bm{x}^{t-1}) = \sum_{l=1}^{t-2} w_l \, f_{U_l\mid V_l}(x_t\mid x_{t-l}) + 
\left( 1 - \sum_{k=1}^{t-2} w_k \right) f_{U_{t-1} \mid V_{t-1}}(x_t\mid x_{1}),
\,\,\,\,\,2 \leq t \leq L.
$$
This time series is first-order strictly
stationary with invariant marginal density $f_X$ if it satisfies 
the invariant condition: $X_1 \sim f_X$, and $f_X(x) =$ $f_{U_l}(x) =$ 
$f_{V_l}(x)$, for all $x \in \mathcal{S}$, and for all $l$.
\end{proposition}

The two different expressions for the transition density allow us to 
establish the stationarity condition for the entire time series.
The relevant form for inference is the one in Equation (\ref{eq:pp1}), 
since we work with the likelihood conditional on the first $L$ time 
series observations. Proposition 1 applies regardless 
of $X_t$ being a continuous, discrete or mixed  random variable.

Regarding strict stationarity, the literature mostly focuses on existence 
of a stationary distribution. Exceptions are \cite{kalliovirta2015gaussian} 
and \cite{meitz2021mixture}, where a stationary marginal distribution 
for a mixture autoregressive model is 
obtained, albeit again under constrained parameter spaces, and 
\cite{mena2007stationary} whose approach is the one most closely 
related to our proposed methods.

\cite{mena2007stationary} use the latent variable method proposed in 
\cite{pitt2002constructing} to construct the conditional density for 
each transition component of the MTD. More specifically,   
$f_l(x_t\,|\,x_{t-l})=$ $\int h_{X|Z}(x_t\,|\,z) h_{Z|X}(z\,|\,x_{t-l})dz$,
where $h_{X|Z}(x\,|\,z) \propto$ $h_{Z|X}(z\,|\,x) f_{X}(x)$, and the integral 
is replaced by a sum if $Z$ is a discrete variable. Then, provided 
$X_1 \sim f_X$, the MTD model is 
first-order strictly stationary with invariant density $f_X$. Under 
this construction, the invariant density $f_X$ can be viewed as the 
prior for likelihood $h_{Z\mid X}$, which is built through latent variable 
$Z$. In practice, this restricts the approach to continuous time series,
and the choices for the invariant density to cases where $f_X$ is 
conjugate to $h_{Z\mid X}$. Even for such cases, the transition component 
will typically have a complex form. In particular, the example explored 
in \cite{mena2007stationary} involves a gamma invariant distribution, 
with $h_{Z\mid X}$ corresponding to a Poisson distribution. In this case, 
$f_l(x_t\,|\,x_{t-l})$ is a countable sum whose evaluation 
requires modified Bessel functions of the first kind. Moreover, 
following \cite{pitt2002constructing}, \cite{mena2007stationary} restrict
attention to choices of $h_{Z\mid X}$ that yield linear conditional expectations
for the transition components, and thus also for the MTD models.

The key feature of our approach is that it builds from the bivariate 
distributions, $f_{U_l,V_l}$, corresponding to the transition components. 
In the next section, we discuss two approaches to specifying those bivariate 
distributions, either directly or via compatible conditionals,
$f_{U_l\mid V_l}$ and $f_{V_l\mid U_l}$.
In conjunction with Proposition 1, we obtain a general framework to 
constructing first-order strictly stationary MTD 
models that can be applied to both discrete and continuous 
time series, while allowing for a wide variety of invariant marginal 
distributions, as well as for both linear and non-linear lag 
dependence in the conditional expectation. 

In general, an explicit expression for the autocorrelation function for general 
MTD models is difficult to derive. However, a recursive 
equation can be obtained for a class of linear MTD models. We say the MTD model is linear 
if $E(U_l\mid V_l = y) = a_l + b_l \, y$ for some $a_l, b_l\in\mathbb{R},\;l = 1,\dots, L$. 
Consider a linear MTD model that satisfies the invariant condition of Proposition 1, 
and assume that the first and second moments of the process, denoted by $\mu$ and 
$\mu^{(2)}$, exist and are finite. Then, for any $L$ and $h\geq L$, we can derive 
\begin{equation}\label{eq:cross_mom}
\begin{aligned}
E(X_{t+h}X_t) 
\, = \, \sum_{l=1}^Lw_la_l\mu \, + \, \sum_{l=1}^L w_l b_l E(X_{t+h-l}X_t).
\end{aligned}
\end{equation}
Assuming that, for any $h\geq 1$, $E(X_{t+h}X_t)$ does not depend on time $t$,
let $r(h)$ be the lag-$h$ autocorrelation function. Then,
\begin{equation}\label{eq:acf}
r(h) = \phi + \sum_{l=1}^Lw_lb_lr(h-l),\;\;\,\,\,h\geq L,
\end{equation}
where $\phi = (\sum_{l=1}^Lw_la_l\mu - (1-\sum_{l=1}^Lw_lb_l)\mu^2)/(\mu^{(2)} - \mu^2)$ is zero if and only if $\mu = 0$ or $a_l = (1-b_l)\mu$. 
When $b_{l} = \rho,\;\rho\in(0,1)$ and 
$a_l = (1 - \rho)\mu$, for all $l$, 
Equation~(\ref{eq:acf}) reduces to $r(h) = \rho\sum_{l=1}^Lw_lr(h-l),\;h\geq L$, 
which is the result in \cite{mena2007stationary}.

In the case of distinct roots, the general solution to
Equation~(\ref{eq:acf}) is 
\begin{equation}
    r(h) = c_1z_1^h + \dots + c_Lz_L^h + \phi\left((1-z_1)\dots(1-z_L)\right)^{-1},
\end{equation}
where $c_1,\dots, c_L$ are determined by the initial conditions $r(0),\dots, r(L-1)$ and $z_1,\dots,z_L$ are the roots of the 
associated polynomial $z^L - w_1b_1z^{L-1} - \dots - w_Lb_L = 0$.
It follows that, as $h\rightarrow\infty$, $r(h)\rightarrow 0$ if 
and only if: (1) $\phi = 0$; (2) $z_1,\dots,z_L$ all lie inside the
unit circle. 

The above discussion provides an approach to obtaining a weakly stationary 
MTD model based on Equation (\ref{eq:pp1}), and is summarized in the following
proposition the proof of which is included in the Supplementary Material.

\begin{proposition} 
The time series defined in Equation (\ref{eq:pp1}) is weakly stationary if: 
(1) the invariant condition of Proposition 1 is satisfied with a 
stationary marginal for which the first two moments exist and are finite;
(2) the conditional expectation with respect to $f_{U_l\mid V_l}$ is
$E(U_l\,|\, V_l = y)=$ $a_l + b_l \, y$, for some $a_l, b_l\in\mathbb{R}$, 
and for all $l$; (3) Equation \eqref{eq:cross_mom} is independent 
of time $t$, and the roots of the equation 
$z^L - w_1b_1z^{L-1} - \dots - w_Lb_L = 0$ all lie inside
the unit circle.
\end{proposition}

Proposition 2 illustrates the construction of a weakly stationary MTD model 
building from the invariant condition of Proposition 1. We focus 
on first-order strictly stationary MTD models. Weak stationarity can be further 
studied if conditions (2) and (3) of Proposition 2 are satisfied. 

\section{Construction of first-order strictly stationary MTD models}\label{sec:construction} 

Here, we present two methods to develop first-order strictly stationary MTD models. 
The bivariate distribution method constructs the transition density given a 
specific marginal distribution. This method may result in analytically intractable 
transition densities. The second method, consisting of directly specifying the 
transition component conditional densities, has estimation advantages, although 
the analytical form of the marginal density may not be readily available. 
Thus, the selection among these methods depends on the modeling objectives. 
In fact, there are special cases where both the transition and marginal densities 
belong to the same family of distributions. 

\subsection{Bivariate distribution method}\label{sec:bivariate}

Under this method, we seek bivariate 
distributions $f_{U_l,V_l}$ whose marginals $f_{U_l}$ and 
$f_{V_l}$ are equal to a given $f_X$, for $l = 1,\dots,L$. 
Consequently, the $l$th transition component density is 
$f_{U_l\mid V_l}(u\,|\, v) =$ $f_{U_l,V_l}(u,v)/f_X(v)$. 
In contrast to the approach in \cite{mena2007stationary},
which is practical when the marginal density is a conjugate prior 
for some likelihood, the bivariate distribution method is applicable
to essentially any discrete or continuous marginal invariant density 
$f_X$. In fact, for most parametric families, there is a rich literature 
defining collections of bivariate distributions with a desired marginal 
distribution, and allowing for a variety of dependence structures. 
The following examples illustrate the method.

\paragraph{\normalfont\itshape{Example 1: Gaussian and continuous mixtures of Gaussians MTD models.}}
Under marginal $f_X(x) = N(x\,|\, \mu,\sigma^2)$, the Gaussian MTD model can be 
constructed via the bivariate Gaussian distribution for $(U_l,V_l)$, 
with mean $(\mu, \mu)^\top$ and covariance matrix 
$\Sigma =$ 
$\sigma^{2} \big(\begin{smallmatrix} 1 & \rho_l \\\rho_l & 1 \end{smallmatrix}\big)$, 
resulting in a Gaussian density for $f_{U_l\mid V_l}$. In particular,
\begin{align}\label{eq:GMTD1}
f(x_t\mid \bm{x}^{t-1}) = \sum_{l=1}^L w_l \, N\left(x_t\mid (1-\rho_l)\mu +
\rho_lx_{t-l}, \sigma^2(1-\rho_l^2)\right).
\end{align}

Let $t(x\,|\, \mu,\sigma,\nu)$
$\propto \left(1 + \nu^{-1}((x-\mu)/\sigma)^2\right)^{-(\nu+1)/2}$
denote the Student-t density, where $\mu$, $\sigma$ and $\nu$ are 
respectively location, scale and tail parameters. 
To construct as a natural extension of the Gaussian MTD model a 
stationary Student-t MTD model, consider the bivariate Student-t distribution,
which can be defined as a scale mixture of a bivariate Gaussian with
mean $(\mu, \mu)^\top$ and covariance matrix $q\Sigma$, with 
$\Sigma$ as previously defined, mixing on $q$ with respect to  
an inverse-gamma, $\mathrm{IG}(\nu/2,\nu/2)$, distribution. 
Under marginal $f_X(x) =$ $t(x\,|\, \mu,\sigma,\nu)$,
the Student-t MTD model is given by
\begin{align}
f(x_t\mid \bm{x}^{t-1}) = \sum_{l=1}^L w_l \, t\left(x_t\mid (1-\rho_l)\mu + \rho_lx_{t-l},
\sigma^2(1-\rho_l^2)(\nu+d_l)/(\nu+1), \nu+1\right),
\end{align}
where $d_l = (x_{t-l} - \mu)^2/\sigma^2$. 
In both the Gaussian and Student-t MTD examples, the transition component densities 
and the invariant density belong to the same family of distributions.

The Student-t MTD model is an example for building MTD models through bivariate 
distributions that admit a location-scale mixture representation. Taking 
an exponential distribution 
for the scale $q$ yields the bivariate Laplace distribution of 
\cite{eltoft2006multivariate}, thus producing an MTD model with an
invariant Laplace marginal density. 
Scaling both the mean $\mu$ and the covariance $\Sigma$ of the bivariate 
Gaussian distribution by a unit rate exponential random variable yields the 
bivariate asymmetric Laplace distribution of \cite{kotz2012laplace}, and 
thus an MTD model with an asymmetric Laplace distribution as the invariant marginal. 
We can further elaborate on this approach using appropriate mixing distributions for the Gaussian 
location and scale to obtain skewed-Gaussian and skewed-t distributions 
\citep{azzalini2013skew} for the bivariate component 
distributions, as well as for the invariant marginal distribution.

\paragraph{\normalfont\itshape{Example 2: Poisson and Poisson mixture MTD models.}}
To model time series of counts taking countably infinite
values, we can construct an MTD model 
with a Poisson marginal by considering the bivariate 
Poisson distribution of \cite{holgate1964estimation} for the 
transition components. This choice has been discussed in
\cite{berchtold2002mixture}, without addressing the stationarity condition.
In particular, we consider the latent variable representation of 
Holgate's bivariate Poisson.
Given a Poisson marginal $f_X(x) = \mathrm{Pois}(x\,|\,\phi)$,
we take $(U_l,V_l) \equiv$ $(U,V) =$ $(Q + Z, W + Z)$, 
for all $l$, where $Q$, $W$ and $Z$
are independent Poisson random variables with means 
$\lambda$, $\lambda$ and $\gamma$, respectively.
It follows that both $U$ and $V$ are Poisson random variables
with rate parameter $\phi = \lambda + \gamma$. Using the latent 
variable representation, the
$l$th component transition density of the Poisson MTD model 
can be sampled 
through $Q_t\sim \mathrm{Pois}(q_t\,|\,\lambda)$
and $Z_t\,|\, X_{t-l} = x_{t-l} \sim \mathrm{Bin}(z_t\,|\,
x_{t-l}, \gamma/\phi)$, with $X_t = Q_t + Z_t$
obtained as the realization from the $l$th component 
conditional distribution $X_t\,|\,X_{t-l} = x_{t-l}$. Here,
$\mathrm{Bin}(x\,|\, n,p) $ denotes the binomial distribution 
with $n$ trials and probability of success $p$. 

A common extension of the Poisson to account for counts that
have excess zeros is a mixture 
of Poisson and a distribution that degenerates at $0$.
A random variable $X$ is zero-inflated Poisson
distributed, denoted as $\mathrm{ZIP}(x\,|\, \phi, q)$, if its
distribution is a mixture of a point mass at zero
and a Poisson distribution with parameter
$\phi$, with respective probabilities $0<q<1$ and $(1-q)$.
Given an invariant marginal $f_X(x) = \mathrm{ZIP}(x\,|\, \phi, q)$,
we use the bivariate zero-inflated Poisson distribution of 
\cite{li1999multivariate} for $(U_l,V_l) \equiv (U,V)$, for all $l$, 
given by a mixture of a point mass at $(0,0)$, two univariate 
Poisson distributions, and a bivariate Poisson distribution; that is
$f_{U,V}(u,v) =$ $q_{0} (0,0) +  
0.5 q_{1}(\mathrm{Pois}(u\,|\, \phi), 0) +  
0.5 q_{1}(0,\mathrm{Pois}(v\,|\, \phi)) + 
q_{2} \mathrm{BP}(u,v\,|\, \phi,\phi)$,
where $\sum_{j=0}^2q_{j} = 1$, $q_0 + 0.5q_1 = q$, 
and $\mathrm{BP}(\cdot,\cdot\mid \phi,\phi)$ denotes Holgate's
bivariate Poisson distribution. Although the 
corresponding component density $f_{U\,|\, V}(u\,|\, v) = f_{U,V}(u,v)/f_X(v)$ 
is complex, this example provides possibilities for modeling 
stationary zero-inflated count time series.

Exploiting the latent variable representation of Holgate's bivariate Poisson, 
we can obtain extensions of the Poisson MTD model that allow for more 
flexible dependence structure and for overdispersion. Following the 
earlier notation, replace the means $\lambda$ and $\gamma$ of the latent 
Poisson random variables with $\alpha\lambda$ and $\alpha\gamma$, 
and mix over $\alpha$ with respect to a $\mathrm{Ga}(\alpha\,|\, k,\eta)$
distribution, where $\mathrm{Ga}(x\,|\, a,b)$ denotes the gamma distribution 
with mean $a/b$. Such mixing yields a bivariate negative binomial distribution  
after $\alpha$ is marginalized out \citep{kocherlakota2006bivariate}.
The conditional distribution of $U$ given $V = v$
admits a convolution representation. Let $Z_1$ and $Z_2$ be conditionally 
independent, given $V = v$, following a 
$\mathrm{Bin}\left(z_1 \,|\,  v, \gamma/(\lambda+\gamma)\right)$ 
and $\mathrm{NB}\left(z_2 \,|\,  k+v,1-\lambda/(2\lambda+\gamma+\eta)\right)$
distribution, respectively, where $\mathrm{NB}(x\,|\, r,p)$ denotes 
the negative binomial distribution with $r$ number of successes and 
probability of success $p$. Then, $U = Z_1 + Z_2$ is a realization 
from the conditional distribution $U \,|\,  V = v$.
Similar to the Poisson case, we can use this convolution representation 
to define a stationary MTD model with a negative binomial marginal 
$f_X(x) = \mathrm{NB}\left(x\,|\, k, \eta/(\lambda+\gamma+\eta)\right)$.

\paragraph{\normalfont\itshape{Example 3: Bernoulli and Binomial MTD models.}}
Assume again $(U_l,V_l) \equiv (U,V)$, for all $l$, and consider the 
bivariate Bernoulli distribution with probability mass function
$p(u,v)=$ $p_{1}^{uv}p_{2}^{u(1-v) + (1-u)v} (1-p_1-2p_2)^{(1-u)(1-v)}$,
where $p_1 > 0$, $p_2 > 0$ and $p_1+2p_2 < 1$.
Then, marginally $U$ and $V$ are both Bernoulli distributed with 
probability of success $p_1+p_2$. The conditional
distribution of $U$ given $V = v$ is also Bernoulli \citep{dai2013multivariate}
with probability of success $p(1,v)/\left(p(1,v) + p(0,v)\right)$.
Using this bivariate Bernoulli distribution, we define a 
stationary Bernoulli MTD model 
\begin{align}
f(x_t\mid \bm{x}^{t-1}) =
\sum_{l=1}^L w_l \, \mathrm{Ber}\left(x_t\mid p(1,x_{t-l})/(p(1,x_{t-l}) + 
p(0,x_{t-l}))\right),
\end{align}
which has a stationary marginal distribution $f_X(x) = \mathrm{Ber}(x\,|\, p_1+p_2)$.

Sequences of independent bivariate Bernoulli random vectors can be used 
as building blocks for various bivarate distributions. In particular, 
a family of bivariate binomial distributions for $(U,V)$ can be constructed
by setting $U = \sum_{i=1}^n \tilde{U}_i$ and $V = \sum_{i=1}^n \tilde{V}_i$,
where $(\tilde{U}_i,\tilde{V}_i),i=1,\dots,n$, are independent from the 
bivariate Bernoulli distribution given above \citep{kocherlakota2006bivariate}.
The conditional distribution of $U$ given $V = v$ 
can be defined through the convolution of two conditionally 
independent, given $V = v$, binomial random variables, one with parameters $n - v$ 
and $p_2/(1-p_1-p_2)$ and the other with parameters $v$ and $p_1/(p_1+p_2)$.
Again, this convolution representation can be used to define a stationary 
binomial MTD model with marginal $f_X(x) = \mathrm{Bin}(x\,|\, n,p_1+p_2)$.

Examples 2 and 3 illustrate MTD models for finite/infinite-range discrete-valued 
time series with high-order dependence, and with stationary marginal distributions 
belonging to a range of families. These can be used, for example, for 
classification of time series data, or for time-varying
counts that exhibit features such as overdispersion or excess of zero values
when compared to a traditional Poisson model.
It is worth mentioning that some of our examples induce non-linear conditional expectations.
For example, the conditional expectation of the Bernoulli MTD model is 
$\sum_{l=1}^Lw_lp(1,x_{t-l})/(p(1,x_{t-l}) + p(0,x_{t-l}))$. 
Building MTD models like the ones we have proposed using the existing methods in the MTD literature is a formidable task.  

\subsection{Conditional distribution method}\label{sec:ccd}

The strategy here is to use compatible conditional densities,
$f_{U_l\mid V_l}$ and $f_{V_l\mid U_l}$, to specify the bivariate density 
of $(U_l,V_l)$ for the $l$th transition component.
Conditional densities $f_{U\mid V}$ and $f_{V\mid U}$ are said to be compatible 
if there exists a bivariate density with its conditionals given by 
$f_{U\mid V}$ and $f_{V\mid U}$; see \cite{arnold1999conditional} for general 
conditions under which candidate families of two conditionals are 
compatible.

We begin with the assumption that $f_{U_l\mid V_l}$ and $f_{V_l\mid U_l}$ 
belong to the same family. This assumption is reasonable, since the 
invariant condition of Proposition 1 requires that all marginals
are the same. Once the family of distributions for the conditionals 
is chosen, we ensure the conditionals are compatible, as well as 
that both marginals of the corresponding bivariate density are given 
by the target invariant density $f_X$. In some special cases, the 
marginal densities are in the same family as the compatible conditionals. 
To demonstrate this method, 
we use a pair of Lomax conditionals and a pair of gamma conditionals;
both cases are considered in \cite{arnold1999conditional} to identify 
compatibility restrictions for their parameters.

\paragraph{\normalfont\itshape{Example 4: Lomax MTD models.}}
The Lomax distribution is a shifted version of the Pareto Type I distribution
such that it is supported on $\mathbb{R}^{+}$.
Denote by $P(x\,|\, \sigma, \alpha) =$
$\alpha \sigma^{-1} \left(1 + x \sigma^{-1} \right)^{-(\alpha+1)}$ the 
Lomax density, where $\alpha > 0$ is the shape parameter, and $\sigma > 0$
the scale parameter. The corresponding tail 
distribution function is $\mathrm{Pr}(X > x) = (1 + x\sigma^{-1})^{-\alpha}$, implying 
a polynomial tail that supports modeling for time series with high levels of 
skewness. We consider a pair of compatible Lomax densities for
$(U_l,V_l) \equiv (U,V)$, for all $l$, such that 
$f_{U\,|\,V}(u\,|\,v) =$ 
$P\left(u\,|\, (\lambda_{0} + \lambda_{1}v)/(\lambda_{1} + \lambda_{2}v), \alpha\right)$,
and $f_{V\,|\, U}(v\,|\, u) =$ 
$P\left(v\,|\, (\lambda_{0} + \lambda_{1}u)/(\lambda_{1} + \lambda_{2}u), \alpha \right)$,
with the restriction that
$\lambda_0,\lambda_1,\lambda_2>0$ if $\alpha = 1$, 
$\lambda_0\geq 0,\lambda_1,\lambda_2>0$ if $0<\alpha<1$, and
$\lambda_0,\lambda_1>0,\lambda_2\geq 0$ if $\alpha > 1$, to guarantee that 
these are proper densities.
Lomax MTD models specified using 
the conditional distributions above
have an invariant marginal 
$f_X(x) \propto(\lambda_1 + \lambda_2x)^{-1}(\lambda_0+\lambda_1x)^{-\alpha}$.
Taking $\alpha > 1$ and $\lambda_2 = 0$ leads to a special case where
both the component transition density and the marginal density 
are Lomax. This particular Lomax MTD model is 
\begin{align}\label{eq:lomaxMTD}
f(x_t\mid \bm{x}^{t-1}) & = \sum_{l=1}^L w_l \, P(x_t\mid \phi + x_{t-l}, \alpha),
\end{align}
where $\phi = \lambda_0/\lambda_1$,
and the invariant marginal is $f_X(x) = P(x\,|\, \phi, \alpha - 1)$.

\paragraph{\normalfont\itshape{Example 5: Gamma MTD models.}} 
We consider a pair of conditional gamma densities for the random vector 
$(U_l,V_l) \equiv (U,V)$, for all $l$, such that 
$f_{U\mid V}(u\,|\, v) =$ $\mathrm{Ga}(u\,|\, m_{0}, m_{1} + m_{2}v)$,
and $f_{V\mid U}(v\,|\, u) =$ $\mathrm{Ga}(v\,|\, m_{0}, m_{1} + m_{2}u)$,
where $m_0,m_1,m_2 > 0$. This pair of conditionals is one of
six choices discussed in \cite{arnold1999conditional} in the context of 
conditional gamma distributions that produce 
proper bivariate densities for $(U,V)$.
The resulting transition density is 
\begin{equation}\label{eq:gammaMTD}
f(x_t\mid \bm{x}^{t-1}) = \sum_{l=1}^L w_l \, \mathrm{Ga}(x_t\mid m_0, m_1 + m_2x_{t-l}),
\end{equation}
and the invariant marginal is $f_X(x) \propto x^{m_0-1}\exp(-m_1 x)(m_1 + m_2x)^{-m_0}$.

Examples 4 and 5 present two stationary MTD models with, respectively, polynomial and exponential 
tail behaviors. They provide alternatives to the existing MTD model literature for positive-valued 
time series, where the only model that has received attention is based on the Weibull distribution.
In addition, 
the general Lomax MTD model with $\lambda_2\neq0$ and the gamma MTD model have non-linear 
conditional expectations. 

\section{Bayesian implementation}

\subsection{Hierarchical model formulation}

Here, we outline an approach to perform posterior inference for the
general MTD model, using a likelihood 
that is conditional on the first $L$ observations of the time series 
realization $\{x_t\}_{t=1}^n$. We introduce a set of latent variables
$\{Z_t\}_{t=L+1}^n$ with $Z_t$ taking values in $\{1,\dots,L\}$ 
such that $p(z_t\,|\, \bm{w}) = \sum_{l=1}^Lw_l\delta_l(z_t)$,
where $\bm{w} = (w_1,\dots,w_L)^\top$, and 
$\delta_l(z_t) = 1$ if $z_t = l$ and 0 otherwise. 
Conditioning on the set of latent variables and 
the first $L$ observations, the
hierarchical representation of the model is:
\begin{equation}\label{eq:hierarchy}
\begin{aligned}
x_t\mid z_t, \bm{\theta} & \stackrel{ind.}{\sim}
f_{z_t}(x_t\mid x_{t-z_t},\bm{\theta}_{z_t}),\;\;z_t\mid \bm{w} \stackrel{i.i.d.}{\sim}
\sum_{l=1}^Lw_l\delta_l(z_t),\;\;t = L+1,\dots,n,\\
\bm{w}  & \sim \pi_{w}(\cdot),\;\;
\bm{\theta}_l\stackrel{ind.}{\sim}\pi_l(\cdot),\;\;l = 1,\dots,L,
\end{aligned}
\end{equation}
where $\bm{\theta}_l$ denotes the
transition component parameters, and 
$\bm{\theta}$ collects all $\bm{\theta}_l$. Any MCMC algorithm for 
finite mixture models is readily adoptable. 
If the transition density of the model is 
sampled via a latent process, such as 
for Example 2 of Section 3, an additional step to sample the 
latent variables needs to be added in Equation \eqref{eq:hierarchy}.

A key component of the 
Bayesian model formulation is the choice of the prior distribution for 
the mixture weights. 
As a point of reference, we consider a uniform Dirichlet prior 
that assumes equal contribution from each lag, denoted by 
$\mathrm{Dir}(\cdot\mid \bm1_L / L)$, where $\bm1_L$ is a unit vector 
of length $L$. We discuss next two priors that assume more structure.

The first prior is a truncated version of the 
stick-breaking prior, which characterizes the weights for random 
discrete distributions generated by the Dirichlet process
\citep{sethuraman1994constructive}.
More specifically, the weights are constructed as follows: 
$w_1 = \zeta_1$, $w_l = \zeta_l \prod_{r=1}^{l-1}(1 - \zeta_r)$, 
$l=2,\dots,L-1$, and $w_L = \prod_{l=1}^{L-1}(1 - \zeta_l)$, where 
$\zeta_l \stackrel{i.i.d.}{\sim} \text{Beta}(1,\alpha_{s})$, 
for $l = 1,\dots, L-1$. 
The resulting joint distribution for 
the mixture weights is a special case of the generalized Dirichlet 
distribution \citep{connor1969concepts}. We denote the truncated 
stick-breaking prior as $\mathrm{SB}(\cdot\,|\, \alpha_{s})$. 
For $l=1,...,L-1$, $E(w_l)=$ 
$\alpha_{s}^{*} (1 - \alpha_{s}^{*})^{l-1}$, where $\alpha_{s}^{*}=$
$(1+\alpha_{s})^{-1}$. Hence, on average, this prior implies geometrically 
decreasing weights, with smaller $\alpha_{s}$ values favoring stronger 
contributions from recent lags. In certain applications, it may be 
natural to expect some directionality in the relevance of the 
weights implied by time, and this prior provides one option to 
incorporate into the model such a property. 

An alternative prior is obtained by assuming that the weights are
increments of a cdf $G$ with
support on $[0, 1]$; that is, $w_l=$ $G(l/L) - G((l-1)/L)$, for 
$l = 1,\dots, L$. 
We place a Dirichlet process prior on $G$, denoted 
as $\text{DP}(\alpha_0, G_0)$, where $G_0 = \text{Beta}(a_0, b_0)$ 
and $\alpha_0 > 0$ is the precision parameter. 
From the Dirichlet process definition \citep{ferguson1973bayesian},
given $\alpha_0$ and $G_0$, the vector of mixture weights follows 
a Dirichlet distribution with shape parameter vector 
$\alpha_0 (a_1, \dots, a_L)^\top$,
where $a_l = G_0(l/L) - G_0((l-1)/L)$, for $l=1,\dots,L$.
We refer to this prior as the cdf-based prior, and denote 
it as $\mathrm{CDP}(\cdot\,|\,\alpha_0, a_0, b_0)$. 
Under this prior, we have that $E(\bm{w}) = (a_1,\dots,a_L)^\top$. 
The nonparametric prior for $G$ supports general distributional 
shapes, and thus allows for flexibility in the estimation of the 
mixture weights.
In particular, multimodal distributions $G$ can produce sparse 
weight vectors, with some/several entries near zero. Hence, this 
prior may be suitable for scenarios where there are inactive
lags between influential lags and the influential lags are not 
necessarily the most recent lags. \cite{heiner2019spv} proposed 
a different prior for sparse probability vectors, which generally 
requires a larger number of prior hyperparameters. 

Overall, the properties of both structured priors support flexible 
inference for the mixture weights, enabling our strategy to specify
a large value of $L$, 
assigning a priori small probabilities to distant lags. The 
contribution of each lag will be induced by the mixing, with 
important lags being assigned large weights a posteriori.

\subsection{Estimation, model checking and prediction}

The posterior distribution of the model parameters, based on 
the conditional likelihood, is 
\begin{align}\label{eq:likhod2}
p(\bm{w},\bm{\theta},\{z_t\}_{t=L+1}^n\mid D_{n} ) \propto
\pi_{w}(\bm{w})\prod_{l=1}^L\pi_{l}(\bm{\theta}_l)\prod_{t=L+1}^n
\left\{f_{z_t}(x_t\mid x_{t-z_t},\bm{\theta}_{z_t})\sum_{l=1}^Lw_l\delta_l(z_t)\right\}
\end{align}
where $D_n = \{x_t\}_{t=L+1}^n$,
and it can be explored using MCMC posterior simulation.

Conditional on $\bm{\theta}$ and $\bm{w}$, the posterior full conditional of 
each $Z_t$ is a discrete distribution 
on $\{ 1,...,L \}$ with probabilities proportional to 
$w_l f_l(x_t\,|\, x_{t-l},\bm{\theta}_l)$.
Conditional on the latent variables and
$w$, the sampling for each $\theta_l$ depends on the particular 
choice of the transition component distributions. 
Details for the models implemented are given in the Supplementary Material.
The sampling for $\bm{w}$, conditional on $\{z_t\}_{t=L+1}^n$ and
$\bm{\theta}$, depends only on $M_l = |\{t:z_t = l\}|$, for $l=1,...,L$, 
where $|\{\cdot\}|$ is the cardinality of the set $\{\cdot\}$. Both
priors for the mixture weights result in ready updates. 
The posterior full conditional of $\bm{w}$ under the truncated stick-breaking 
prior can be sampled through latent variables $\zeta^{*}_{l}$, which are
conditionally independent 
$\mathrm{Beta}(1 + M_l, \alpha_{s} + \sum_{r=l+1}^LM_r)$, for 
$l = 1,\dots,L-1$, such that $w_1 = \zeta^*_1$, 
$w_l = \zeta^*_l \prod_{r=1}^{l-1}(1 - \zeta^*_r)$, for $l=2,\dots,L-1$,
and $w_L = \prod_{l=1}^{L-1} (1 - \zeta^*_l)$. 
Under the cdf-based prior, the posterior full conditional of $\bm{w}$ is 
Dirichlet with parameter vector 
$(\alpha_0 a_1 + M_1, \dots, \alpha_0 a_L + M_L)^\top$. 

We assess the model's validity using randomized quantile residuals 
\citep{dunn1996randomized,escarela2006flexible}. 
Such residuals are calculated by
inverting the fitted conditional cdf
for the time series. Posterior samples of
these quantile sets can then be compared with the standard Gaussian
distribution, providing a measure of goodness-of-fit with
uncertainty quantification. 
Specifically, the randomized quantile residual for continuous $x_t$ 
is defined as $r_t = \Phi^{-1}\left(F(x_t\,|\, \bm{x}^{t-1})\right)$
where $\Phi(\cdot)$ is the cdf of the
standard Gaussian distribution. If $x_t$ is discrete, $r_t =
\Phi^{-1}(u_t)$, where $u_t$ is generated from a uniform
distribution on the interval $(a_t,b_t)$ with $a_t =
F(x_t-1\,|\, \bm{x}^{t-1})$ and $b_t = F(x_t\,|\, \bm{x}^{t-1})$. If $F$ is correctly
specified, the residuals $r_t$, $t = L+1,\dots,n$, will be independently 
and identically distributed as a standard Gaussian distribution.

Finally, we consider prediction for future observations. 
The posterior predictive density of $X_{n+1}$, corresponding to the 
first out-of-sample observation, is obtained by marginalizing the
transition density with respect to the posterior distribution of
model parameters:
\begin{align}\label{eq:1stepPP}
p(x_{n+1}\mid D_n) =
\int \int 
\left\{ \sum_{l=1}^L w_l \, f_l(x_{n+1} \mid x_{n+1-l},\bm{\theta}_{l}) \right\}
\, p(\bm{\theta},\bm{w}\mid D_n) \, d\bm{\theta}d\bm{w}.
\end{align}
Exploiting the structure of the conditional distributions of the 
MTD model, we can sample from the 
$k$-step-ahead posterior predictive density using a straightforward 
extension of Equation~(\ref{eq:1stepPP}). 
Note that the $k$-step-ahead posterior predictive uncertainty incorporates
both the uncertainty from the parameter estimation, and the uncertainty 
from the predictions of the previous $(k-1)$ out-of-sample observations.

\section{Data illustrations}

\subsection{Simulation example}

We generated $2000$ observations from the Gaussian MTD model 
specified in Equation~(\ref{eq:GMTD1}) with $\mu = 10, 
\sigma^2 = 100$, under two scenarios for the mixture weights, 
one with exponentially decreasing weights and the other one 
with an uneven arrangement of the relevant
lags.  In Scenario 1, we took $\bm{\rho} =
(0.7, 0.3, 0.1, 0.05, 0.05)^\top$ and $w_i \propto \exp(-i), i = 1,\dots,5$. 
In Scenario 2, we took $\bm{\rho} = (0.4, 0.1, 0.7, 0.1, 0.5)^\top$ and 
$\bm{w} = (0.2, 0.05, 0.45, 0.05, 0.25)^\top$. 
We consider these two scenarios to examine the effectiveness of 
structured priors for the mixture weights.

We applied the Gaussian MTD model with three different orders 
$L = 5, 15, 25$. In each case, we considered
three priors for the weights: the Dirichlet prior, the
truncated stick-breaking prior, and the cdf-based prior.
The shape parameter of the Dirichlet prior was $\bm1_L/L$ for each $L$.
The precision parameter $\alpha_s$ for the truncated stick-breaking prior
was taken to be $1, 2, 3$, corresponding to the three $L$ values.
For the cdf-based prior, we chose 
$\alpha_0 = 5$ as the precision parameter, and used as base distribution 
a beta with shape parameter $a_0 = 1$, and $b_0 = 3, 6, 7$ respectively
for the three orders considered. Thus, this prior elicited
a decreasing pattern similar to the truncated stick-breaking prior.
For all models, the mean $\mu$ and the variance $\sigma^2$ received
conjugate priors $N(\mu\,|\, 0, 100)$ and $\mathrm{IG}(\sigma^2\,|\, 2, 0.1)$,
respectively, and the component-specific correlation coefficient
$\rho_l$ was assigned a uniform prior $\mathrm{Unif}(-1,1)$
independently for all $l$.

\begin{figure*}[t!]
    \centering
    \includegraphics[width=.32\textwidth]{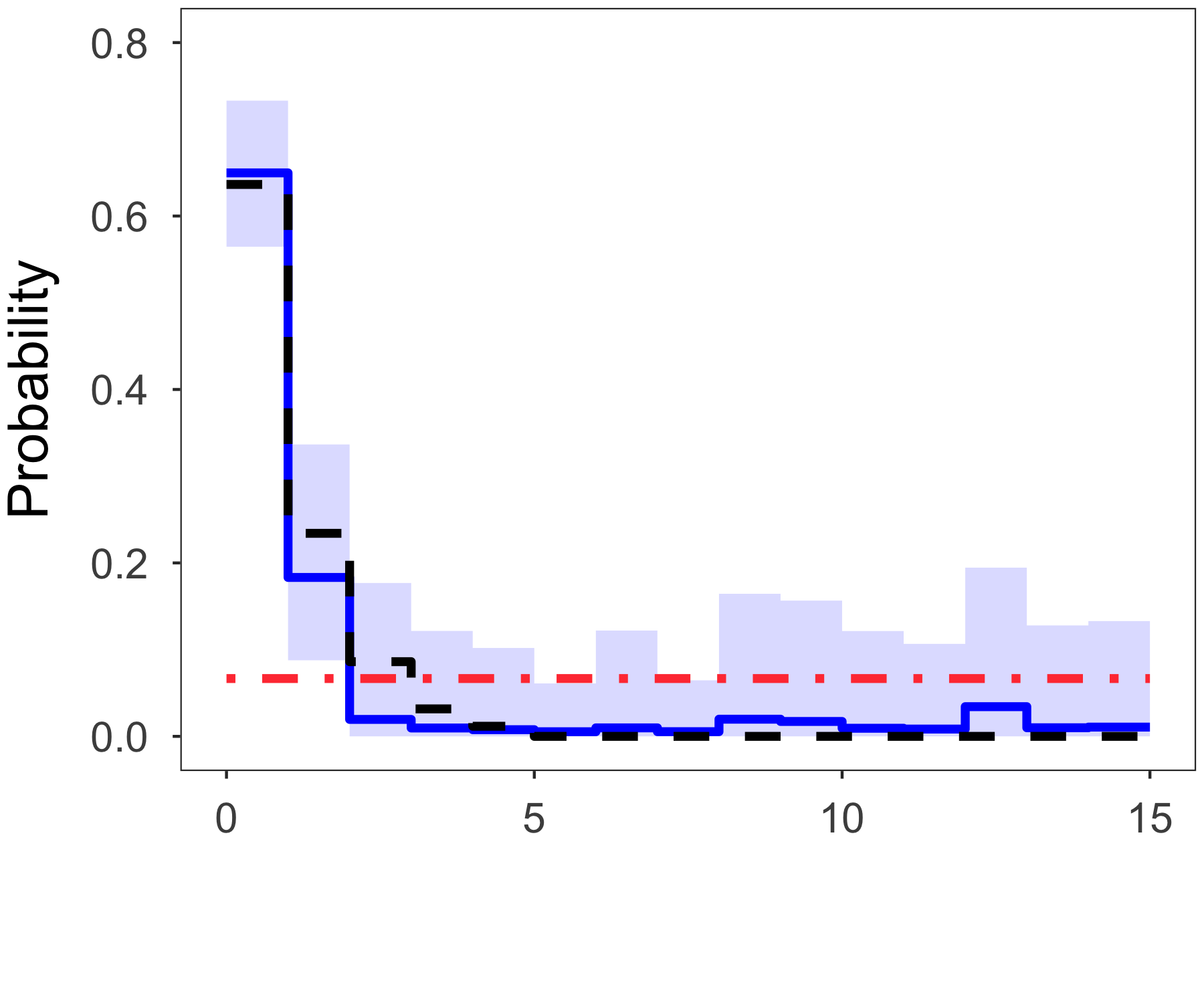}
    \includegraphics[width=.32\textwidth]{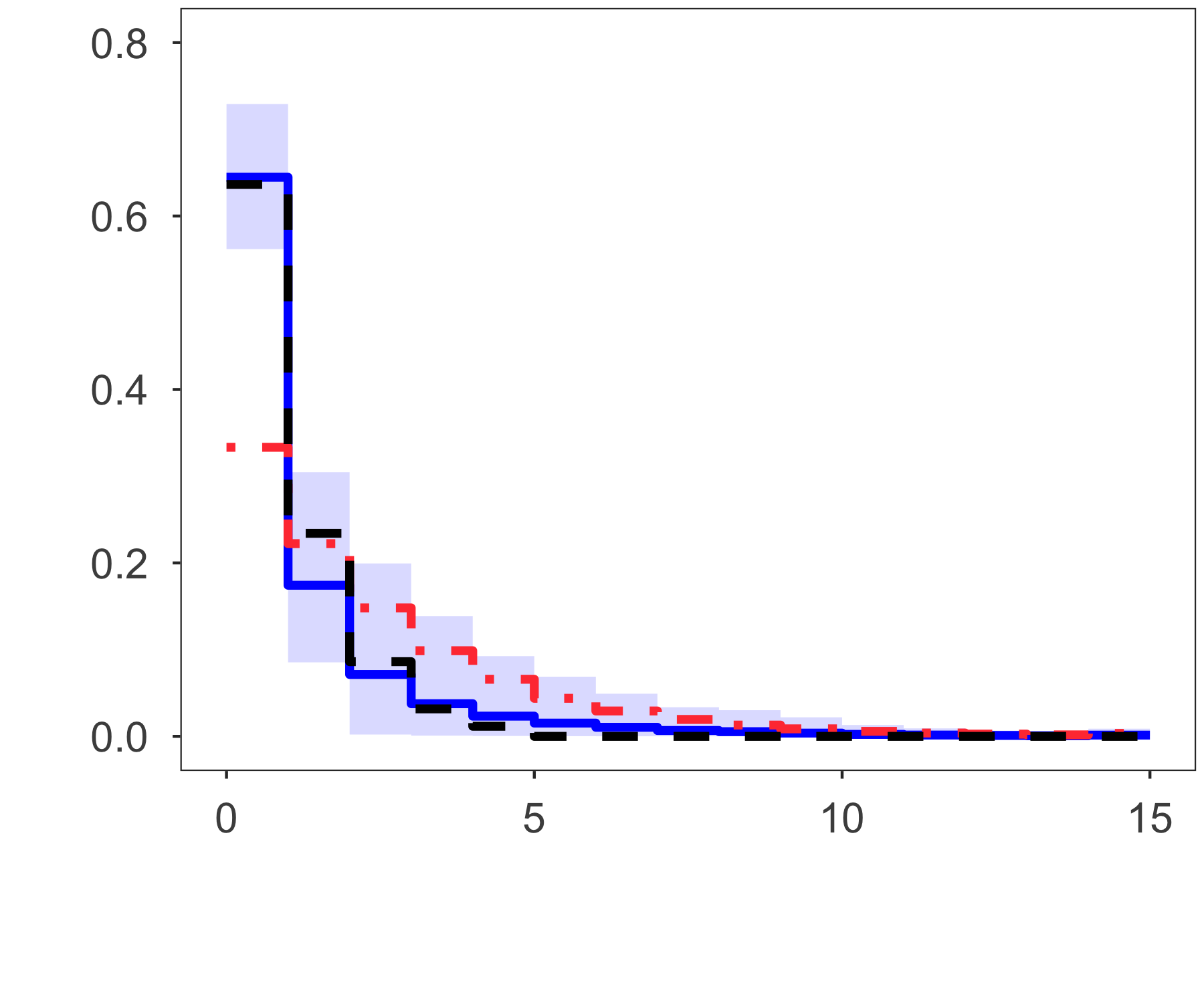}
    \includegraphics[width=.32\textwidth]{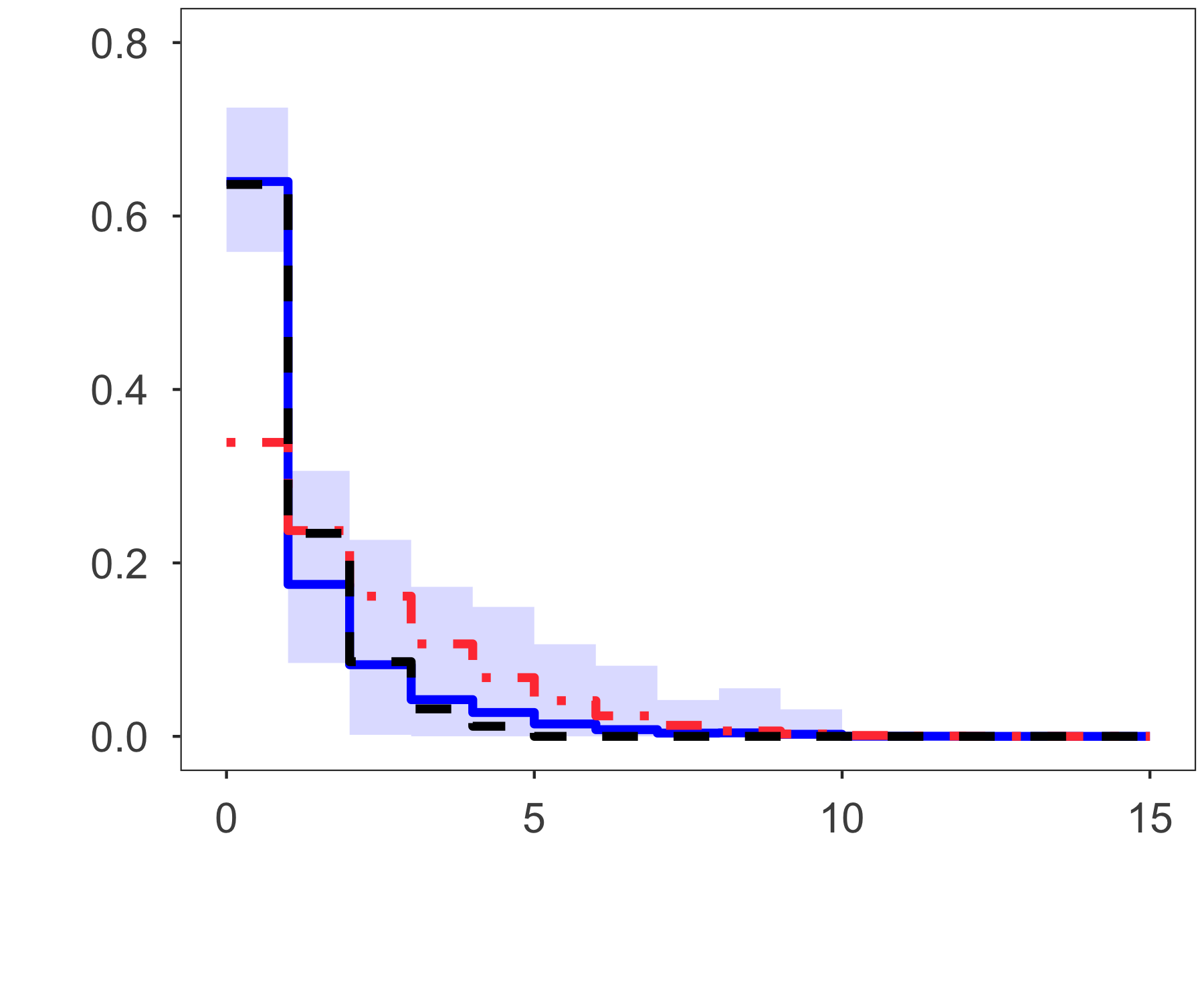}
    \medskip
    \includegraphics[width=.32\textwidth]{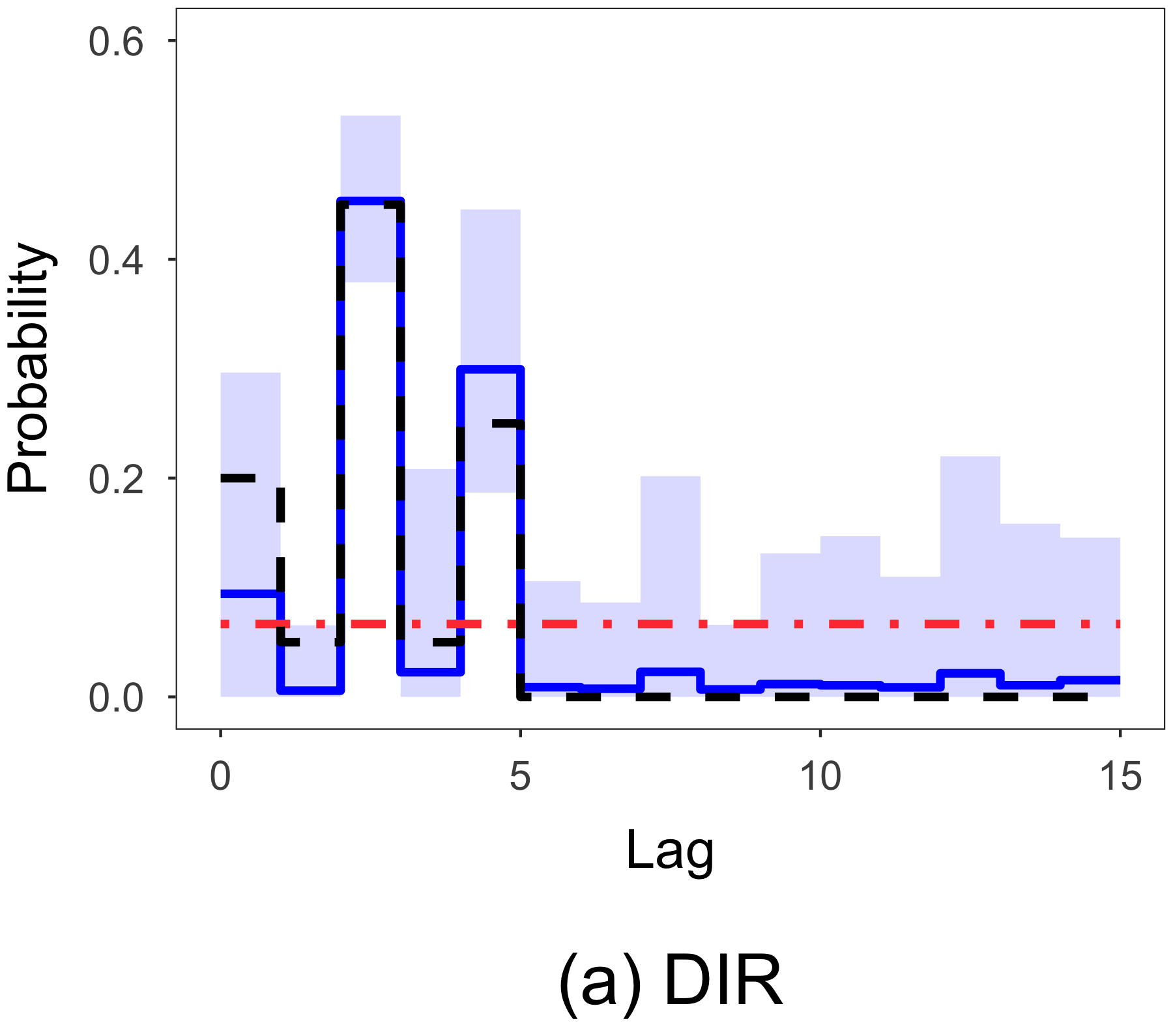}
    \includegraphics[width=.32\textwidth]{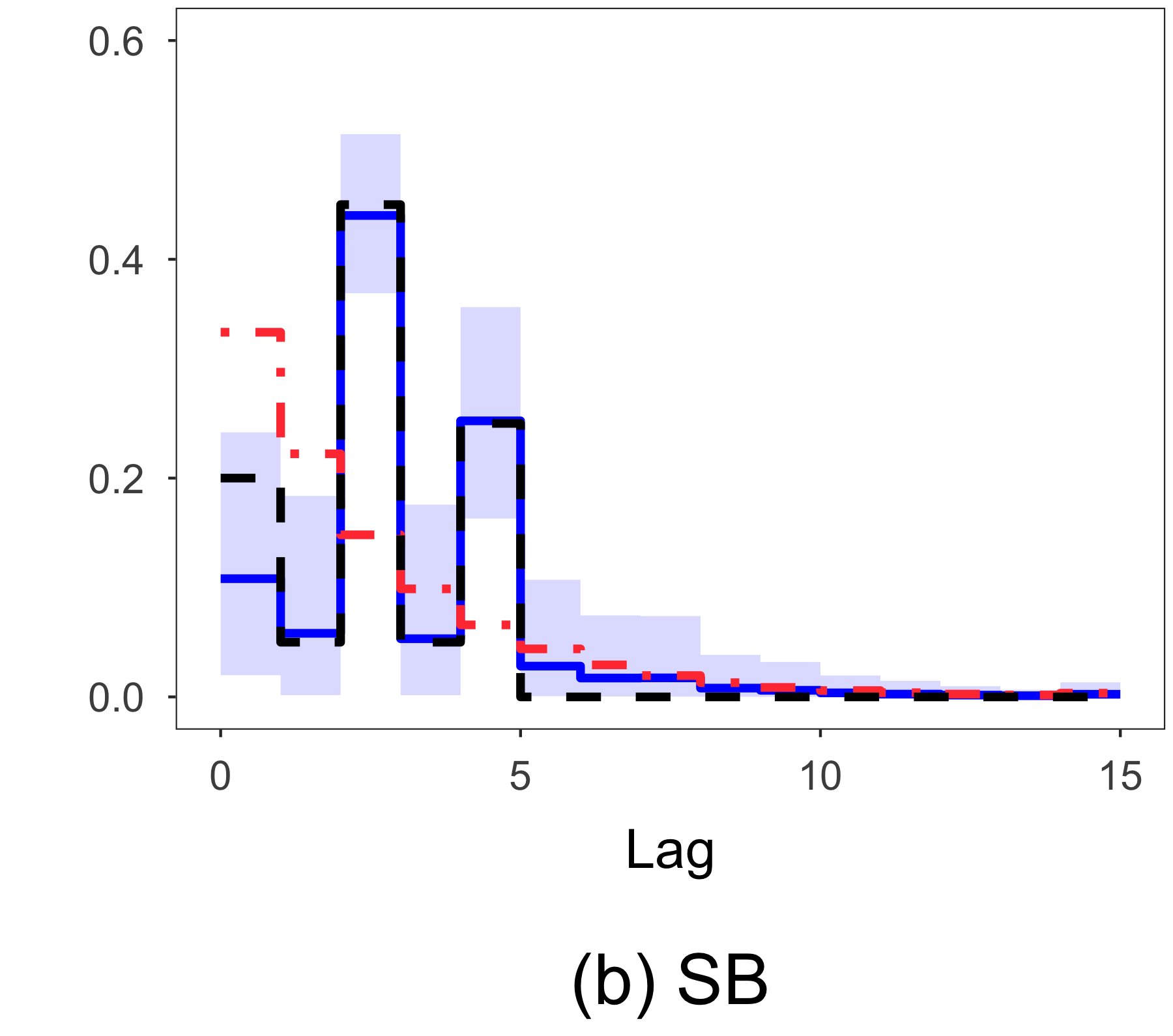}
    \includegraphics[width=.32\textwidth]{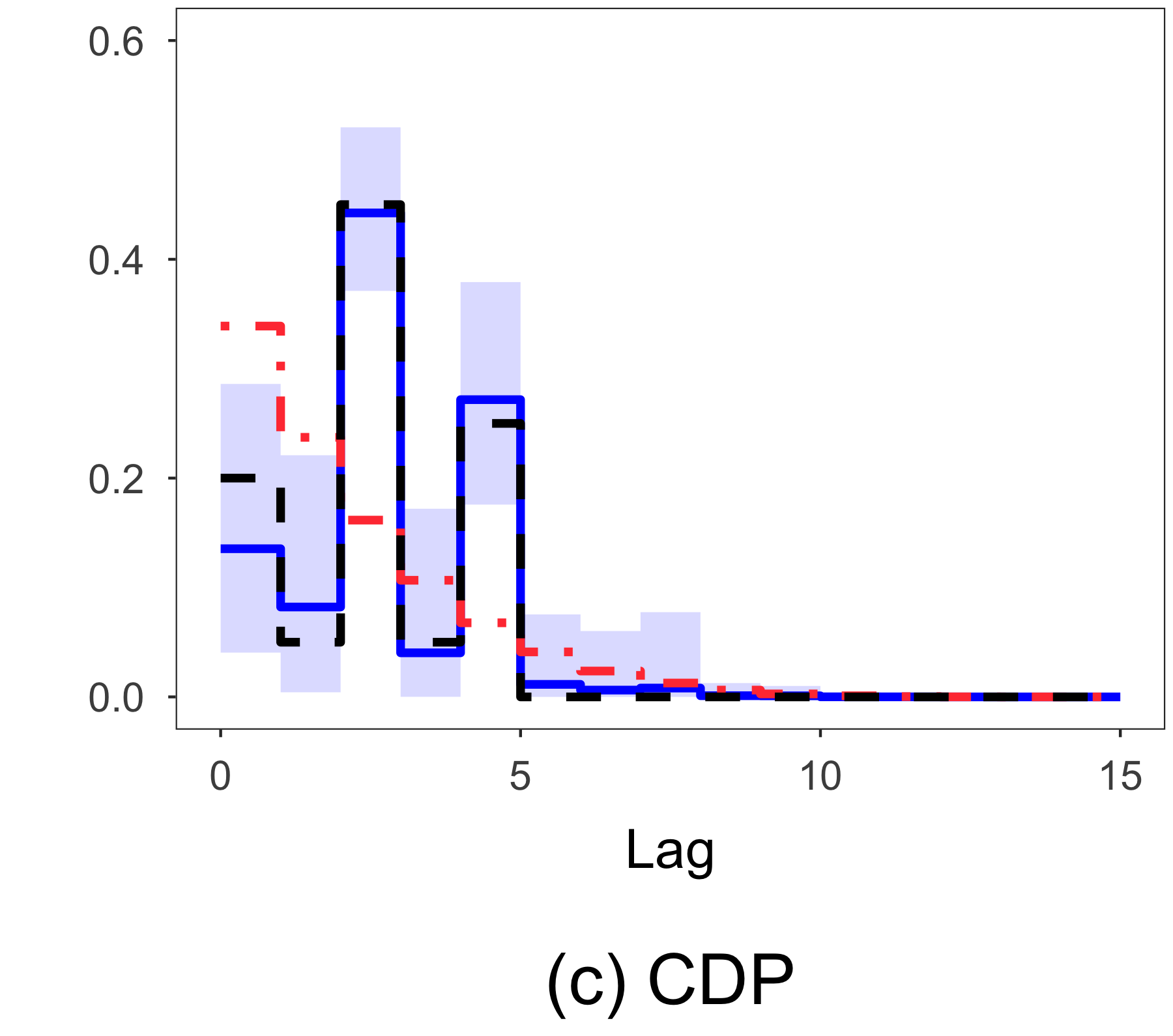}
    \caption{
    Simulation study. Inference results for the weights under 
    Scenarios 1 (top) and 2 (bottom), based on the Gaussian MTD model ($L=15$) 
    with the Dirichlet (column (a)), the truncated stick-breaking (column (b)), 
    and the cdf-based (column (c)) priors.
    Dashed lines are true weights, dot-dashed lines 
    are prior means, solid lines are posterior means, and polygons are 95\%
    posterior credible intervals.}
    \label{fig:s1}
\end{figure*}

We ran the Gibbs sampler for 165000 iterations, discarding the first 
5000 samples as burn-in, and collected samples every 20 iterations.
Focusing on inference results for the mixture weights, when the order 
was correctly specified, that is, $L = 5$, all three models provided 
good estimates.
Figure~\ref{fig:s1} provides a visual inspection on the posterior 
estimates for the mixture weights when $L = 15$ (the weight patterns 
estimated from the three models were similar when $L = 25$).
In Scenario 1, all models underestimated the weight for lag 2. 
Models with the proposed priors produced accurate estimates for the 
rest of the lags, while the model that used the Dirichlet prior 
systematically overestimated
the weight for the first lag, and underestimated all other
weights. In Scenario 2, all models underestimated the weight
for the first lag. For the other non-zero weights, the model with the
Dirichlet prior tended to underestimate the weights for lag 2, 4 
and overestimated the weight for lag 5, while the other two models 
estimated the weights quite well. 
In both scenarios, the proposed priors had a parsimonious behavior in 
that, given the data, distant lags were assigned almost zero probability 
mass with low posterior uncertainty. Overall, we note that, under an 
over-specified order $L$, the proposed priors offer inferential advantages 
when compared to the Dirichlet prior.

We conducted an additional simulation to demonstrate
the ability of the negative binomial MTD model to accommodate over-dispersed 
count data, including comparison with the Poisson MTD model. Details of this 
simulation example are presented in the Supplementary Material.

\subsection{Chicago crime data}

The first real data example involves the 1090 daily reported incidents of
domestic-related theft that have occurred in Chicago from 2015 to 2017, 
extracted online from the Chicago Data Portal
(https://data.cityofchicago.org/). The data exhibits some flat
stretches, without evidence of overdispersion.
The empirical mean and variance are 6.05 and 6.39.

We applied the Poisson MTD model 
discussed in Example 2 of Section 3, with order $L = 20$,
selected based on the autocorrelation and partial autocorrelation functions. 
We reparameterize the model in terms of rate parameter
$\lambda$, and binomial probability $\theta = \gamma/\phi$ 
for $Z_t\,|\,X_{t-l}$. This allows Gibbs updates for $\lambda$
and $\theta$ with posterior full conditionals available in closed form. 
The prior for $(\lambda,\theta)$ was taken to be 
$\mathrm{Ga}(\lambda\,|\,2,1)\mathrm{Beta}(\theta\,|\,2,2)$, implying a 
$\mathrm{Ga}(4,1)$ prior for $\phi$.
Two priors, SB($w\,|\,2$) and CDP($w\,|\,5, 1, 8$), were considered for the mixture weights.
Both models were fitted to the entire data set. After fitting the model, we
obtained the one-step posterior predictive distribution at each
time $t$ and the corresponding posterior predictive intervals.

\begin{figure*}[t!]
    \centering
    \includegraphics[width=.96\textwidth]{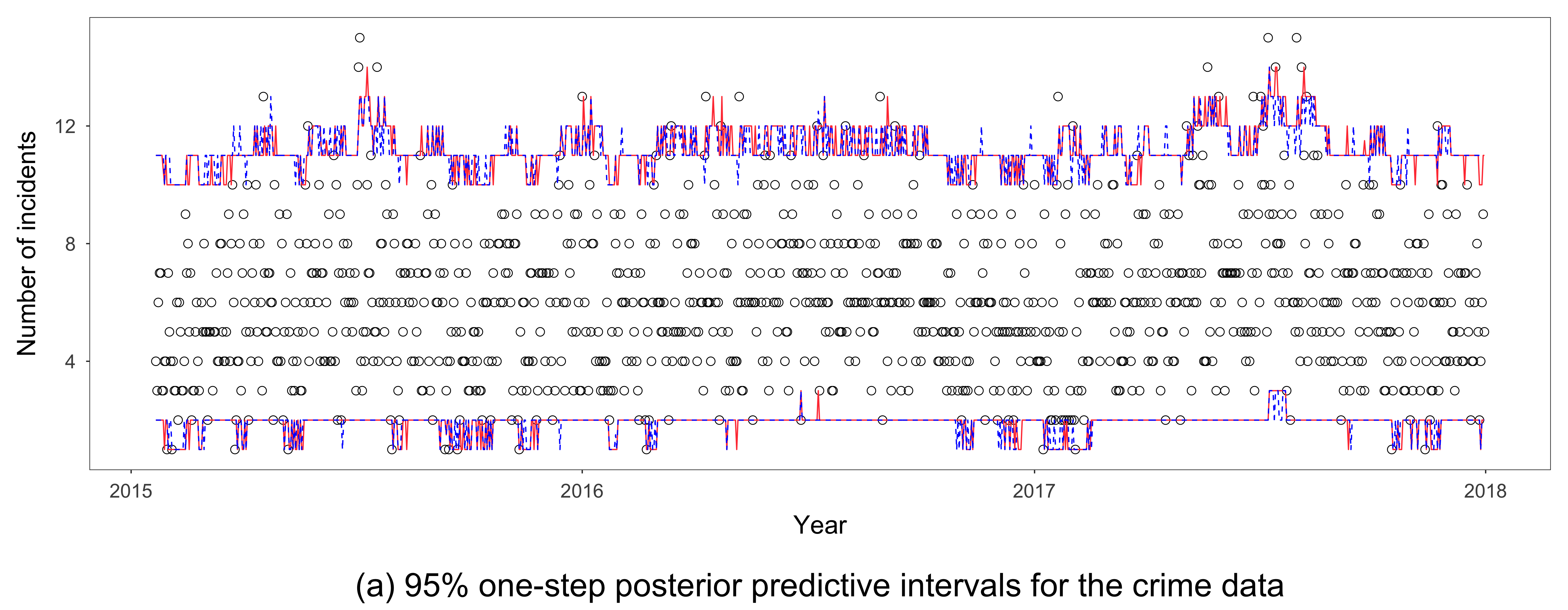}
    \medskip
    \vspace{5pt}
    \includegraphics[width=.48\textwidth]{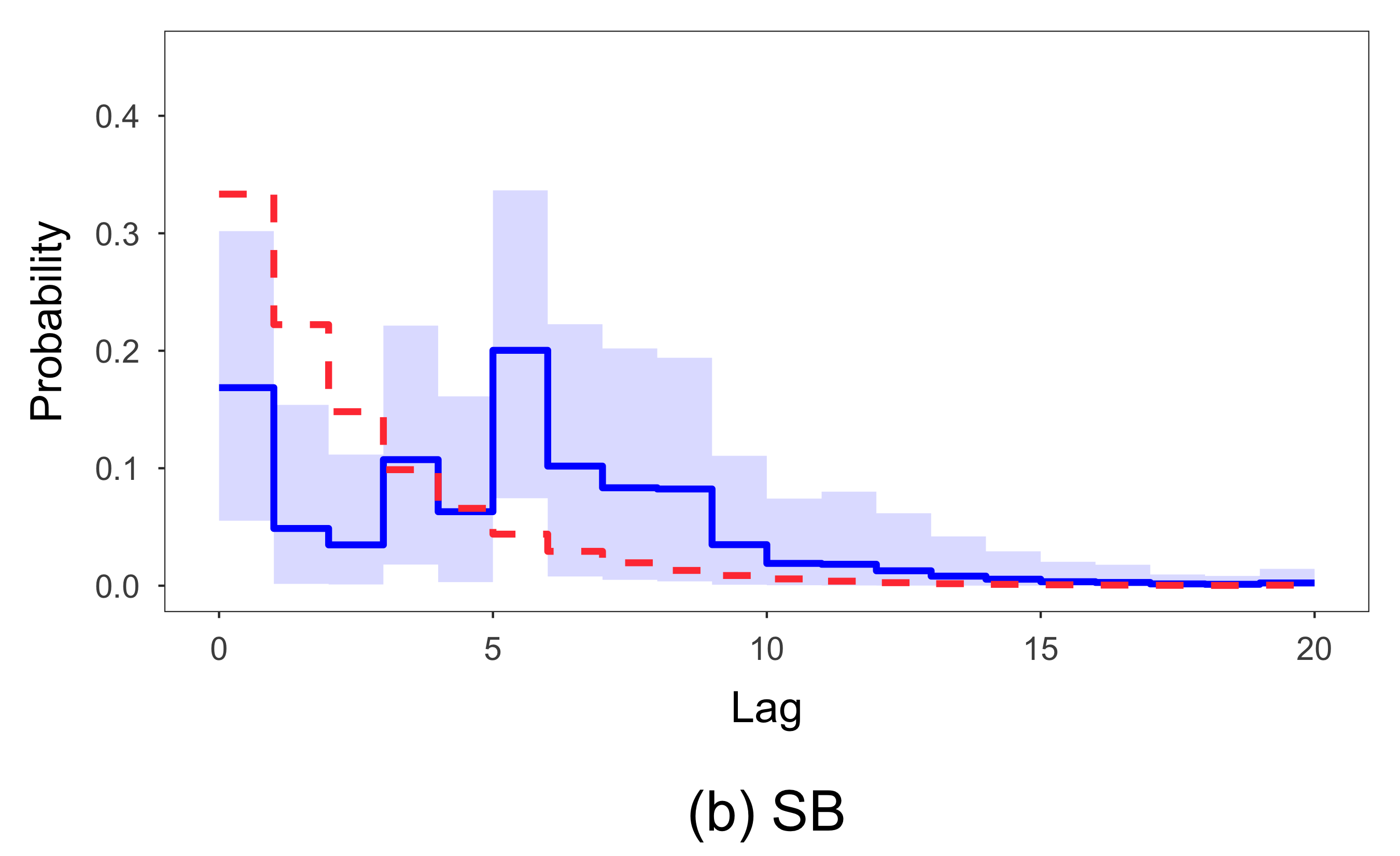}
    \includegraphics[width=.48\textwidth]{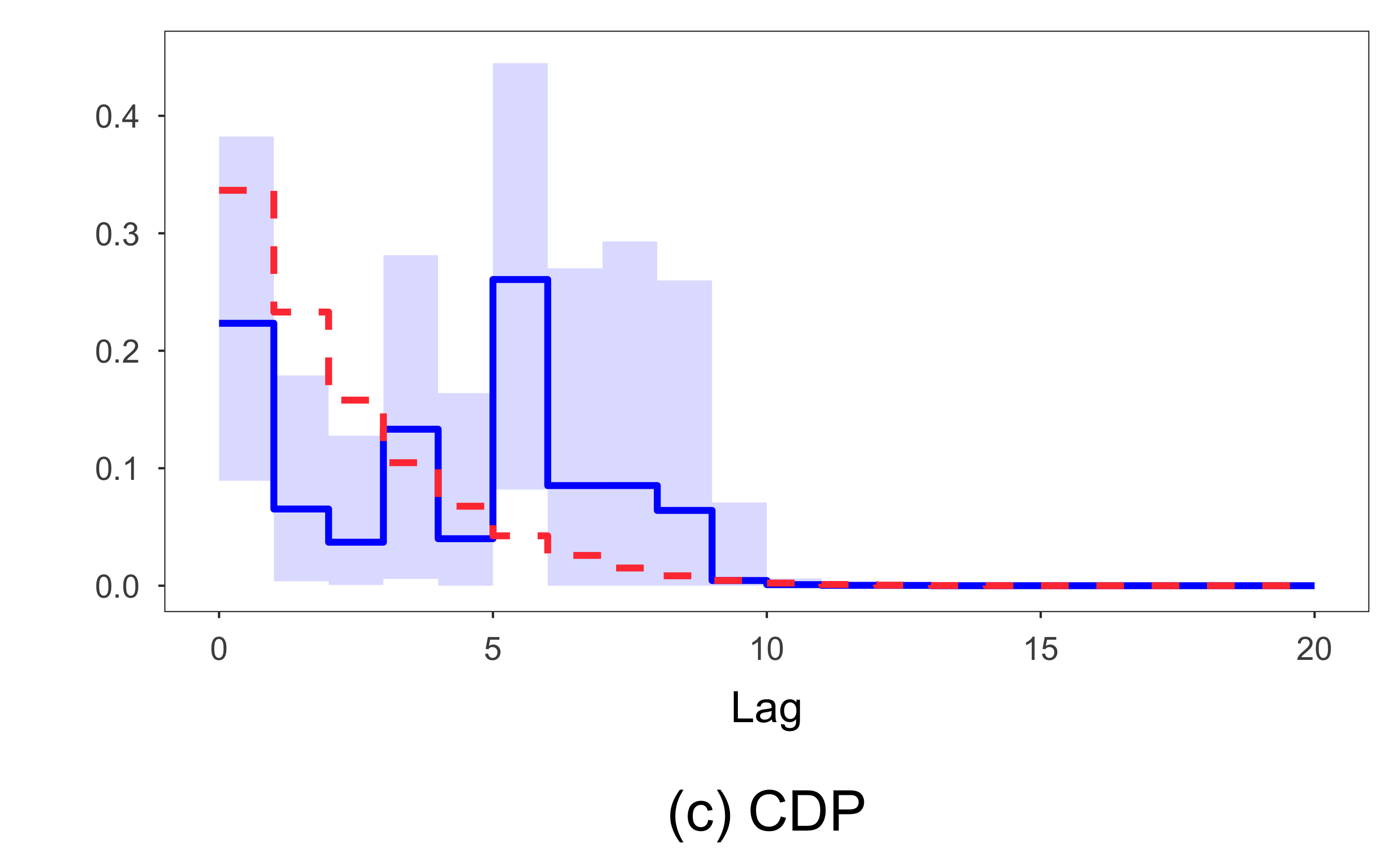}
    \caption{Chicago crime data analysis. 
    In panel (a), the circles denote the data, and solid and dashed lines correspond to 
    the model with the SB and CDP prior, respectively.
    Panels (b) and (c): prior means (dashed line), posterior means (solid line) 
    and 95\% credible intervals (polygon) of the weights under the SB and CDP prior, respectively.}
    \label{fig:chi_crime}
\end{figure*}

We obtained a thinned sample retaining every 10th iteration, from 
a total of 85000 samples with the first 5000 as burn-in.
The posterior mean and 95\% interval for 
$\phi$ are $6.04\,(5.79, 6.30)$ and $6.05\,(5.82, 6.29)$ for models with
SB($\bm{w}\,|\,2$) and CDP($\bm{w}\,|\,5, 1, 8$) priors.
This indicates an average of around six incidents of domestic-related theft per day.
Multiple influential lags, with gaps in between, are suggested by the results in 
Fig.~\ref{fig:chi_crime}(b)-\ref{fig:chi_crime}(c). 
Both models agree on the pattern for the weights, as well as on lags 1, 4, 6 
being the most relevant ones. 
Compared to the truncated stick-breaking prior, 
the cdf-based prior suggests a weight pattern that decreases slightly faster, 
and it assigns relatively larger weights to important lags, albeit with 
higher uncertainty.
Figure ~\ref{fig:chi_crime}(a) shows that both models produce 
similar one-step predictive intervals. 

\begin{figure*}[t!]
    \centering
    \includegraphics[width=.96\textwidth]{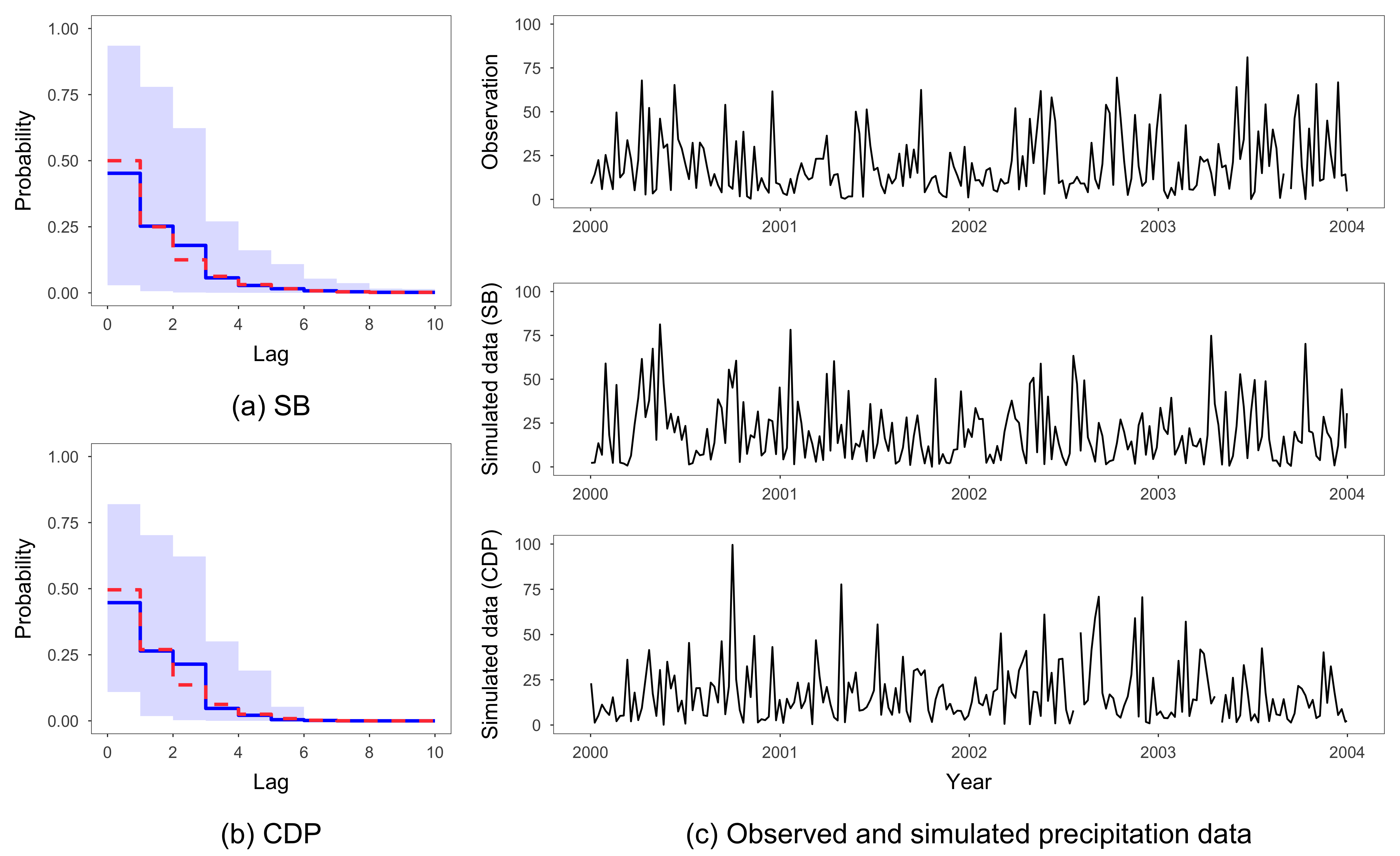}
    \caption{Precipitation data analysis.
    Panels (a) and (b): prior means (dashed line), posterior means (solid line) 
    and 95\% intervals (polygons) of the weights under two priors. The
    top row of panel (c) plots the observed precipitation amounts from 2000
    to 2004, and the middle and bottom rows show sample paths generated from 
    the fitted models with SB and CDP priors, respectively.}
    \label{fig:LMTDsp}
\end{figure*}

\subsection{Tunkhannock Creek precipitation data}

Our second example involves 22 years of rainfall data from January 1982 
to December 2003.
The data consists of 1149 mean areal precipitation amounts ranging from 
0.01 to 128.87 millimeters, aggregated to a weekly time scale 
from the daily data for the Tunkhannock Creek near Tunkhannock, Pennsylvania.
The data was extracted through R package hddtools \citep{hddtools}.

We consider a multiplicative model $y_t = \mu_t\epsilon_t$, where $\mu_t$ is a 
seasonal factor and $\epsilon_t$ is 
generated by a Lomax MTD model specified in Equation~(\ref{eq:lomaxMTD}),
with polynomial tails that can accommodate large precipitation events. 
More specifically, the model is given by
\begin{equation}\label{eq:SLMTD}
\begin{aligned}
y_t & = \mu_t\epsilon_t,\;\;
\mu_t = \exp(\bm{x}_t^\top\bm{\beta}),\;\;t = 1,\dots, n,\\
\epsilon_t\mid \bm{\epsilon}^{t-1},w, \phi, \alpha & \sim \sum_{l=1}^Lw_l
\, P(\epsilon_t\mid \phi + \epsilon_{t-l}, \alpha),\;\;t = L+1,\dots, n,
\end{aligned}
\end{equation}
with $\bm{x}_t = \left(\cos(\omega t), \sin(\omega t), \cos(2\omega t),
\sin(2\omega t), \cos(3\omega t), \sin(3\omega t)\right)^\top$ and $\omega =
2\pi/T$ where $T = 52$ is the period for weekly data.
On the basis of the autocorrelation and partial
autocorrelation functions, we chose model order $L = 10$. The
regression coefficients vector $\bm{\beta}=(\beta_1,\dots,\beta_6)^\top$
was assigned a flat prior. The shape parameter $\alpha$ was assigned a 
$\mathrm{Ga}(\alpha\,|\,6,1)$ prior, and the scale parameter $\phi$ 
an $\mathrm{IG}(\phi\,|\,3,20)$ prior.
Note that the invariant marginal of the process $\{\epsilon_t\}$
is $P(\epsilon\,|\,\phi,\alpha-1)$ and its tail distribution function is 
$(1 + \epsilon/\phi)^{-(\alpha-1)}$. A small value of $\alpha$ indicates a heavy tail,
while a large value of $\alpha$ ensures the existence of finite high moments. 
Under the priors above, $E(\alpha) = 6$, implying the expectation that the first 
four moments are finite with respect to both the component and marginal distributions
of the Lomax MTD for $\{\epsilon_t\}$. We fit the model with
$\mathrm{SB}(\bm{w}\,|\,1)$ and $\mathrm{CDP}(\bm{w}\,|\,5,1 ,6.5)$ 
priors for the weights.

\begin{figure*}[t!]
    \centering
    \includegraphics[width=.96\textwidth]{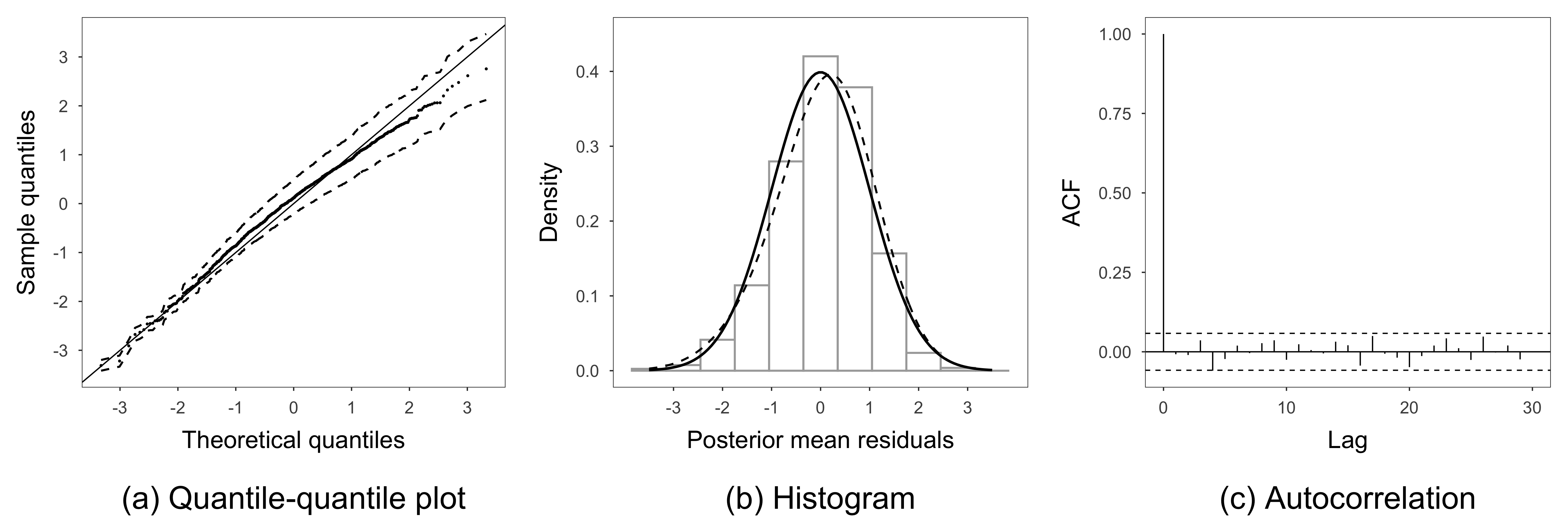}
    \caption{Precipitation data analysis.
    Randomized quantile residual analysis for 
    the fitted model with the SB($w\,|\,1$) prior. 
    In panel (a), the circles and dashed lines correspond to the posterior mean 
    and 95\% interval bands, respectively. 
    In panel (b), the solid and dashed line are the standard Gaussian density
    and the kernel density estimate of the posterior means of the residuals, respectively. 
    Panel (c) is based on the posterior means of the residuals.}
    \label{fig:mc}
\end{figure*}

We ran the algorithm for 85000 iterations
and collected samples every 10 iterations after the first 5000 was discarded.
The inference results were almost the same for the two models. 
Here we report the ones under the SB$(\bm{w}\,|\,1)$ prior.
The posterior mean and 95\% credible interval of 
the shape parameter $\alpha$ are $14.80\,(10.30, 20.91)$, 
indicating a moderately heavy tail.
The corresponding estimates for the scale parameter $\phi$ 
are $254.33\,(166.36, 370.04)$, indicating substantial dispersion.  
Among the harmonic component coefficients, the first and the fourth 
have 95\% posterior credible intervals that indicate statistical
significance; the estimates are 
$-0.14 \, (-0.23, -0.05)$ for $\beta_1$, and 
$-0.13 \,(-0.22, -0.03)$ for $\beta_4$,
implying the presence of semiannual and annual seasonality in
the data. 
Figure \ref{fig:LMTDsp}(a)-\ref{fig:LMTDsp}(b) show 
that both models suggest a decreasing weight pattern,
with the first three lags being the most influential.
As shown in Fig. \ref{fig:LMTDsp}(c),
the sample paths generated from the models
resemble the observed precipitation time series.

Randomized quantile residual analysis results were similar 
for both models; in Fig.~\ref{fig:mc}, we present the ones 
under the SB$(\bm{w}\,|\,1)$ prior. 
The figure shows posterior mean 
and interval estimates for the Gaussian quantile-quantile plot, 
and the histogram and autocorrelation function for the posterior means 
of the residuals. The results suggest reasonably good model fit, providing 
an illustration of the flexibility of the proposed 
MTD model to capture non-Gaussian tails.

\section{Discussion}

We have developed a broad class of stationary MTD models focusing on 
attaining stationarity from the
perspective of a distributional formulation. The advantage of our
proposed approach over more traditional methods is that no constraints
on the parameter space are needed. This facilitates inference for
model parameters, as the need for constrained optimization or
sampling is avoided. We further proposed structured
priors to support flexible inference on the weights, which accommodate 
non-standard scenarios that a model with a Dirichlet prior may fail to capture.

The proposed constructive framework brings several 
options for alternative parametric families that were formidable 
to tackle for the MTD model and its extensions, 
when stationarity is a desirable property. 
A limitation of our approach is that, if the stationary marginal
distribution shares all the parameters with the bivariate component distribution,
the resulting transition component lacks component-varying parameters.
One solution is to specify the bivariate distribution using a copula \citep{joe2014dependence},
which we regard as a special case of the bivariate distribution method. 
Given a pre-specified marginal, the construction boils down to the selection of a copula.
The copula function, which brings additional component parameters, allows specifying 
dependence in the bivariate distribution,
separately from modeling the marginal distribution. On the other hand, some properties 
of the resulting model, including the conditional expectation, may be intractable,
and the computational cost may increase, especially in the discrete case.

The class of models proposed in this paper can be easily extended
for non-stationary time series that exhibit trends and seasonality, 
by incorporating corresponding factors into the model, either
multiplicatively or additively. This is illustrated in our second
real data example. A similar approach can be applied
to incorporate covariates. Therefore this class of models is quite
general, and is useful as an alternative to the existing time
series models, especially when traditional models fail to capture 
non-Gaussian features suggested by the data.

\section*{Acknowledgements}
This research was supported in part by the National Science Foundation under 
award SES 1631963, and SES 2050012. 
The authors wish to thank three reviewers and an Associate Editor for useful comments.

\section*{Supplementary material}

\textbf{MTD\_SM.pdf} includes the proof for Proposition 2, sampling algorithm details, 
additional simulation results and model checking for the data examples.
\textbf{Code.zip} contains code and data necessary for data analysis in Section 5.
\textbf{README} contains descriptions and instructions.

\section*{Appendix}
\subsection*{Proof of Proposition 1}

\begin{proof} 
Without loss of generality, we consider the case where $X_t$ has a continuous 
distribution for all $t$. Moreover, for the argument that follows to apply to 
any $t \geq 2$, we express the transition density as 
$f(x_t\mid x^{t-1}) =$ $\sum_{l=1}^{t_L} w_{l}^{*} \, f_{U_l\mid V_l}(x_t\mid x_{t-l})$,
for $t \geq 2$, where $t_L = \min\{t-1,L\}$. When $t > L$, 
$w_{l}^{*} \equiv w_{l}$, for $l=1,...,L$, whereas for $2 \leq t \leq L$,
$w_{l}^{*} = w_{l}$, for $l=1,...,t_{L} - 1$, and 
$w^{*}_{t_L} =$ $1 - \sum_{k=1}^{t_{L}-1} w_k$. With this notational
convention, we have $\sum_{l=1}^{t_L} w_{l}^{*} = 1$.

Using the proposition assumptions, 
$$
g_{2}(x_{2}) = \int_{\mathcal{S}} f(x_2\mid x_1) f_X(x_1) dx_1 =
\int_{\mathcal{S}} f_{U_1\mid V_1}(x_2\mid x_1) f_{V_{1}}(x_1) dx_1 = 
f_{U_{1}}(x_2) = f_X(x_2)
$$
and thus the result is valid for $t=2$. 
To prove the proposition by induction, 
assume the result holds true for generic $t-1$, that is, 
$g_{t^{\prime}}(x_{t^{\prime}}) =$ $f_{X}(x_{t^{\prime}})$, for all  
$x_{t^{\prime}} \in \mathcal{S}$, and for all $t^{\prime} \leq t-1$.
Denote by $p(x_{1},\dots,x_{t-1})$ and $p(x_{t-t_{L}},\dots,x_{t-1})$ the 
joint density for random vector $(X_{1},\dots,X_{t-1})$ and 
$(X_{t-t_{L}},\dots,X_{t-1})$, respectively. Then, the marginal density 
for $X_{t}$ can be derived as follows:
$$
\begin{aligned}
g_{t}(x_{t}) & =  
\int_{\mathcal{S}^{t-1}} f(x_t\mid x^{t-1}) \, p(x_{1},\dots, x_{t-1}) \, 
dx_{1} \dots dx_{t-1} \\
& = \sum_{l=1}^{t_L} w_{l}^{*} \int_{\mathcal{S}^{t_{L}}} f_{U_l\mid V_l}(x_t\mid x_{t-l})
\, p(x_{t-t_{L}},\dots,x_{t-1}) \, dx_{t-t_{L}}\dots dx_{t-1} \\
& = \sum_{l=1}^{t_L} w_{l}^{*} \int_{\mathcal{S}} f_{U_l\mid V_l}(x_t\mid x_{t-l})
\, g_{t-l}(x_{t-l}) \, dx_{t-l} \\
& = \sum_{l=1}^{t_L} w_{l}^{*} \int_{\mathcal{S}} f_{U_l\mid V_l}(x_t\mid x_{t-l})
\, f_{V_{l}}(x_{t-l}) \, dx_{t-l} \\
& = f_X(x_t),
\end{aligned}
$$
where for the second-to-last equation we used $g_{t-l} = f_{X}$, for  
$l=1,...,t_{L}$, obtained from the induction argument, as well as the 
proposition assumption, $f_{X} = f_{V_{l}}$, for all $l$. Finally, the last 
equation is based on the proposition assumption that $f_{U_{l}} =$ $f_{X}$, 
for all $l$. 
\end{proof}

\bibliographystyle{jasa3}

\bibliography{ref}

\begin{thebibliography}{47}
\newcommand{\enquote}[1]{``#1''}
\expandafter\ifx\csname natexlab\endcsname\relax\def\natexlab#1{#1}\fi
\expandafter\ifx\csname url\endcsname\relax
  \def\url#1{{\tt #1}}\fi
\expandafter\ifx\csname urlprefix\endcsname\relax\def\urlprefix{URL }\fi

\bibitem[\protect\citeauthoryear{Arnold, Castillo, Sarabia, and Sarabia}{Arnold
  et~al.}{1999}]{arnold1999conditional}
Arnold, B.~C., Castillo, E., Sarabia, J.-M., and Sarabia, J.~M. (1999), {\em
  Conditional specification of statistical models\/}, Springer Science \&
  Business Media.

\bibitem[\protect\citeauthoryear{Azzalini}{Azzalini}{2013}]{azzalini2013skew}
Azzalini, A. (2013), {\em The skew-normal and related families\/}, volume~3,
  Cambridge University Press.

\bibitem[\protect\citeauthoryear{Bartolucci and Farcomeni}{Bartolucci and
  Farcomeni}{2010}]{bartolucci2010note}
Bartolucci, F. and Farcomeni, A. (2010), \enquote{A note on the mixture
  transition distribution and hidden {M}arkov models,} {\em Journal of Time
  Series Analysis\/}, 31, 132--138.

\bibitem[\protect\citeauthoryear{Berchtold}{Berchtold}{2001}]{berchtold2001estimation}
Berchtold, A. (2001), \enquote{Estimation in the mixture transition
  distribution model,} {\em Journal of Time Series Analysis\/}, 22, 379--397.

\bibitem[\protect\citeauthoryear{Berchtold}{Berchtold}{2003}]{berchtold2003mixture}
--- (2003), \enquote{Mixture transition distribution ({MTD}) modeling of
  heteroscedastic time series,} {\em Computational statistics \& data
  analysis\/}, 41, 399--411.

\bibitem[\protect\citeauthoryear{Berchtold and Raftery}{Berchtold and
  Raftery}{2002}]{berchtold2002mixture}
Berchtold, A. and Raftery, A. (2002), \enquote{The mixture transition
  distribution model for high-order {M}arkov chains and non-{G}aussian time
  series,} {\em Statistical Science\/}, 17, 328--356.

\bibitem[\protect\citeauthoryear{Bolano and Berchtold}{Bolano and
  Berchtold}{2016}]{bolano2016general}
Bolano, D. and Berchtold, A. (2016), \enquote{General framework and model
  building in the class of Hidden Mixture Transition Distribution models,} {\em
  Computational Statistics \& Data Analysis\/}, 93, 131--145.

\bibitem[\protect\citeauthoryear{Cervone, Pillai, Pati, Berbeco, Lewis
  et~al.}{Cervone et~al.}{2014}]{cervone2014location}
Cervone, D., Pillai, N.~S., Pati, D., Berbeco, R., Lewis, J.~H., et~al. (2014),
  \enquote{A location-mixture autoregressive model for online forecasting of
  lung tumor motion,} {\em The Annals of Applied Statistics\/}, 8, 1341--1371.

\bibitem[\protect\citeauthoryear{Connor and Mosimann}{Connor and
  Mosimann}{1969}]{connor1969concepts}
Connor, R.~J. and Mosimann, J.~E. (1969), \enquote{Concepts of independence for
  proportions with a generalization of the {D}irichlet distribution,} {\em
  Journal of the American Statistical Association\/}, 64, 194--206.

\bibitem[\protect\citeauthoryear{Dai, Ding, Wahba et~al.}{Dai
  et~al.}{2013}]{dai2013multivariate}
Dai, B., Ding, S., Wahba, G., et~al. (2013), \enquote{Multivariate {B}ernoulli
  distribution,} {\em Bernoulli\/}, 19, 1465--1483.

\bibitem[\protect\citeauthoryear{Dunn and Smyth}{Dunn and
  Smyth}{1996}]{dunn1996randomized}
Dunn, P.~K. and Smyth, G.~K. (1996), \enquote{Randomized quantile residuals,}
  {\em Journal of Computational and Graphical Statistics\/}, 5, 236--244.

\bibitem[\protect\citeauthoryear{Eltoft, Kim, and Lee}{Eltoft
  et~al.}{2006}]{eltoft2006multivariate}
Eltoft, T., Kim, T., and Lee, T.-W. (2006), \enquote{On the multivariate
  Laplace distribution,} {\em IEEE Signal Processing Letters\/}, 13, 300--303.

\bibitem[\protect\citeauthoryear{Escarela, Mena, and Castillo-Morales}{Escarela
  et~al.}{2006}]{escarela2006flexible}
Escarela, G., Mena, R.~H., and Castillo-Morales, A. (2006), \enquote{A flexible
  class of parametric transition regression models based on copulas:
  application to poliomyelitis incidence,} {\em Statistical Methods in Medical
  Research\/}, 15, 593--609.

\bibitem[\protect\citeauthoryear{Ferguson}{Ferguson}{1973}]{ferguson1973bayesian}
Ferguson, T.~S. (1973), \enquote{A {B}ayesian analysis of some nonparametric
  problems,} {\em The Annals of Statistics\/}, 1, 209--230.

\bibitem[\protect\citeauthoryear{Fong, Li, Yau, and Wong}{Fong
  et~al.}{2007}]{fong2007mixture}
Fong, P.~W., Li, W.~K., Yau, C., and Wong, C. (2007), \enquote{On a mixture
  vector autoregressive model,} {\em Canadian Journal of Statistics\/}, 35,
  135--150.

\bibitem[\protect\citeauthoryear{Hassan and Lii}{Hassan and
  Lii}{2006}]{hassan2006modeling}
Hassan, M.~Y. and Lii, K.-S. (2006), \enquote{Modeling marked point processes
  via bivariate mixture transition distribution models,} {\em Journal of the
  American Statistical Association\/}, 101, 1241--1252.

\bibitem[\protect\citeauthoryear{Heiner and Kottas}{Heiner and
  Kottas}{2019}]{heiner2019estimation}
Heiner, M. and Kottas, A. (2019), \enquote{Estimation and selection for
  high-order {M}arkov chains with {B}ayesian mixture transition distribution
  models,} {\em arXiv preprint arXiv:1906.10781\/}.

\bibitem[\protect\citeauthoryear{Heiner and Kottas}{Heiner and
  Kottas}{2021}]{GP-MTD}
--- (2021), \enquote{Autoregressive density modeling with the {G}aussian
  process mixture transition distribution,} {\em Journal of Time Series
  Analysis\/}, To appear.

\bibitem[\protect\citeauthoryear{Heiner, Kottas, and Munch}{Heiner
  et~al.}{2019}]{heiner2019spv}
Heiner, M., Kottas, A., and Munch, S. (2019), \enquote{Structured priors for
  sparse probability vectors with application to model selection in {M}arkov
  chains,} {\em Statistics and Computing\/}, 29, 1077--1093.

\bibitem[\protect\citeauthoryear{Holgate}{Holgate}{1964}]{holgate1964estimation}
Holgate, P. (1964), \enquote{Estimation for the bivariate {P}oisson
  distribution,} {\em Biometrika\/}, 51, 241--287.

\bibitem[\protect\citeauthoryear{Joe}{Joe}{2014}]{joe2014dependence}
Joe, H. (2014), {\em Dependence modeling with copulas\/}, CRC press.

\bibitem[\protect\citeauthoryear{Kalliovirta, Meitz, and Saikkonen}{Kalliovirta
  et~al.}{2015}]{kalliovirta2015gaussian}
Kalliovirta, L., Meitz, M., and Saikkonen, P. (2015), \enquote{A {G}aussian
  mixture autoregressive model for univariate time series,} {\em Journal of
  Time Series Analysis\/}, 36, 247--266.

\bibitem[\protect\citeauthoryear{Kalliovirta, Meitz, and Saikkonen}{Kalliovirta
  et~al.}{2016}]{kalliovirta2016gaussian}
--- (2016), \enquote{{G}aussian mixture vector autoregression,} {\em Journal of
  Econometrics\/}, 192, 485--498.

\bibitem[\protect\citeauthoryear{Khalili, Chen, and Stephens}{Khalili
  et~al.}{2017}]{khalili2017regularization}
Khalili, A., Chen, J., and Stephens, D.~A. (2017), \enquote{Regularization and
  selection in {G}aussian mixture of autoregressive models,} {\em Canadian
  Journal of Statistics\/}, 45, 356--374.

\bibitem[\protect\citeauthoryear{Kocherlakota and Kocherlakota}{Kocherlakota
  and Kocherlakota}{2006}]{kocherlakota2006bivariate}
Kocherlakota, S. and Kocherlakota, K. (2006), \enquote{Bivariate discrete
  distributions,} {\em Encyclopedia of Statistical Sciences\/}.

\bibitem[\protect\citeauthoryear{Kotz, Kozubowski, and Podgorski}{Kotz
  et~al.}{2012}]{kotz2012laplace}
Kotz, S., Kozubowski, T., and Podgorski, K. (2012), {\em The Laplace
  distribution and generalizations: a revisit with applications to
  communications, economics, engineering, and finance\/}, Springer Science \&
  Business Media.

\bibitem[\protect\citeauthoryear{Lanne and Saikkonen}{Lanne and
  Saikkonen}{2003}]{lanne2003modeling}
Lanne, M. and Saikkonen, P. (2003), \enquote{Modeling the {US} short-term
  interest rate by mixture autoregressive processes,} {\em Journal of Financial
  Econometrics\/}, 1, 96--125.

\bibitem[\protect\citeauthoryear{Lau and So}{Lau and
  So}{2008}]{lau2008bayesian}
Lau, J.~W. and So, M.~K. (2008), \enquote{{B}ayesian mixture of autoregressive
  models,} {\em Computational Statistics \& Data Analysis\/}, 53, 38--60.

\bibitem[\protect\citeauthoryear{Le, Martin, and Raftery}{Le
  et~al.}{1996}]{le1996modeling}
Le, N.~D., Martin, R.~D., and Raftery, A.~E. (1996), \enquote{Modeling flat
  stretches, bursts outliers in time series using mixture transition
  distribution models,} {\em Journal of the American Statistical
  Association\/}, 91, 1504--1515.

\bibitem[\protect\citeauthoryear{Li, Lu, Park, Kim, Brinkley, and Peterson}{Li
  et~al.}{1999}]{li1999multivariate}
Li, C.-S., Lu, J.-C., Park, J., Kim, K., Brinkley, P.~A., and Peterson, J.~P.
  (1999), \enquote{Multivariate zero-inflated {P}oisson models and their
  applications,} {\em Technometrics\/}, 41, 29--38.

\bibitem[\protect\citeauthoryear{Li, Zhu, Liu, and Li}{Li
  et~al.}{2017}]{li2017mixture}
Li, G., Zhu, Q., Liu, Z., and Li, W.~K. (2017), \enquote{On mixture double
  autoregressive time series models,} {\em Journal of Business \& Economic
  Statistics\/}, 35, 306--317.

\bibitem[\protect\citeauthoryear{Luo and Qiu}{Luo and
  Qiu}{2009}]{luo2009parameter}
Luo, J. and Qiu, H.-b. (2009), \enquote{Parameter estimation of the {WMTD}
  model,} {\em Applied Mathematics-A Journal of Chinese Universities\/}, 24,
  379.

\bibitem[\protect\citeauthoryear{MacDonald and Zucchini}{MacDonald and
  Zucchini}{1997}]{macdonald1997hidden}
MacDonald, I.~L. and Zucchini, W. (1997), {\em Hidden {M}arkov and other models
  for discrete-valued time series\/}, volume 110, CRC Press.

\bibitem[\protect\citeauthoryear{Meitz, Preve, and Saikkonen}{Meitz
  et~al.}{2021}]{meitz2021mixture}
Meitz, M., Preve, D., and Saikkonen, P. (2021), \enquote{A mixture
  autoregressive model based on Student’st--distribution,} {\em
  Communications in Statistics-Theory and Methods\/}, 1--76.

\bibitem[\protect\citeauthoryear{Mena and Walker}{Mena and
  Walker}{2007}]{mena2007stationary}
Mena, R.~H. and Walker, S.~G. (2007), \enquote{Stationary Mixture Transition
  Distribution ({MTD}) models via predictive distributions,} {\em Journal of
  statistical planning and inference\/}, 137, 3103--3112.

\bibitem[\protect\citeauthoryear{Nguyen, McLachlan, Ullmann, and Janke}{Nguyen
  et~al.}{2016}]{nguyen2016laplace}
Nguyen, H.~D., McLachlan, G.~J., Ullmann, J.~F., and Janke, A.~L. (2016),
  \enquote{Laplace mixture autoregressive models,} {\em Statistics \&
  Probability Letters\/}, 110, 18--24.

\bibitem[\protect\citeauthoryear{Pitt, Chatfield, and Walker}{Pitt
  et~al.}{2002}]{pitt2002constructing}
Pitt, M.~K., Chatfield, C., and Walker, S.~G. (2002), \enquote{Constructing
  first order stationary autoregressive models via latent processes,} {\em
  Scandinavian Journal of Statistics\/}, 29, 657--663.

\bibitem[\protect\citeauthoryear{Raftery and Tavar{\'e}}{Raftery and
  Tavar{\'e}}{1994}]{raftery1994estimation}
Raftery, A. and Tavar{\'e}, S. (1994), \enquote{Estimation and modelling
  repeated patterns in high order {M}arkov chains with the mixture transition
  distribution model,} {\em Journal of the Royal Statistical Society: Series C
  (Applied Statistics)\/}, 43, 179--199.

\bibitem[\protect\citeauthoryear{Raftery}{Raftery}{1985}]{raftery1985model}
Raftery, A.~E. (1985), \enquote{A model for high-order {M}arkov chains,} {\em
  Journal of the Royal Statistical Society: Series B (Methodological)\/}, 47,
  528--539.

\bibitem[\protect\citeauthoryear{Raftery}{Raftery}{1994}]{raftery1994change}
--- (1994), \enquote{Change point and change curve modeling in stochastic
  processes and spatial statistics,} {\em Journal of Applied Statistical
  Science\/}, 1, 403--423.

\bibitem[\protect\citeauthoryear{Sethuraman}{Sethuraman}{1994}]{sethuraman1994constructive}
Sethuraman, J. (1994), \enquote{A constructive definition of {D}irichlet
  priors,} {\em Statistica Sinica\/}, 4, 639--650.

\bibitem[\protect\citeauthoryear{Vitolo}{Vitolo}{2017}]{hddtools}
Vitolo, C. (2017), \enquote{hddtools: Hydrological Data Discovery Tools,} {\em
  The Journal of Open Source Software\/}, 2.

\bibitem[\protect\citeauthoryear{Wong, Chan, and Kam}{Wong
  et~al.}{2009}]{wong2009student}
Wong, C., Chan, W., and Kam, P. (2009), \enquote{A Student t-mixture
  autoregressive model with applications to heavy-tailed financial data,} {\em
  Biometrika\/}, 96, 751--760.

\bibitem[\protect\citeauthoryear{Wong and Li}{Wong and
  Li}{2000}]{wong2000mixture}
Wong, C.~S. and Li, W.~K. (2000), \enquote{On a mixture autoregressive model,}
  {\em Journal of the Royal Statistical Society: Series B (Statistical
  Methodology)\/}, 62, 95--115.

\bibitem[\protect\citeauthoryear{Wong and Li}{Wong and
  Li}{2001{\natexlab{a}}}]{wong2001logistic}
--- (2001{\natexlab{a}}), \enquote{On a logistic mixture autoregressive model,}
  {\em Biometrika\/}, 88, 833--846.

\bibitem[\protect\citeauthoryear{Wong and Li}{Wong and
  Li}{2001{\natexlab{b}}}]{wong2001mixture}
--- (2001{\natexlab{b}}), \enquote{On a mixture autoregressive conditional
  heteroscedastic model,} {\em Journal of the American Statistical
  Association\/}, 96, 982--995.

\bibitem[\protect\citeauthoryear{Zhu, Li, and Wang}{Zhu
  et~al.}{2010}]{zhu2010mixture}
Zhu, F., Li, Q., and Wang, D. (2010), \enquote{A mixture integer-valued {ARCH}
  model,} {\em Journal of Statistical Planning and inference\/}, 140,
  2025--2036.

\end{thebibliography}

\clearpage\pagebreak\newpage

\spacingset{1.5} 

\begin{center}
\LARGE\bf Supplementary Material
\end{center}
\setcounter{section}{0}
\setcounter{equation}{0}
\setcounter{figure}{0}
\setcounter{table}{0}

\renewcommand{\thesection}{\Alph{section}}  

\section{Proof of Proposition 2}

    \begin{proof}
    We refer to the definition of weak stationarity from 
    Brockwell and Davis (1991). A time series $\{X_t:t\in\mathbb{N}\}$,
    with index set $\mathbb{N}=\{1,2,\dots\}$, is said to be weakly stationary if
    i) $E(X_t^2)<\infty$ for all $t\in\mathbb{N}$; ii) $E(X_t) = m$ for some 
    finite $m$ and for all $t\in\mathbb{N}$; iii) 
    $\mathrm{Cov}(X_{t+h},X_t) = \gamma(h)$ for all $t,h\in\mathbb{N}$.
    Under condition (1) of Proposition 2, if an MTD time series has 
    a stationary marginal distribution such that its corresponding first 
    and second moments exist and are finite, 
    then $\mu = E(X_t)$ and $\mu^{(2)} = E(X_t^2)$ are finite for all $t\in\mathbb{N}$.
    Thus, the weak stationarity conditions (i) and (ii) are satisfied.
    
    Under condition (2) of Proposition 2, the cross moment
    $$
    \begin{aligned}
    E(X_{t+h}X_t) & = E(X_tE(X_{t+h}\,|,X_{t+h-1},\dots,X_{t+h-L}))\\
    & = E(X_t\sum_{l=1}^Lw_l(a_l+b_lX_{t+h-l})) 
    = \sum_{l=1}^Lw_la_l\mu + \sum_{l=1}^Lw_lb_lE(X_{t+h-l}X_t),
    \end{aligned}
    $$
    for all $t\in\mathbb{N}$ and $h\geq L$.
    Assuming that the cross moment is independent of $t$ for $h\geq 1$, 
    we can obtain the following non-homogeneous difference equation for the 
    autocovariance function:
    $$
    \begin{aligned}
    \gamma(h) & = E(X_{t+h}X_t) - \mu^2
    = \sum_{l=1}^Lw_la_l\mu - (1-\sum_{l=1}^Lw_lb_l)\mu^2 + \sum_{l=1}^Lw_lb_l\gamma(h-l),\;\;h\geq L.
    \end{aligned}
    $$
    
    With regard to the autocorrelation function, we have
    \begin{equation}\label{eq:acf}
    r(h) = \gamma(h)/(\mu^{(2)}-\mu^2) = \phi + \sum_{l=1}^Lw_lb_lr(h-l),\;\;h\geq L,
    \end{equation}
    where $\phi = (\sum_{l=1}^Lw_la_l\mu - (1-\sum_{l=1}^Lw_lb_l)\mu^2)/(\mu^{(2)}-\mu^2)$.
    
    The necessary and sufficient condition for the non-homogeneous difference
    equation \eqref{eq:acf} to have a stable solution is that the roots 
    $z_1,\dots, z_L$ of the equation
    $$
    z^L - w_1b_1z^{L-1} - \dots - w_Lb_L = 0
    $$
    all lie inside the unit circle. This condition, with the assumption that the 
    cross moment is independent of $t$, forms condition (3) of Proposition 2.
    Under condition (3), the weak stationarity condition (iii) is satisfied.
    \end{proof}

\section{Simulation study}

\subsection{First experiment}

We generated $2000$ observations from the Gaussian MTD model with 
$\mu = 10, \sigma^2 = 100$, under two scenarios for the weights.
Scenario 1 considered exponentially decreasing weights namely 
$w_i\propto e^{-i}, i = 1,\dots, 5$, with corresponding correlations
$\rho = (0.7, 0.3, 0.1, 0.05, 0.05)^{\mathrm{T}}$. Scenario 2 considered
non-standard weight pattern such that 
$w = (0.2, 0.05, 0.45, 0.05, 0.25)$ with correlations
$\rho = (0.4, 0.1, 0.7, 0.1, 0.5)^{\mathrm{T}}$.

\begin{figure*}[htbp]
    \centering
    \includegraphics[width=.32\textwidth]{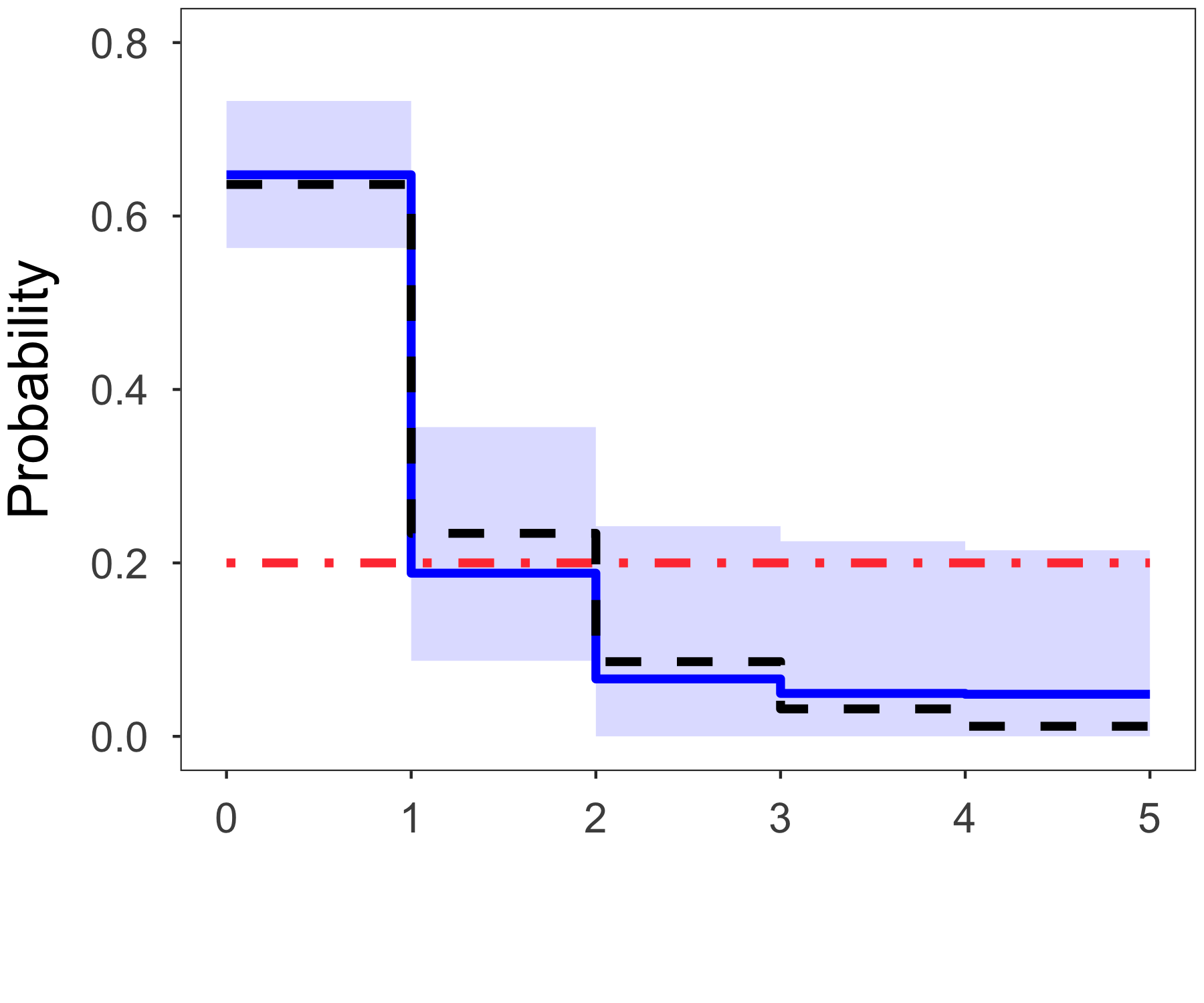}
    \includegraphics[width=.32\textwidth]{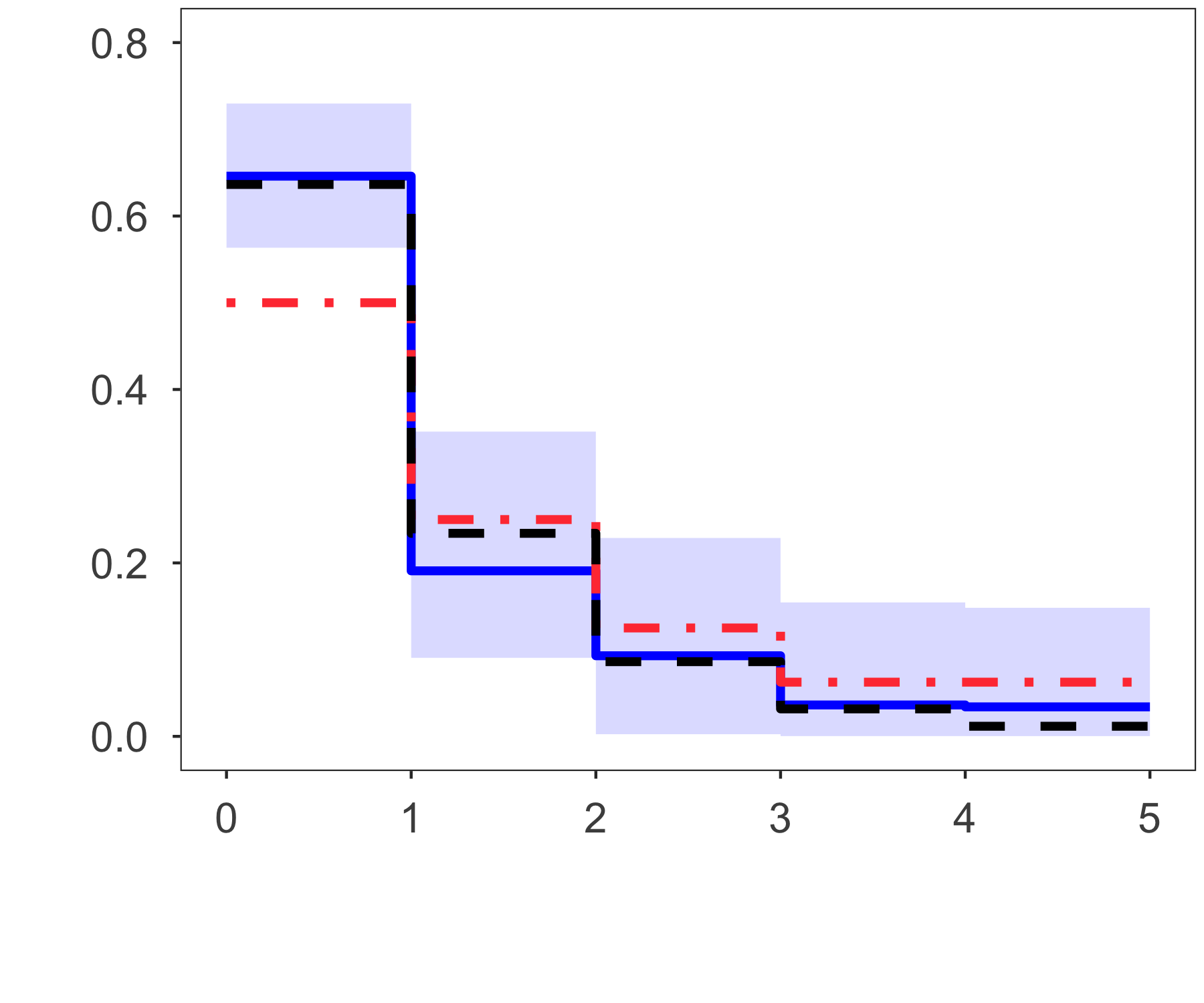}
    \includegraphics[width=.32\textwidth]{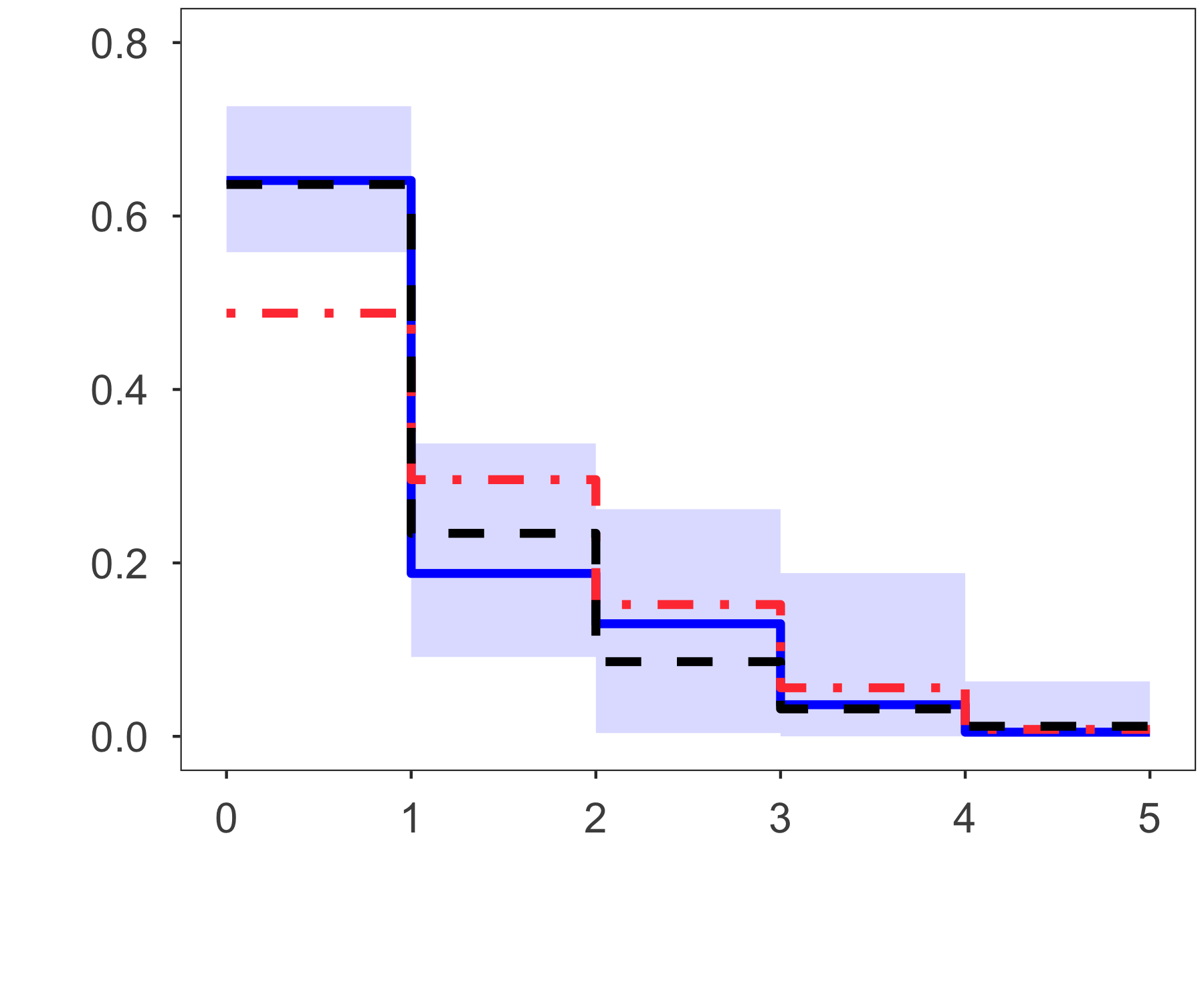}
    \medskip
    \includegraphics[width=.32\textwidth]{Sim1/15_DIR.png}
    \includegraphics[width=.32\textwidth]{Sim1/15_SB.png}
    \includegraphics[width=.32\textwidth]{Sim1/15_CDP.png}
    \medskip
    \includegraphics[width=.32\textwidth]{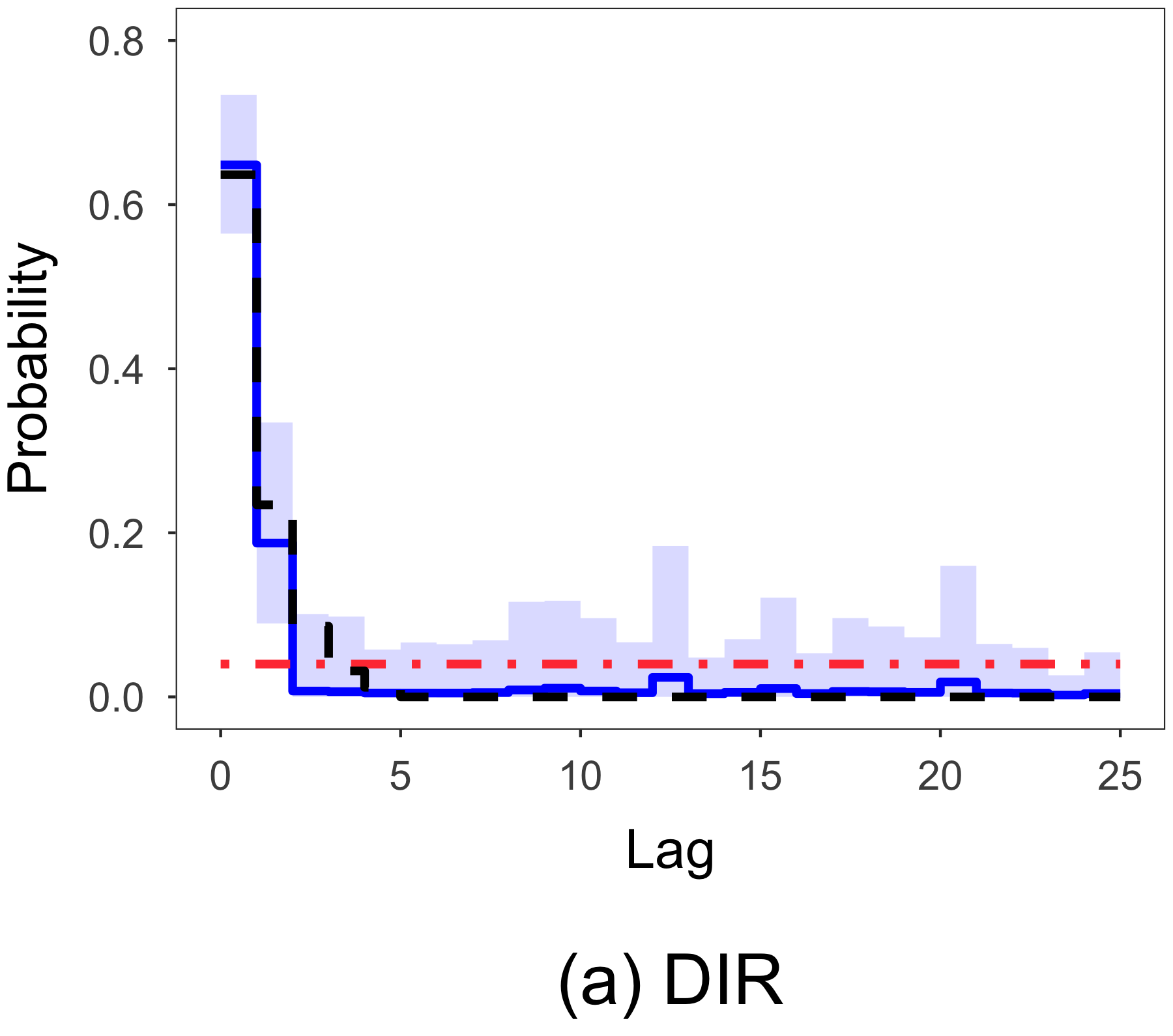}
    \includegraphics[width=.32\textwidth]{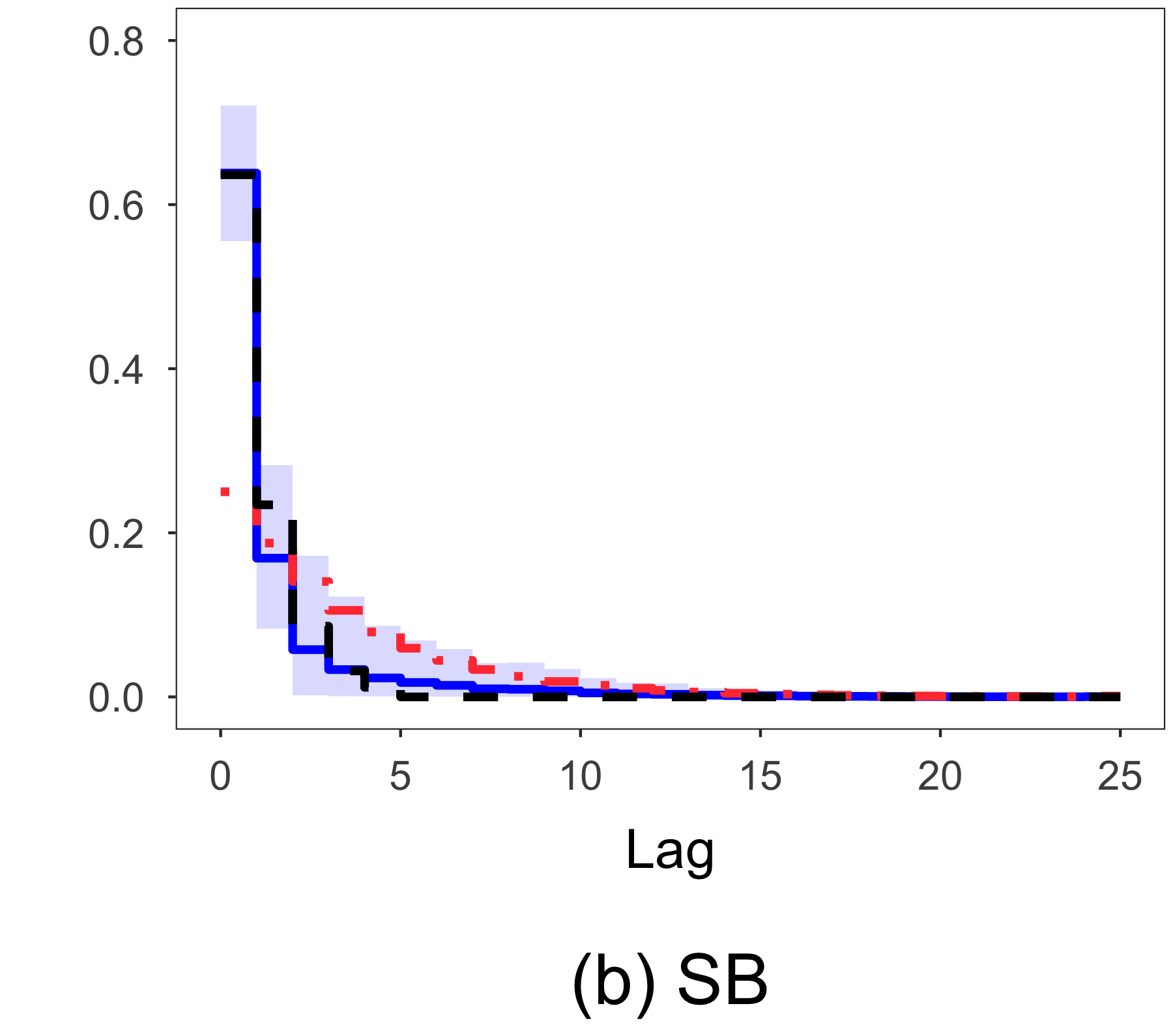}
    \includegraphics[width=.32\textwidth]{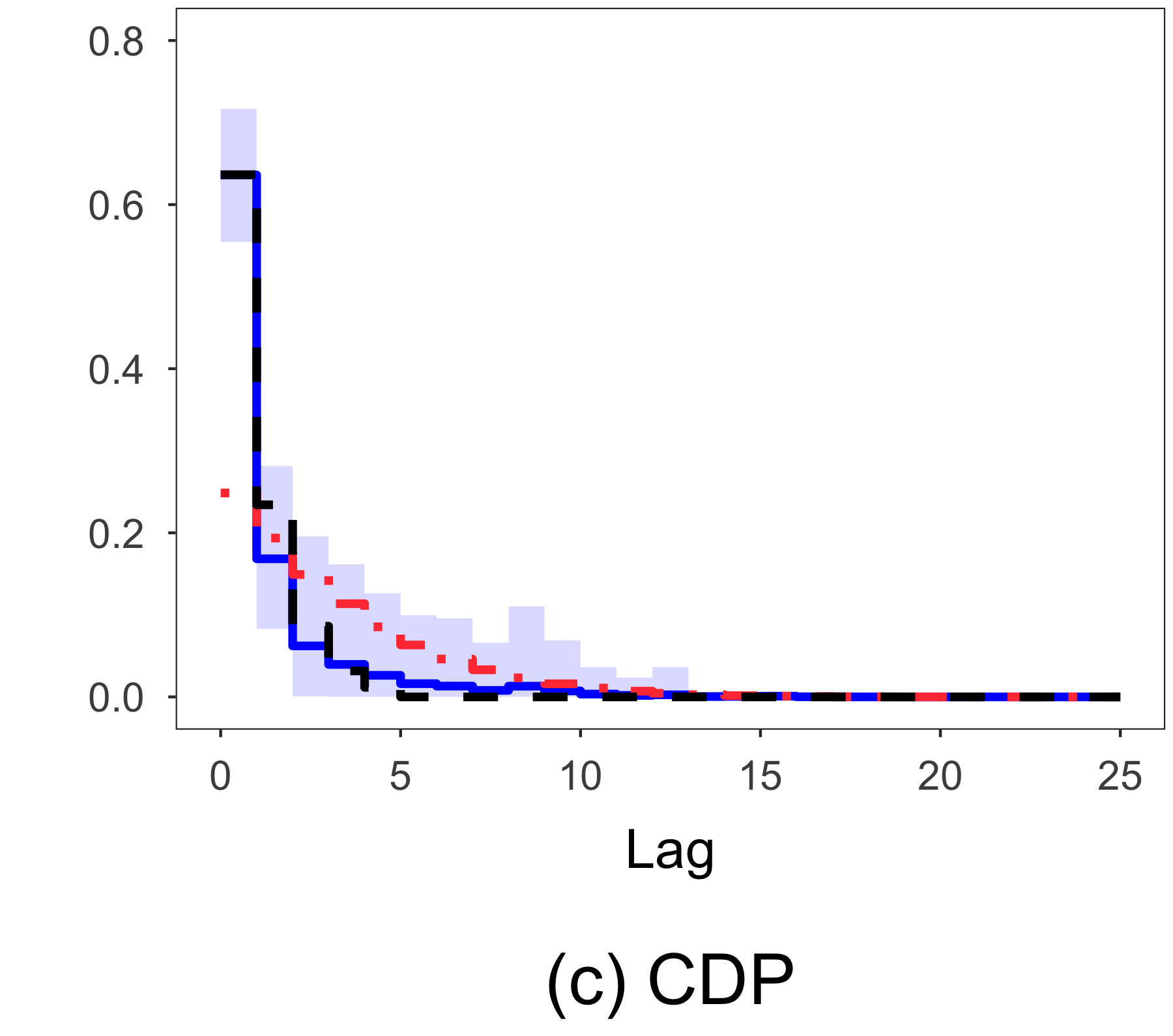}    
    \caption{
    Simulation study 1: mixture weight estimates under Scenario 1
    when $L = 5$ (top), $L = 15$ (middle) and $L = 25$ (bottom).
    Black dashed lines are true weights, red dot-dahsed lines are prior
    means, blue solid lines are posterior means, and blue polygons are 95\%
    credible intervals. 
    }
    \label{fig:s1}
\end{figure*}

\begin{figure*}[htbp]
    \centering
    \includegraphics[width=.32\textwidth]{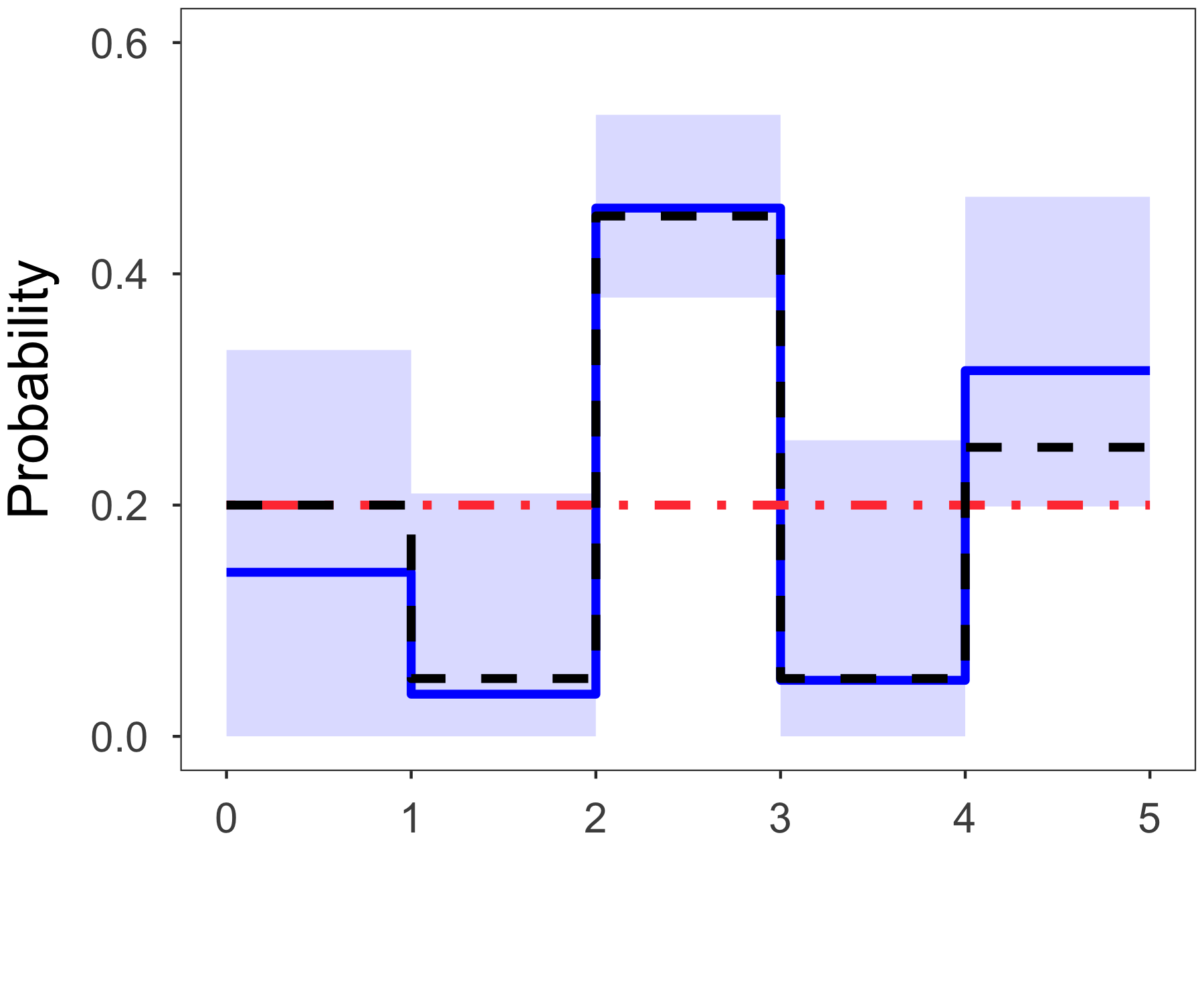}
    \includegraphics[width=.32\textwidth]{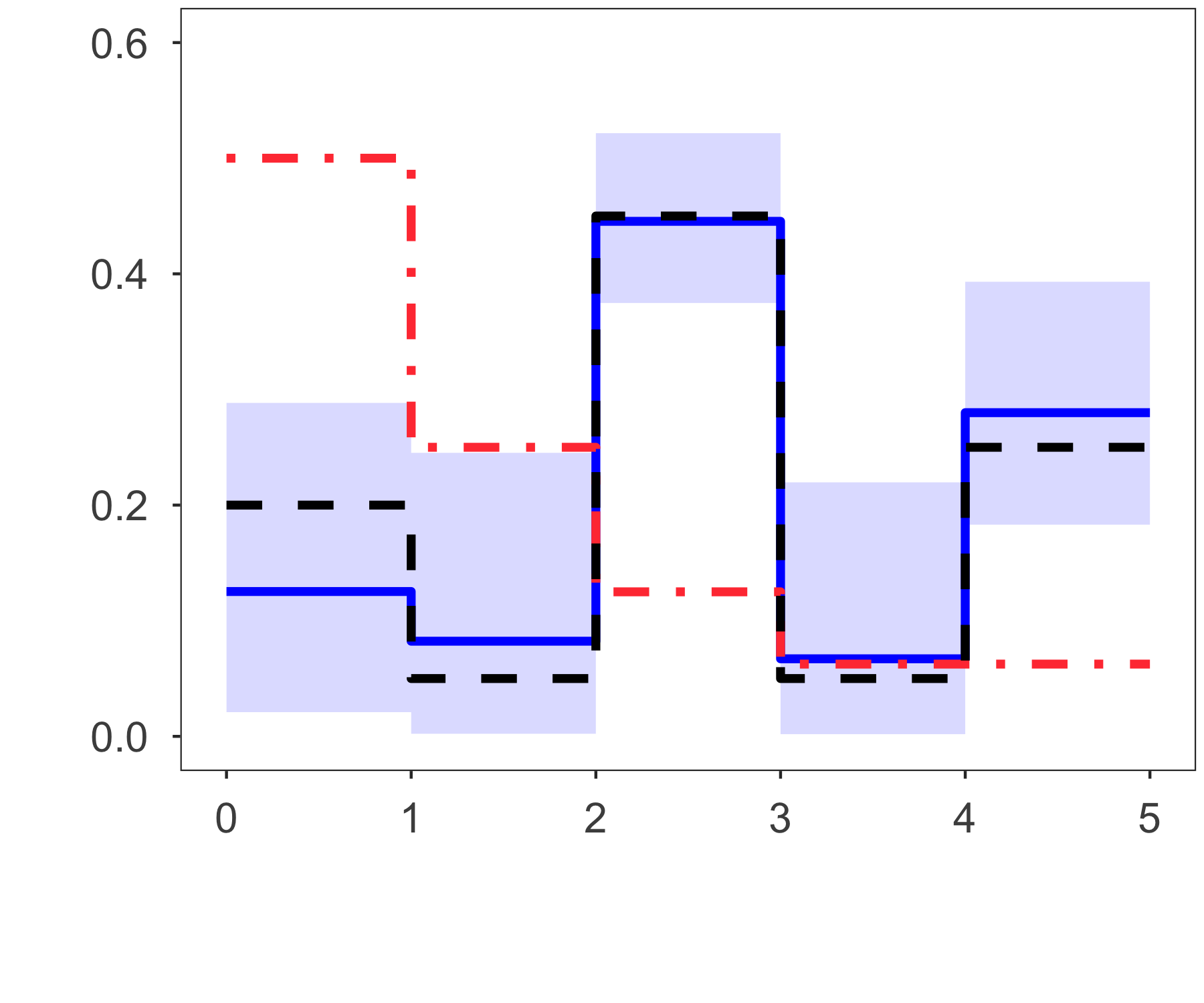}
    \includegraphics[width=.32\textwidth]{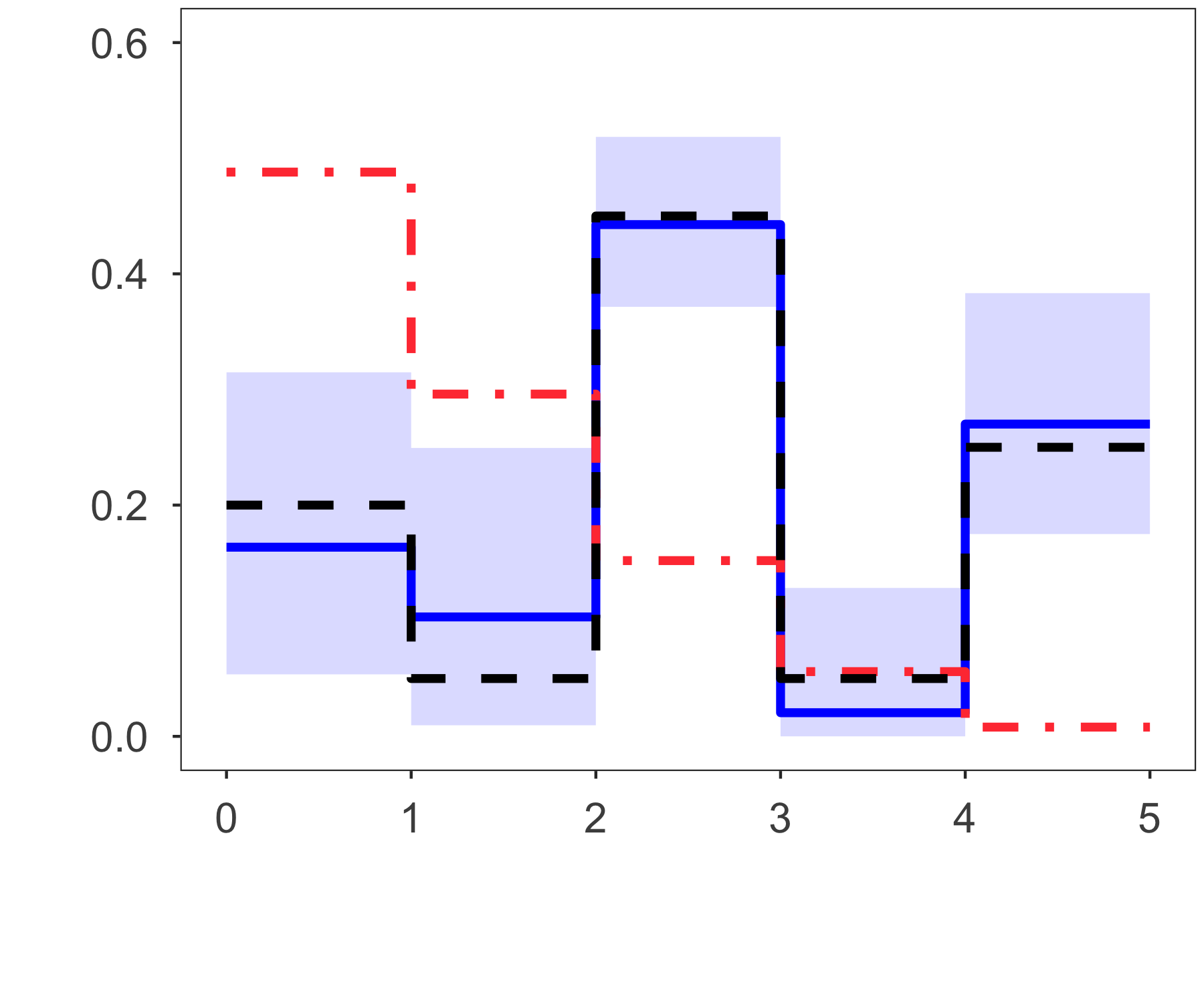}
    \medskip
    \includegraphics[width=.32\textwidth]{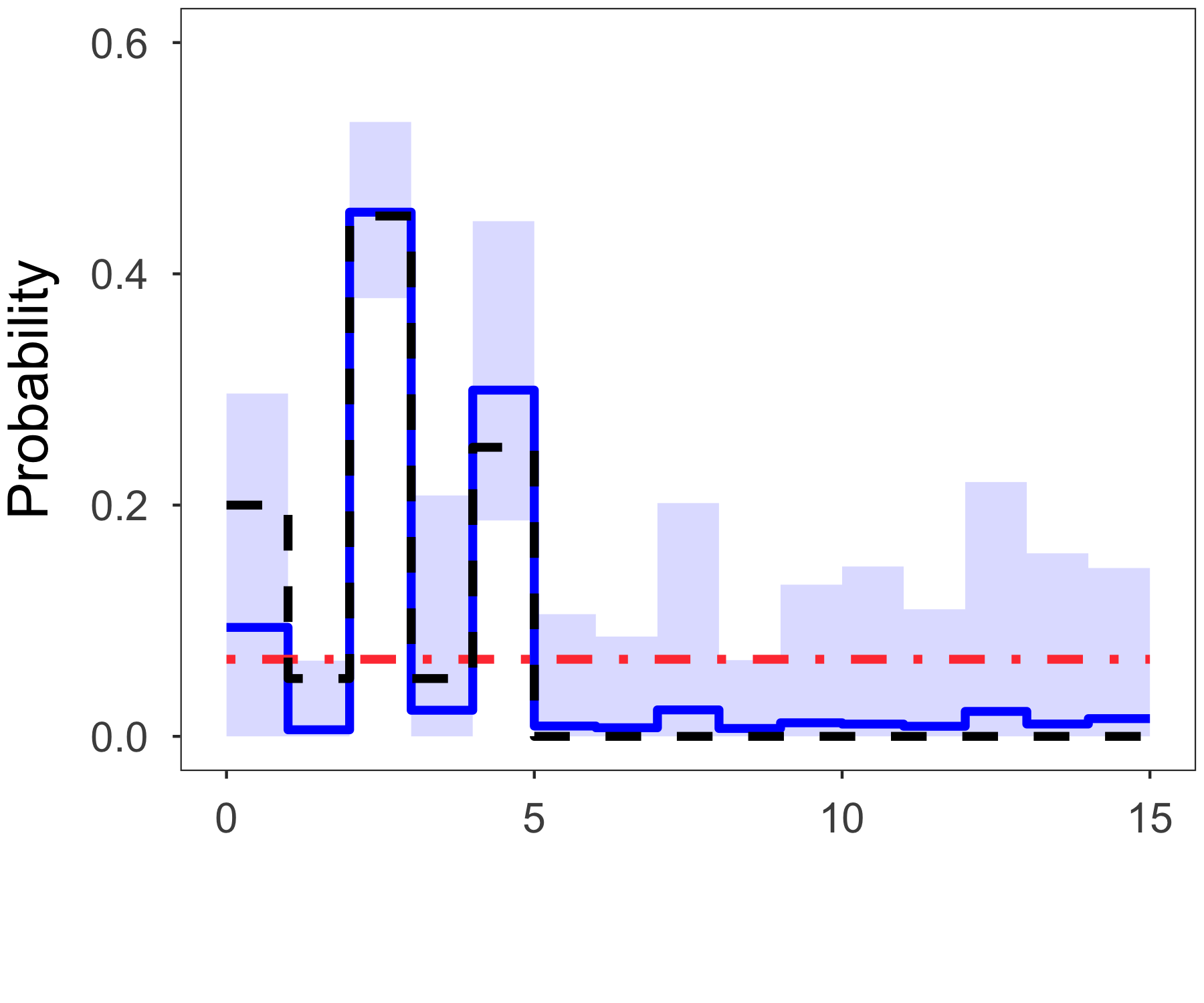}
    \includegraphics[width=.32\textwidth]{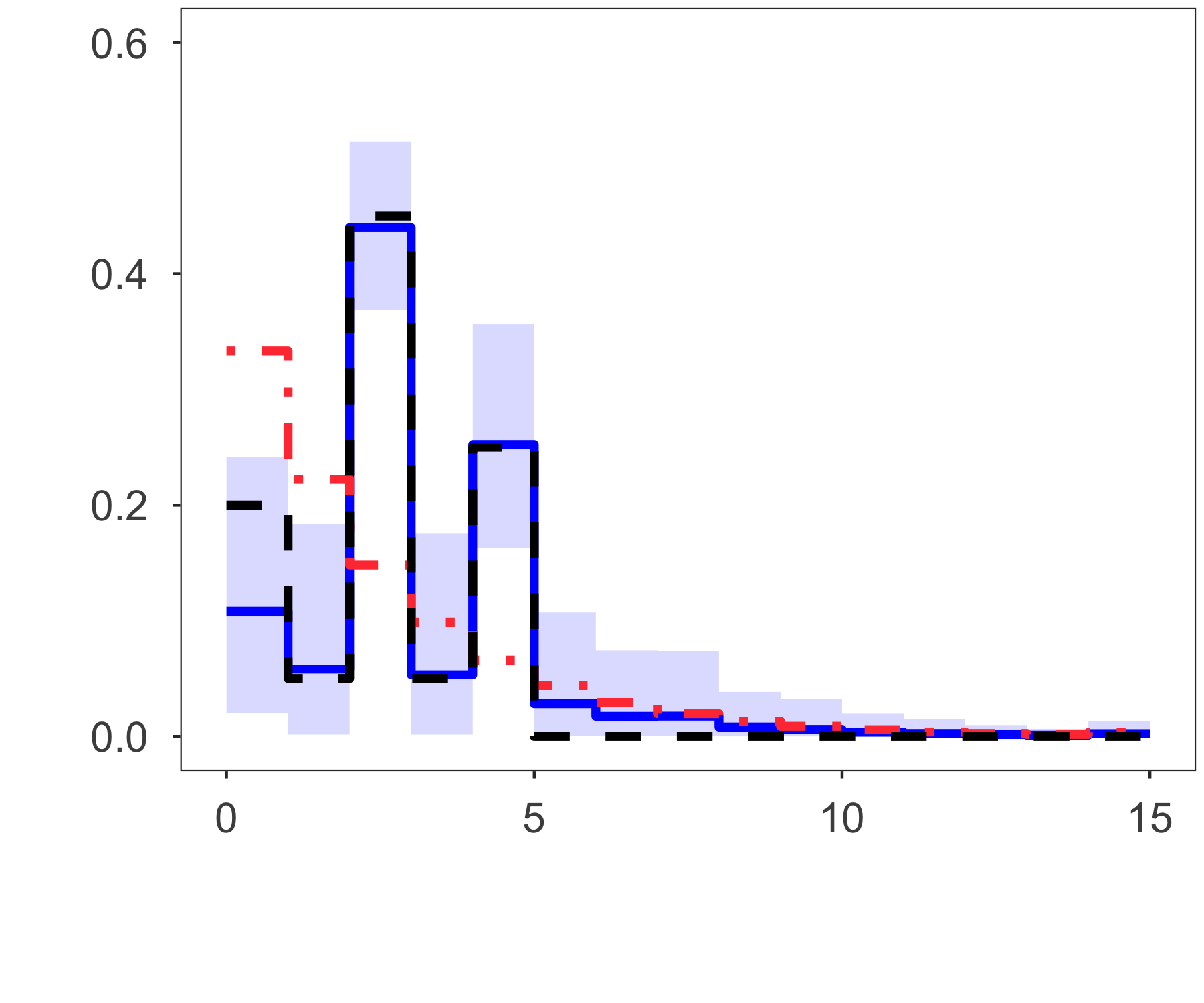}
    \includegraphics[width=.32\textwidth]{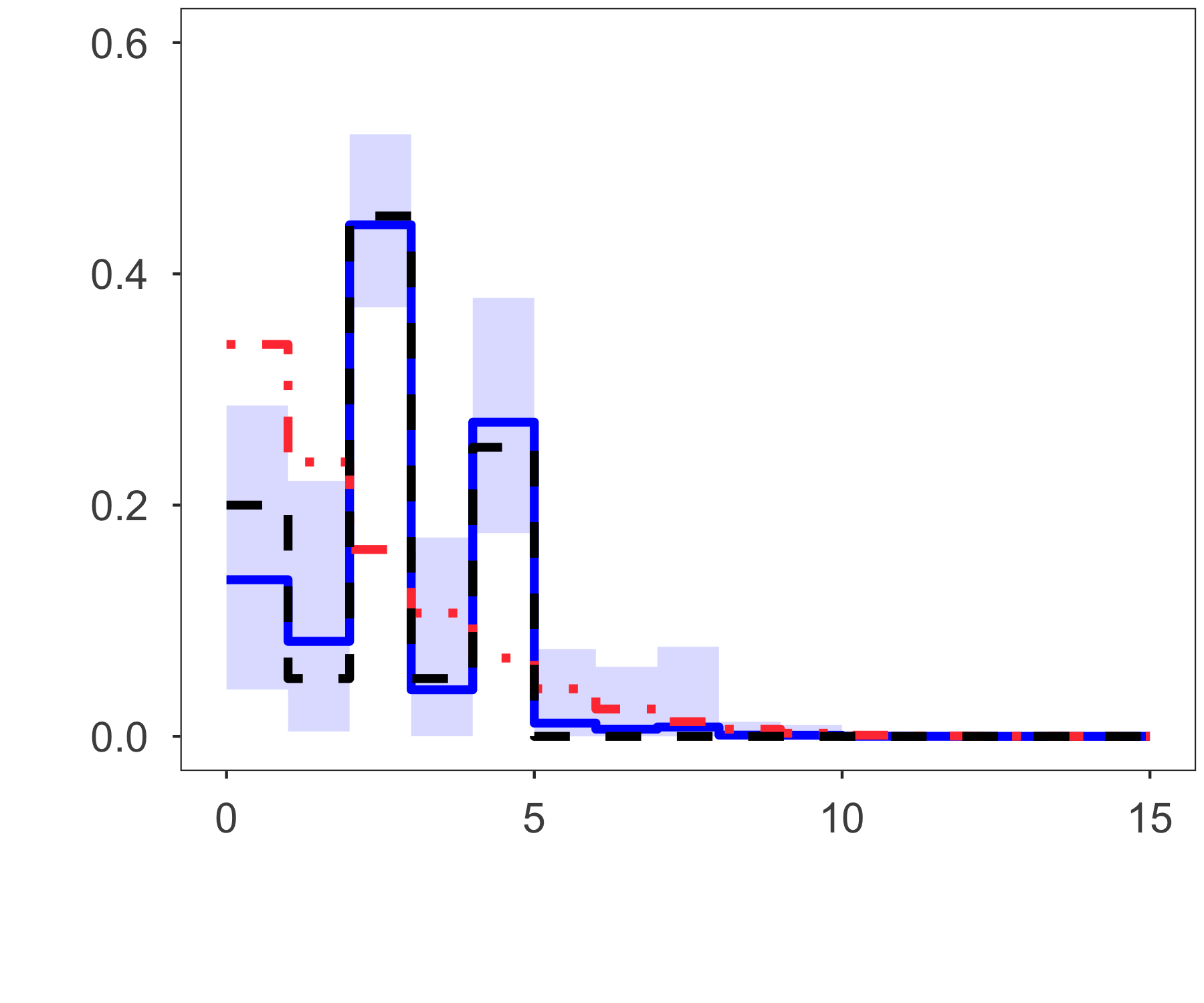}
    \medskip
    \includegraphics[width=.32\textwidth]{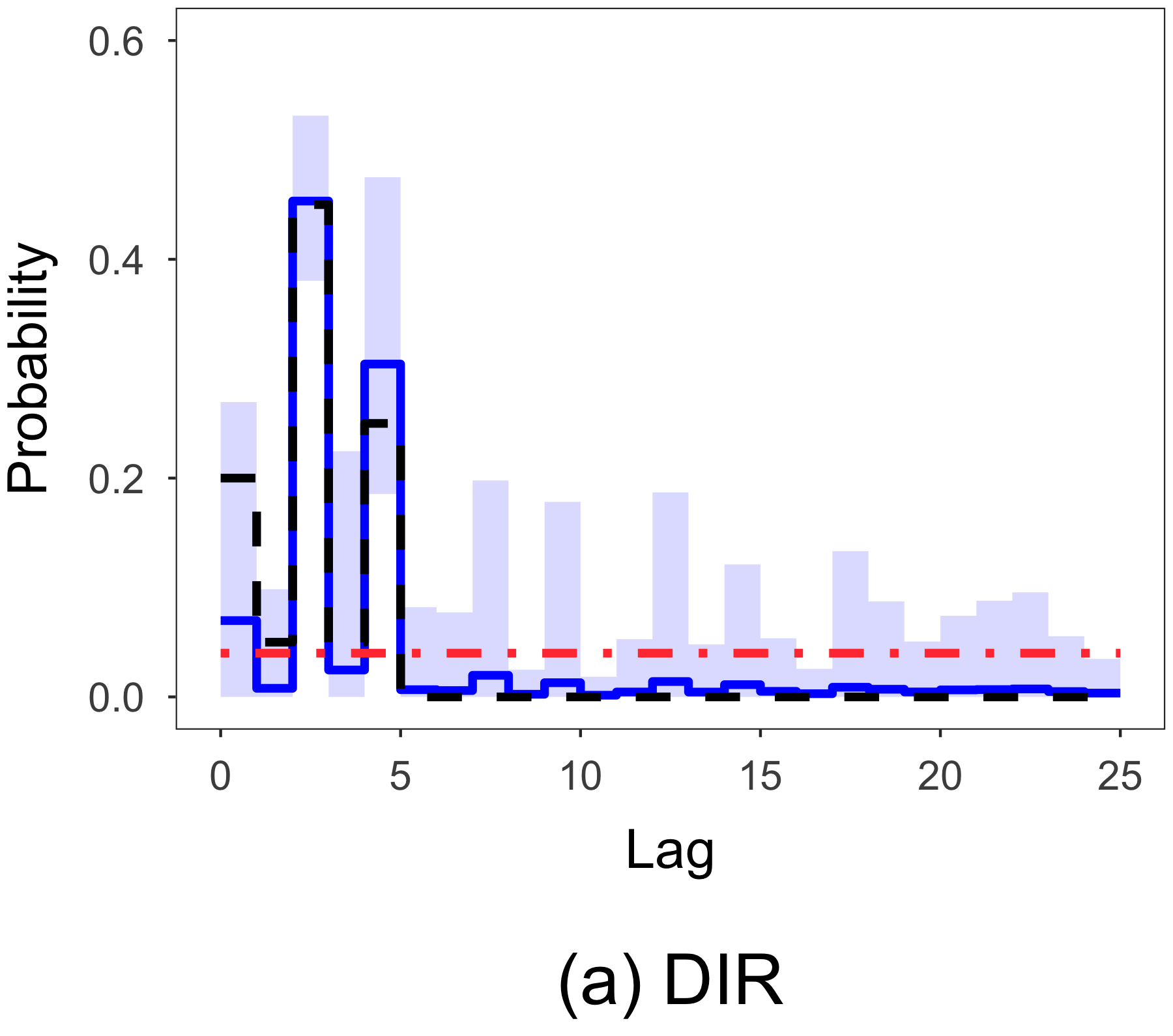}
    \includegraphics[width=.32\textwidth]{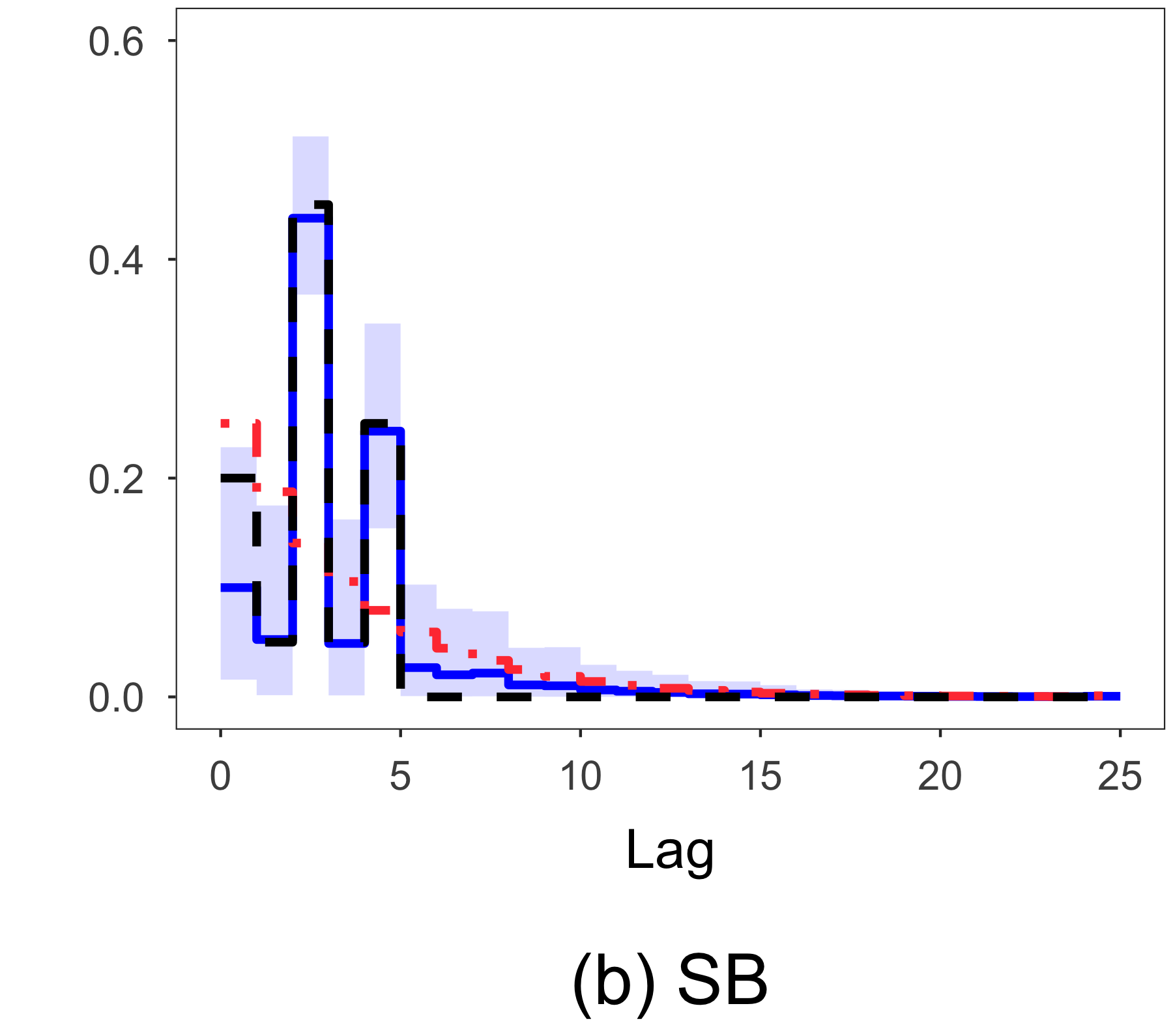}
    \includegraphics[width=.32\textwidth]{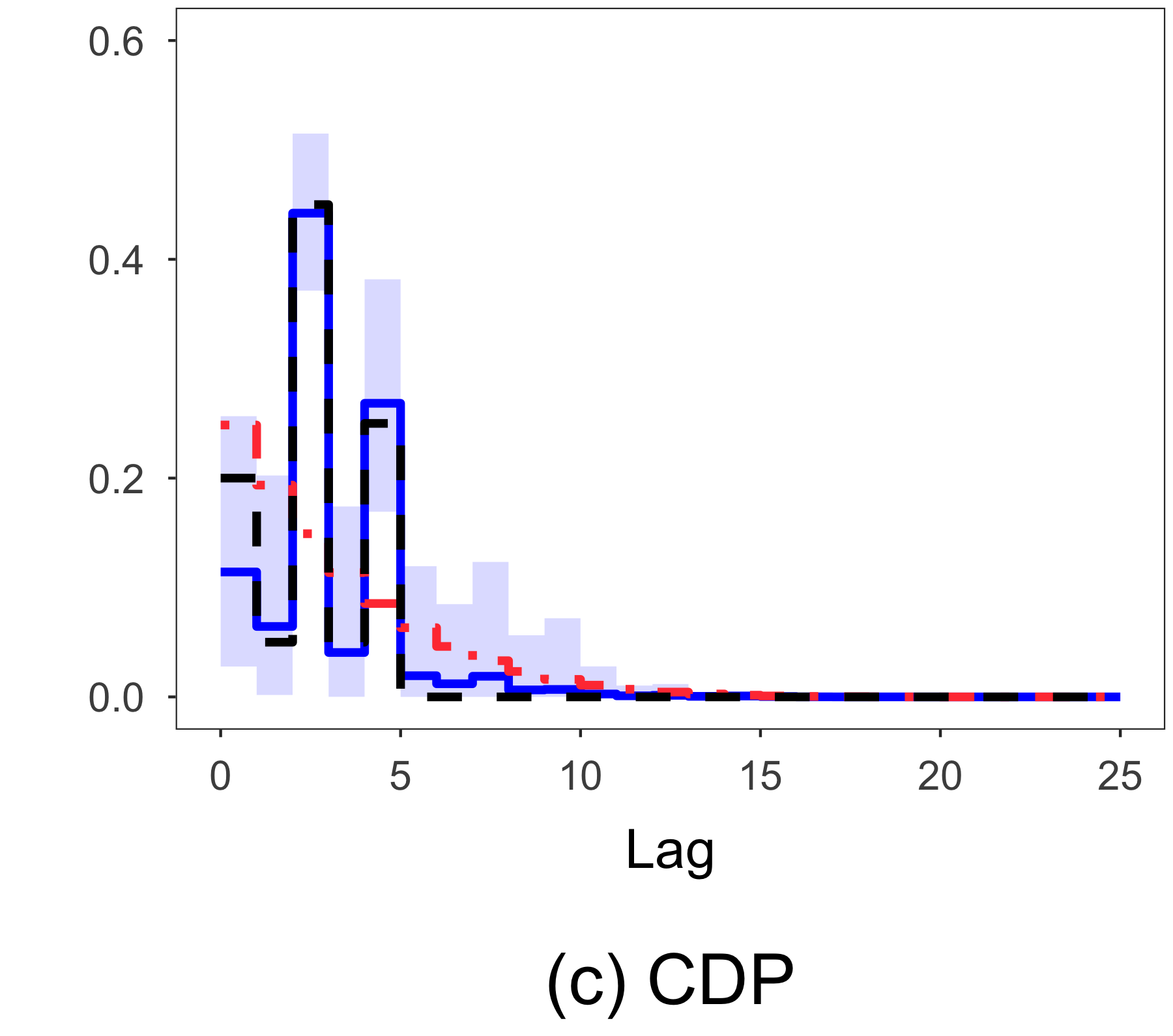}        
    \caption{
    Simulation study 1: mixture weight estimates under Scenario 2
    when $L = 5$ (top), $L = 15$ (middle) and $L = 25$ (bottom).
    Black dashed lines are true weights, red dot-dahsed lines are prior
    means, blue solid lines are posterior means, and blue polygons are 95\%
    credible intervals. 
    }
    \label{fig:s2}
\end{figure*}

We applied the Gaussian MTD models
with three different orders $L = 5, 15, 25$. In each case, we considered
three priors for the weights: the Dirichlet prior, the
truncated stick-breaking prior, and the cdf-based prior.
The shape parameter of the Dirichlet prior was $\bm1_L/L$ for each $L$.
The precision parameter $\alpha_s$ for the truncated stick-breaking prior
was taken to be $1, 2, 3$ corresponding to three orders.
For the cdf-based prior, we chose 
$\alpha_0 = 5$ as the precision parameter, and used as base distribution 
a beta with shape parameter $a_0 = 1$, and $b_0 = 3, 6, 7$ respectively
for the three orders considered. Thus a priori this prior elicited
a decreasing pattern similar to the truncated stick-breaking prior.
For all models, the mean $\mu$ and the variance $\sigma^2$ received
conjugate priors $\mathrm{N}(\mu\mid 0, 100)$ and $\mathrm{IG}(\sigma^2\mid 2, 0.1)$,
respectively, and the component-specific correlation coefficient
$\rho_l$ was assigned a uniform prior $\mathrm{Unif}(-1,1)$
independently for all $l$. To obtain the estimates, we ran a Gibbs 
sampler for 165000 iterations with the first 5000 as burn-in and collected 
samples every 20 iterations.
Figures~\ref{fig:s1} and \ref{fig:s2} show posterior summaries of 
the weights under Scenarios 1 and 2, respectively.

In Scenario 1, regardless of the selected model order, models with all three priors were
capable of capturing the first weight which corresponds to the most important lag.
On the other hand, for the rest of the weights, the models with the truncated stick-braking 
prior and the cdf-based prior captured the pattern much better. 

In Scenario 2, we can see that when $L$ is correctly specified, 
models with all three priors provided good estimates of the mixture weights. 
The model with the Dirichlet prior correctly recovered the middle three weights, 
while it failed to capture the first and the last weights.
On the other hand, the model with the cdf-based prior instead captured well the 
three most important weights. 
The performance of the model with the truncated stick-breaking prior was between the former two. 
When $L$ was over-specified, models with the truncated stick-breaking prior 
and the cdf-based prior captured much better the mixture weights, compared to the 
the model with the Dirichlet prior, though all the models systematically underestimated
the fist weight. Overall, both scenarios indicate the proposed priors are
more suitable under our modeling strategy.

\subsection{Second experiment}

In the second experiment, we demonstrate the effectiveness of 
using a negative binomial MTD (NBMTD) model for over-dispersed count data,
compared to a Poisson MTD (PMTD) model. Both models are discussed in Example 2 of Section
3 in the paper. In particular, the bivariate distribution that 
defines the NBMTD model is obtained
by replacing the rate parameters $\lambda$ and $\gamma$ of the bivariate 
Poisson distribution with $\alpha\lambda$ and $\alpha\gamma$ with 
$\alpha\sim\mathrm{Ga}(\kappa,\eta)$. Marginalizing out $\alpha$, we 
obtain a bivariate distribution with negative binomial marginal
$\mathrm{NB}(\kappa, \eta/(\lambda+\gamma+\eta))$.

\begin{figure*}[t!]
    \centering
    \includegraphics[width=.24\textwidth]{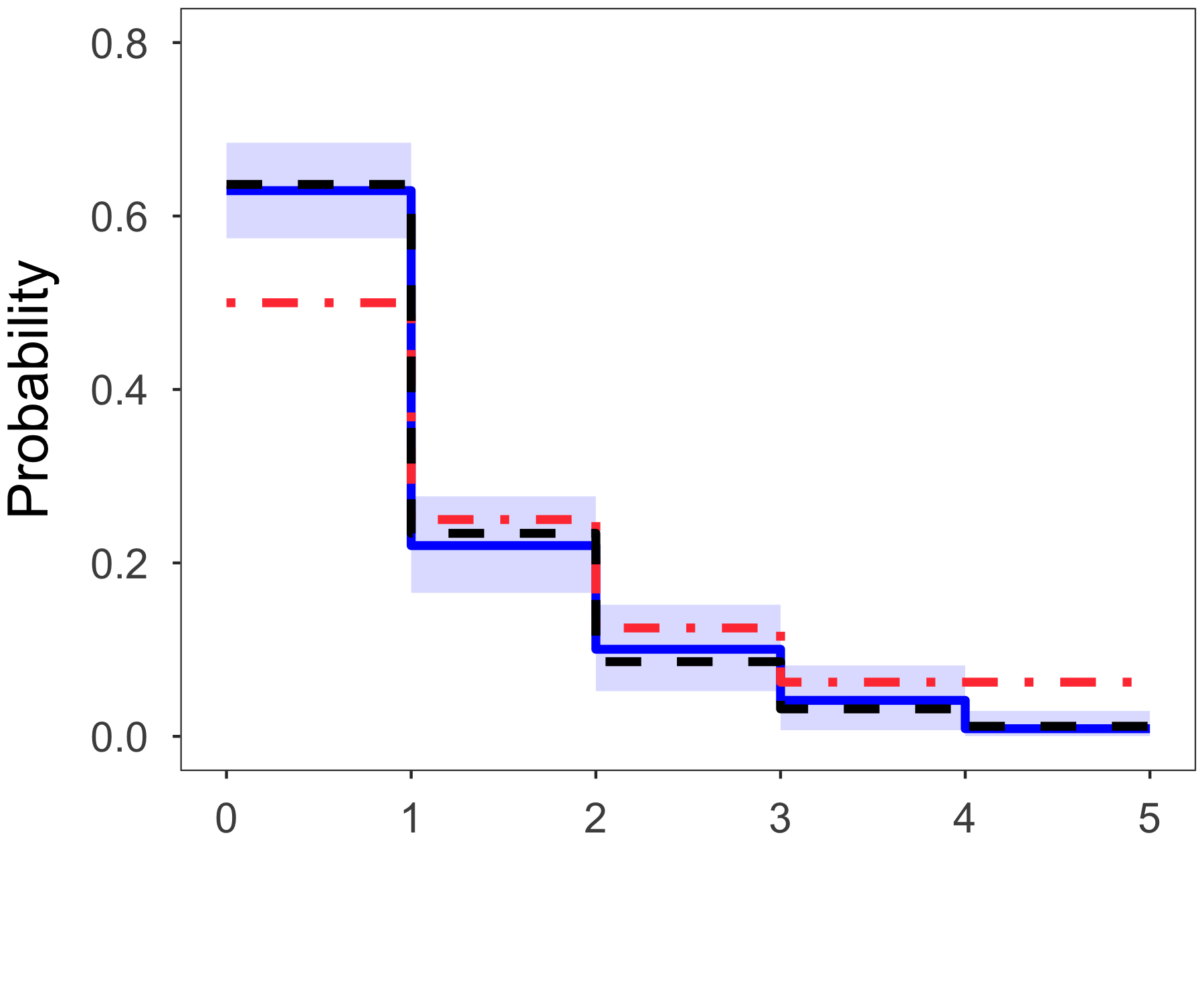}
    \includegraphics[width=.24\textwidth]{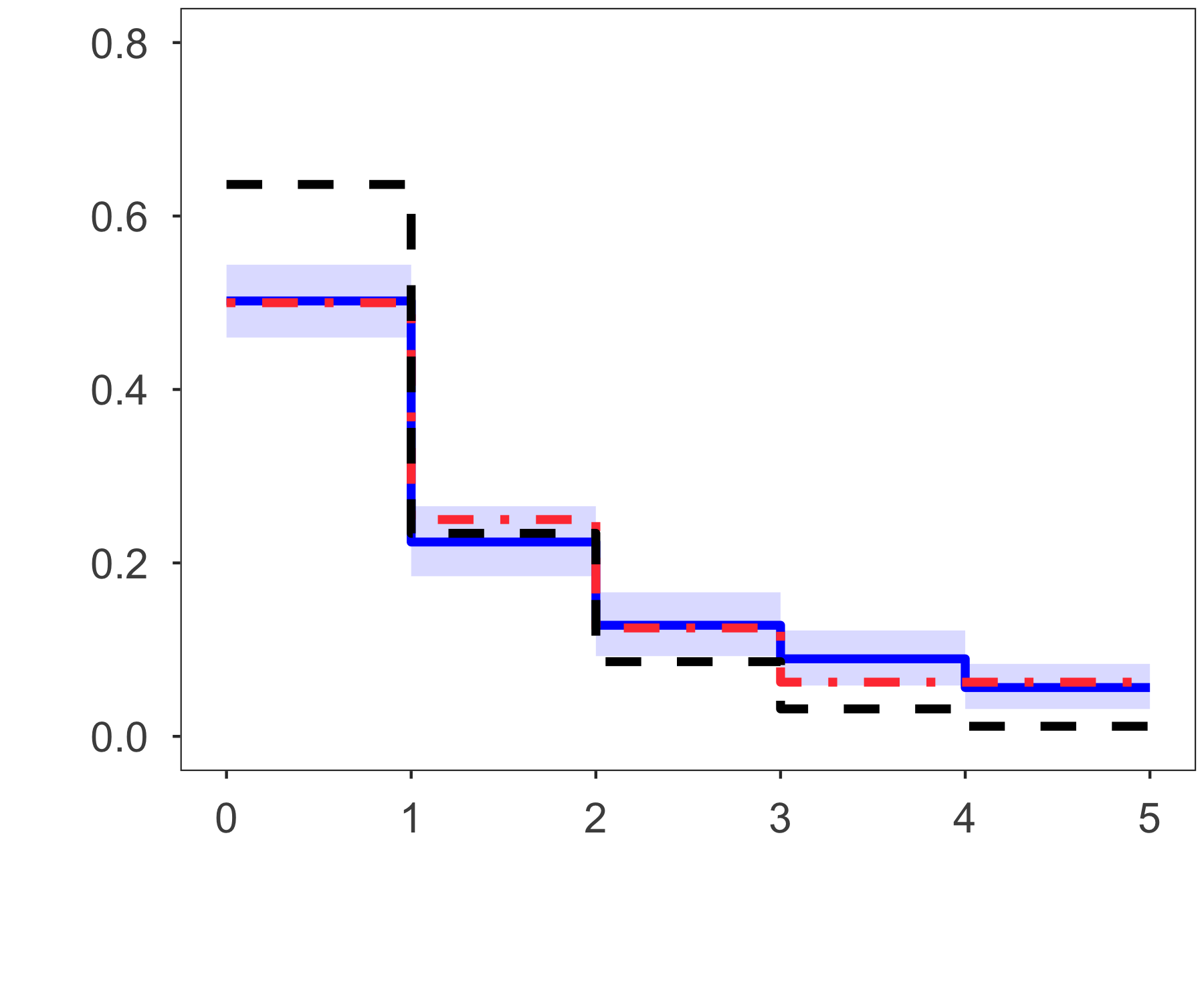}
    \includegraphics[width=.24\textwidth]{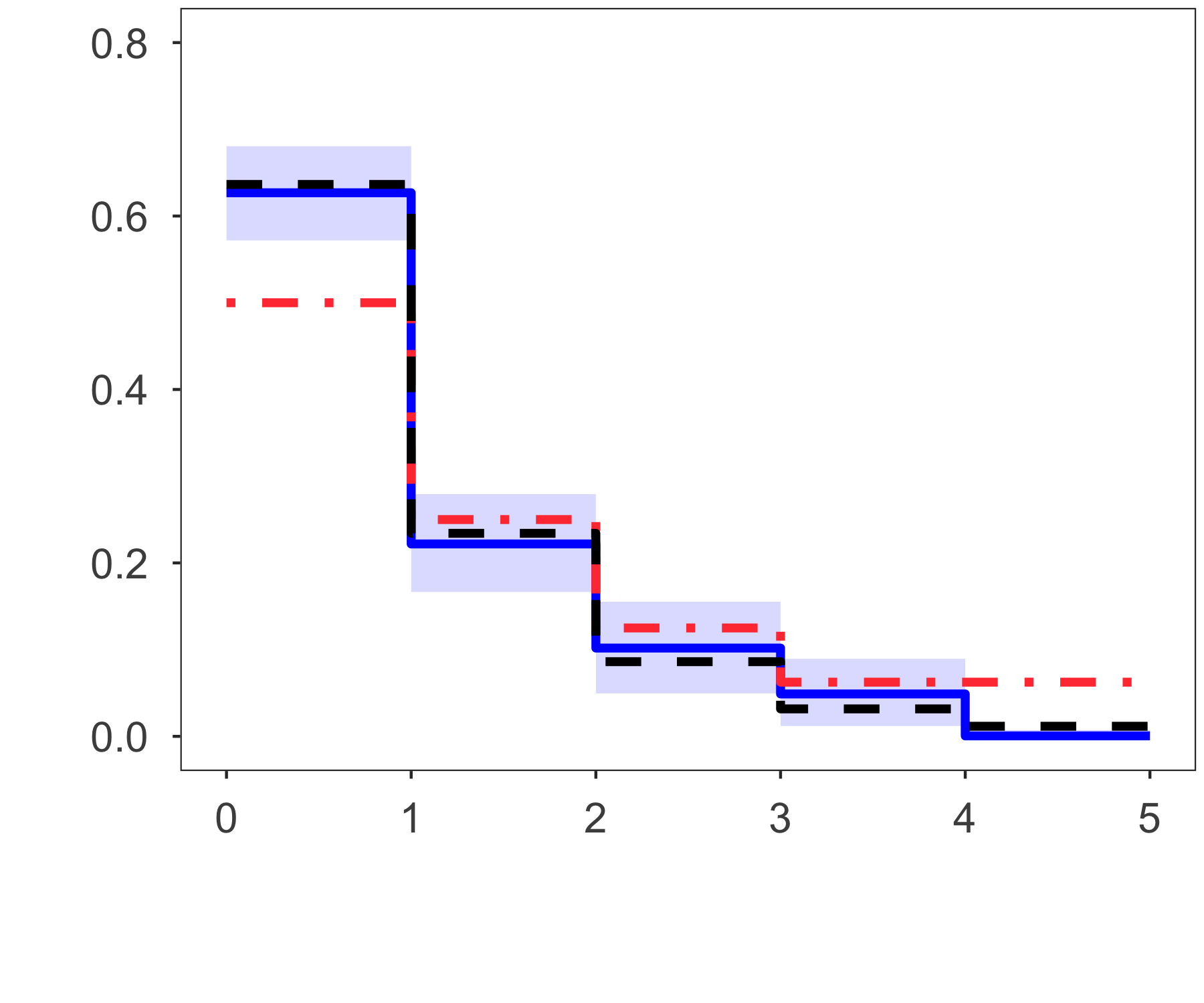}
    \includegraphics[width=.24\textwidth]{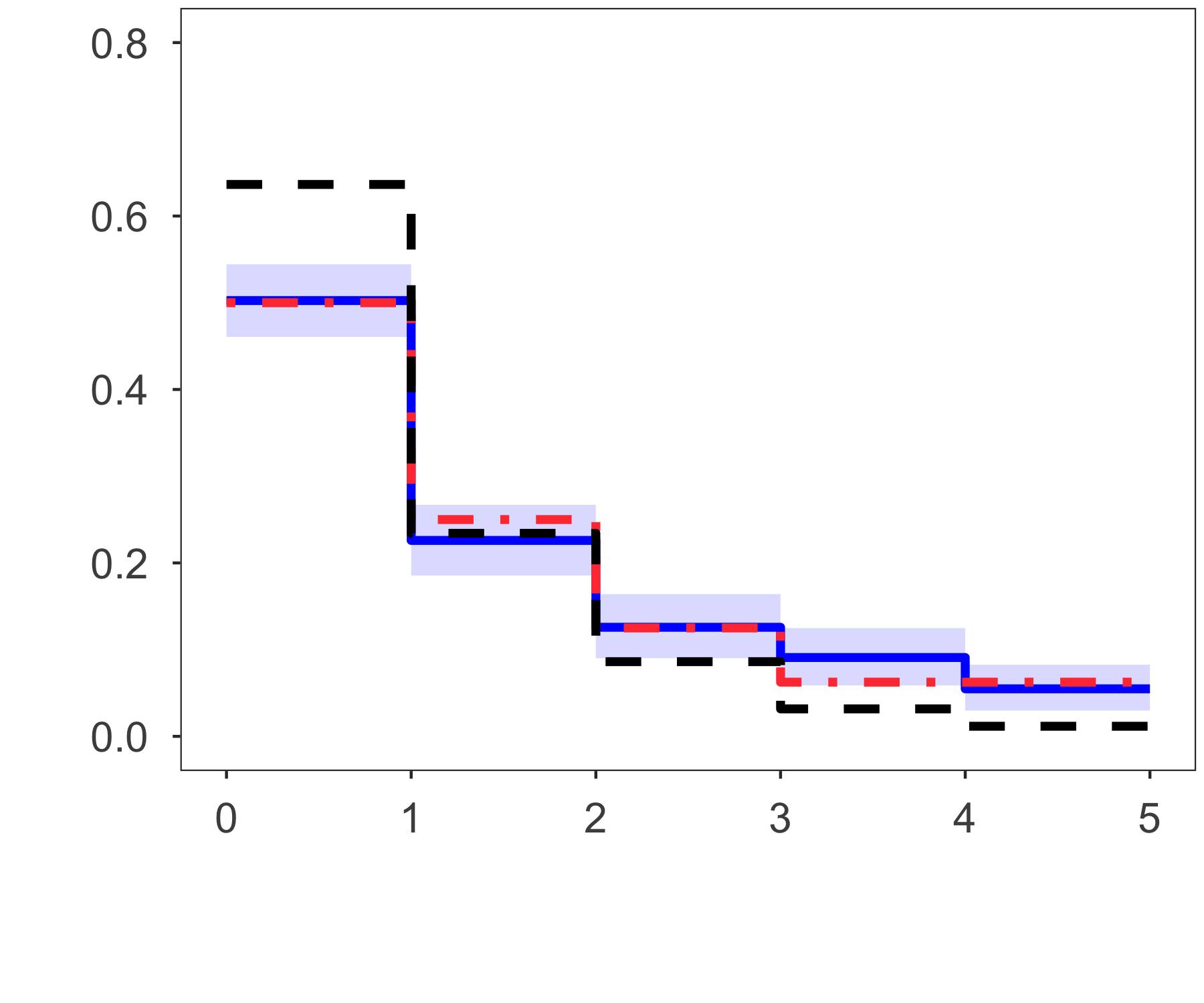}
    \medskip
    \includegraphics[width=.24\textwidth]{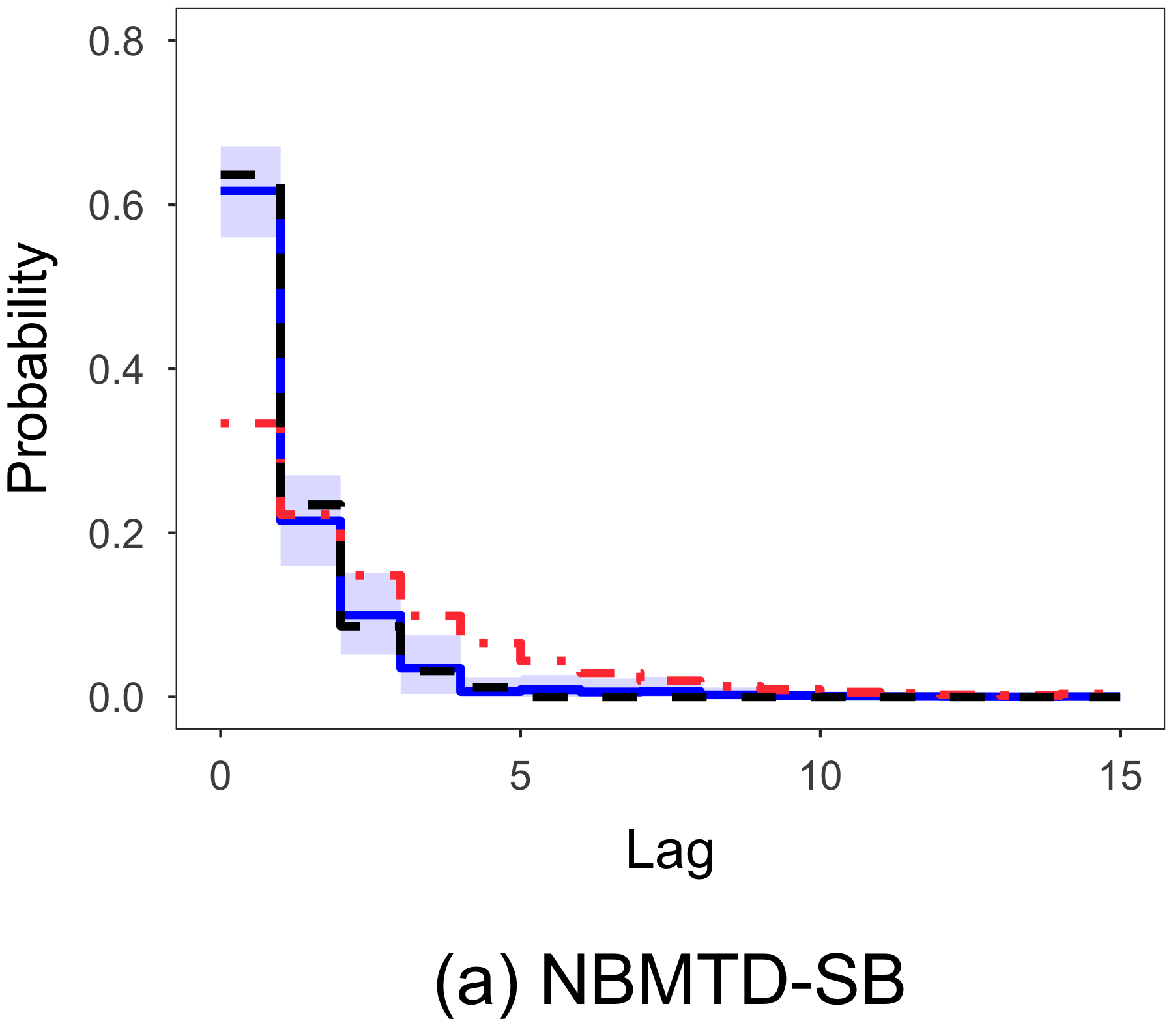}
    \includegraphics[width=.24\textwidth]{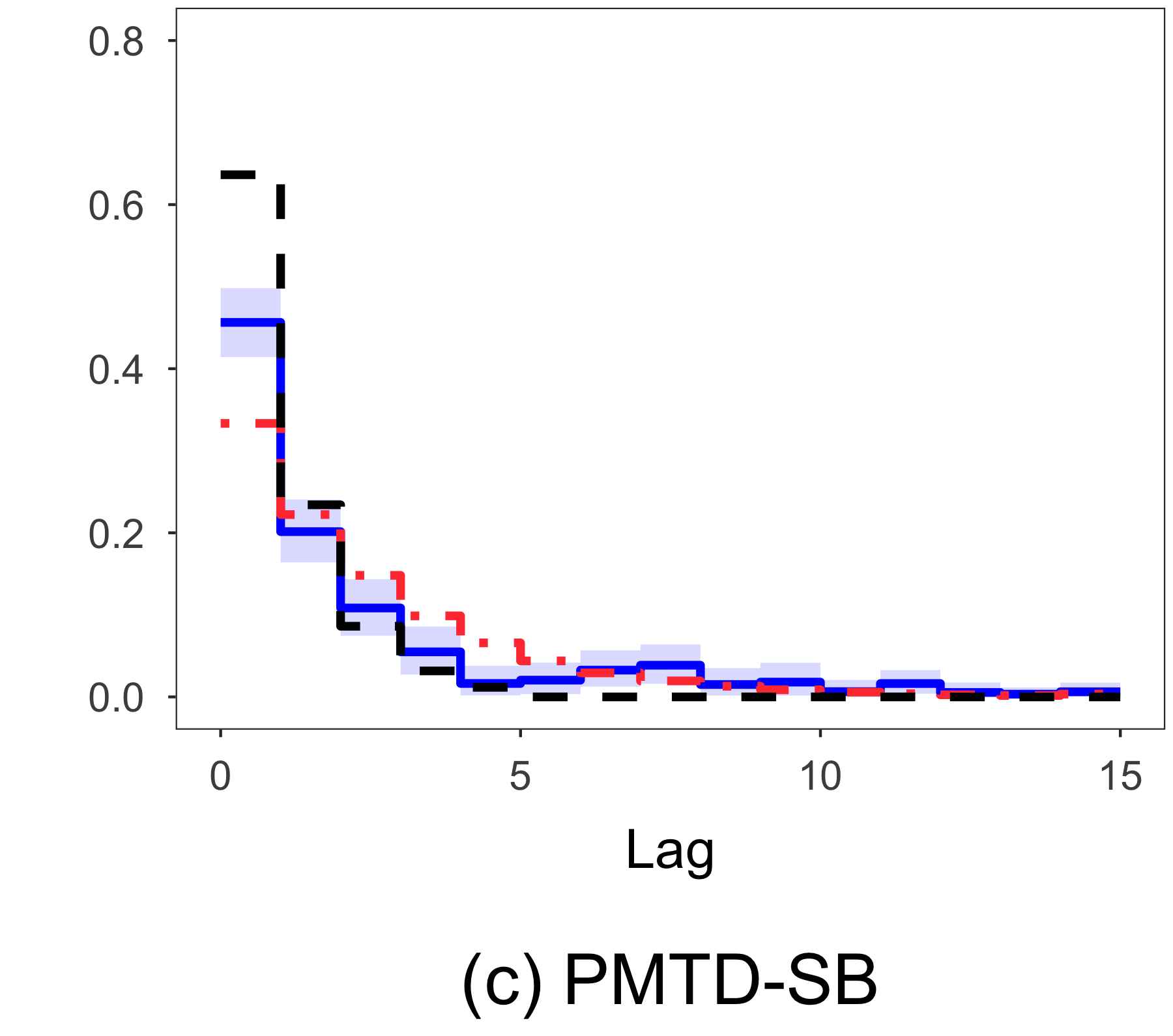}
    \includegraphics[width=.24\textwidth]{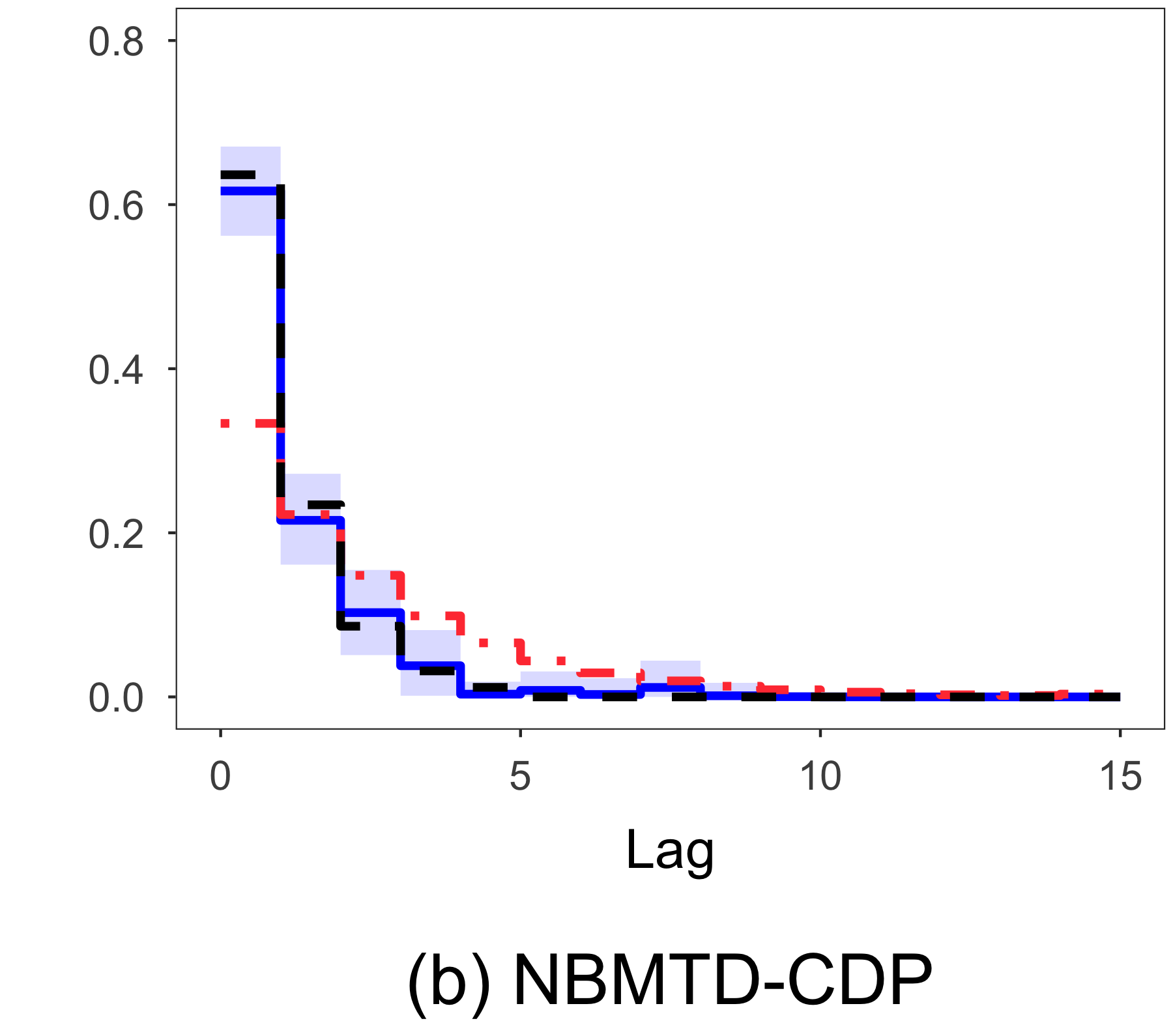}
    \includegraphics[width=.24\textwidth]{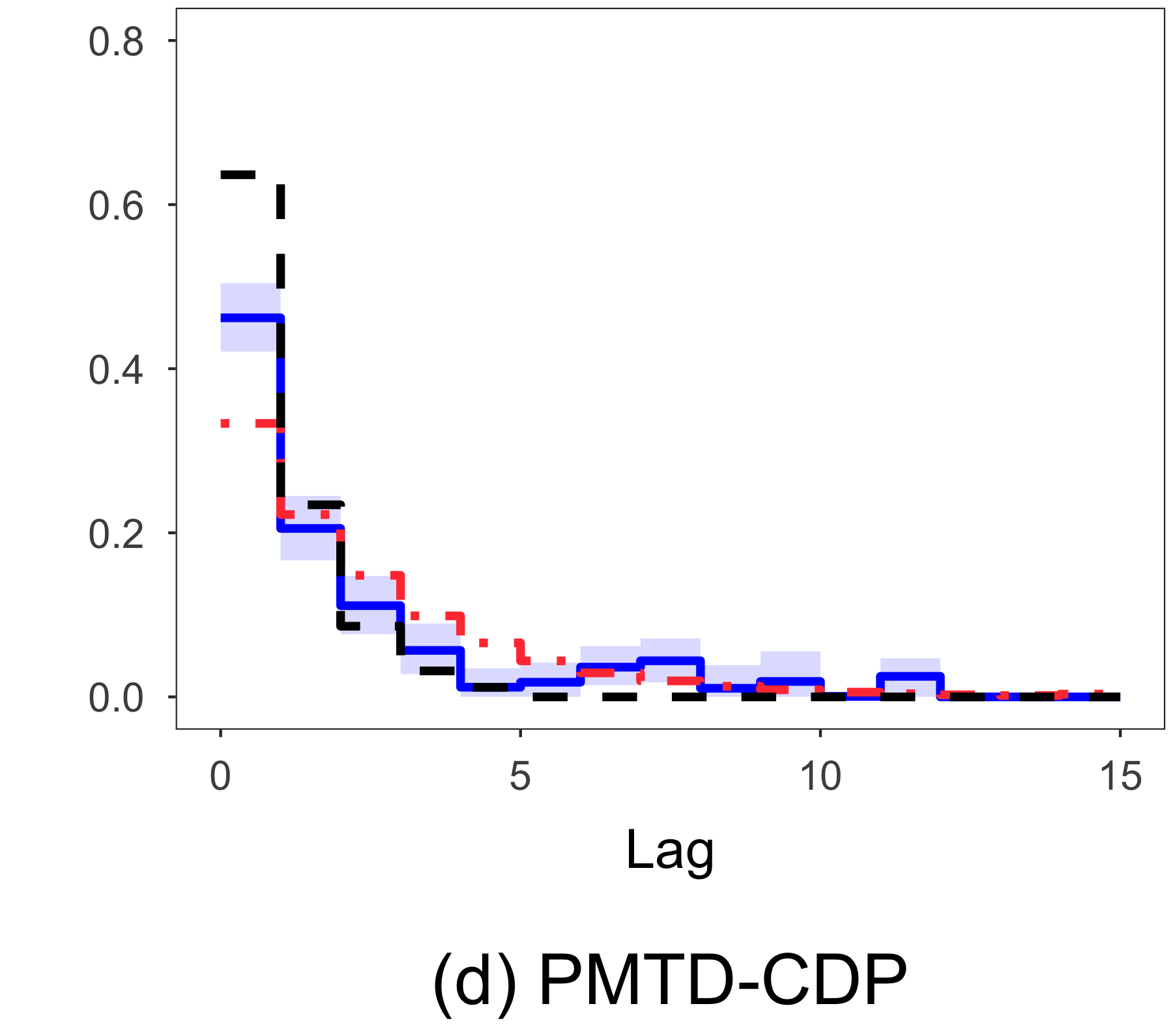}
    \caption{
    Simulation study 2: mixture weight estimates when $L = 5$ (top) and $L = 15$ (bottom).
    Black dashed lines are true weights, red dot-dahsed lines are prior
    means, blue solid lines are posterior means, and blue polygons are 95\%
    credible intervals. 
    }
    \label{fig:nbmtd_weight}
\end{figure*}

We generated 800 observations from the NBMTD model by the following scheme,
$$
\begin{aligned}
x_t\,|\,q_t,x_{t-z_t},z_t, \theta & \stackrel{ind.}{\sim} 
\mathrm{Bin}(x_t-q_t\,|\,x_{t-z_t},\gamma/(\lambda+\gamma),\\
q_t\,|\,x_{t-z_t},\kappa,\psi & \stackrel{ind.}{\sim} 
\mathrm{NB}(q_t\,|\,\kappa+x_{t-z_t}, 1-\lambda/(2\lambda+\gamma+\eta)),\\
z_t\,|\,\bm{w} & \stackrel{i.i.d.}{\sim} \sum_{l=1}^Lw_l\delta_l(\cdot),
\end{aligned}
$$
for $t = L+1,\dots, n$, given the first $L$ pre-specified observations.
We took $\lambda = 5, \gamma = 3, \kappa = 3, \eta = 2$, and specified
exponentially decreasing weights such that $w_i\propto \exp(-i),\;i=1,\dots,5$.
As a result, the synthetic data was over-dispersed, with empirical mean and variance 
being $12.95$ and $67.46$, respectively.

We applied the PMTD and the NBMTD models to the synthetic data.
For efficient posterior simulation, we reparameterized both models.
In particular, for both models, we used $\theta = \gamma/(\lambda+\gamma)$ 
as the probability of success of the binomial distribution for $X_t$. 
Furthermore, for the negative binomial model, 
we took $\psi = 1 - \lambda/(2\lambda+\gamma+\eta)$ as the probability of 
success of the negative binomial distribution for $Q_t$. 
Implementation details of the two models are given in Section D. 

\begin{figure*}[t!]
    \centering
    \includegraphics[width=.40\textwidth]{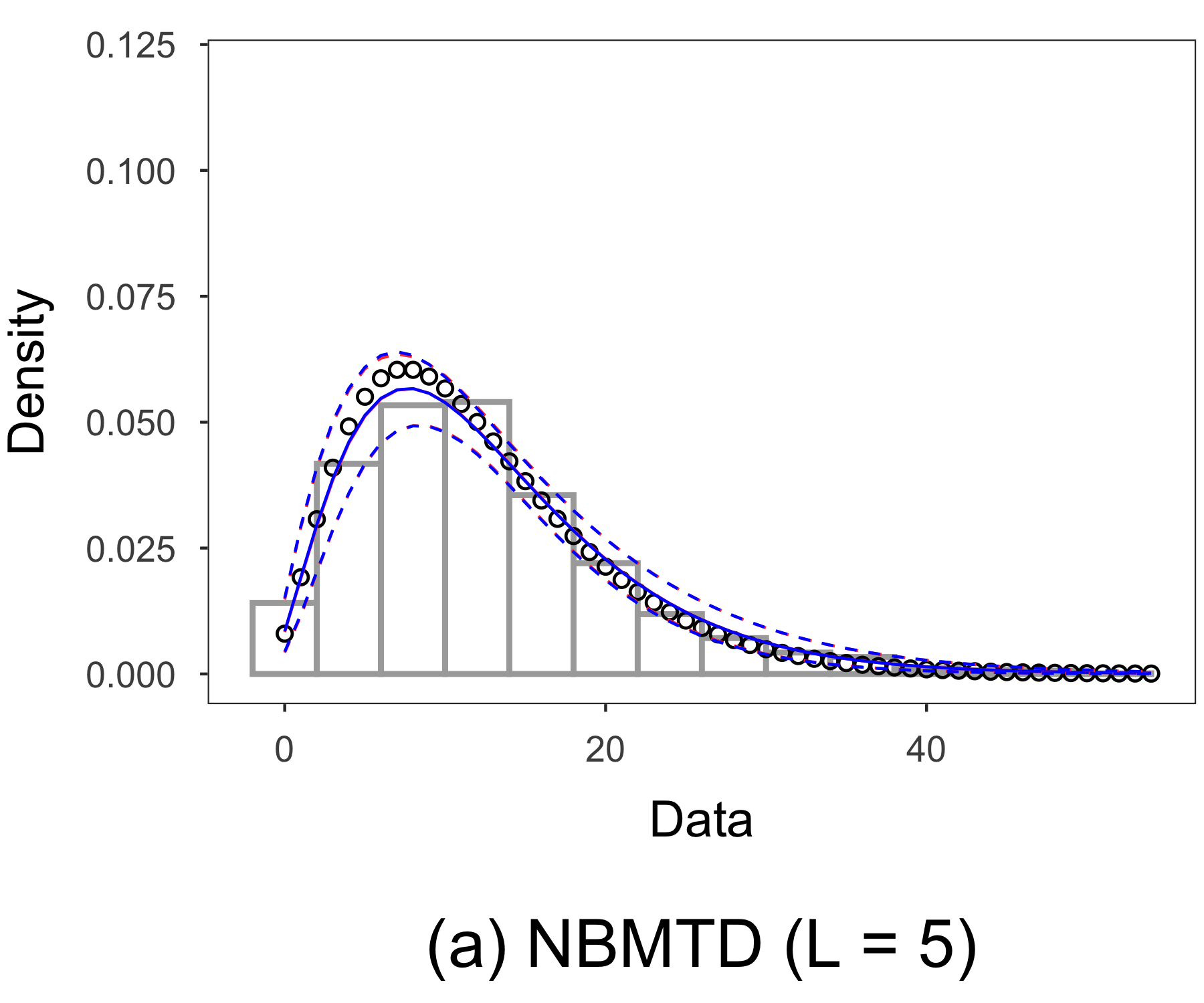}
    \includegraphics[width=.40\textwidth]{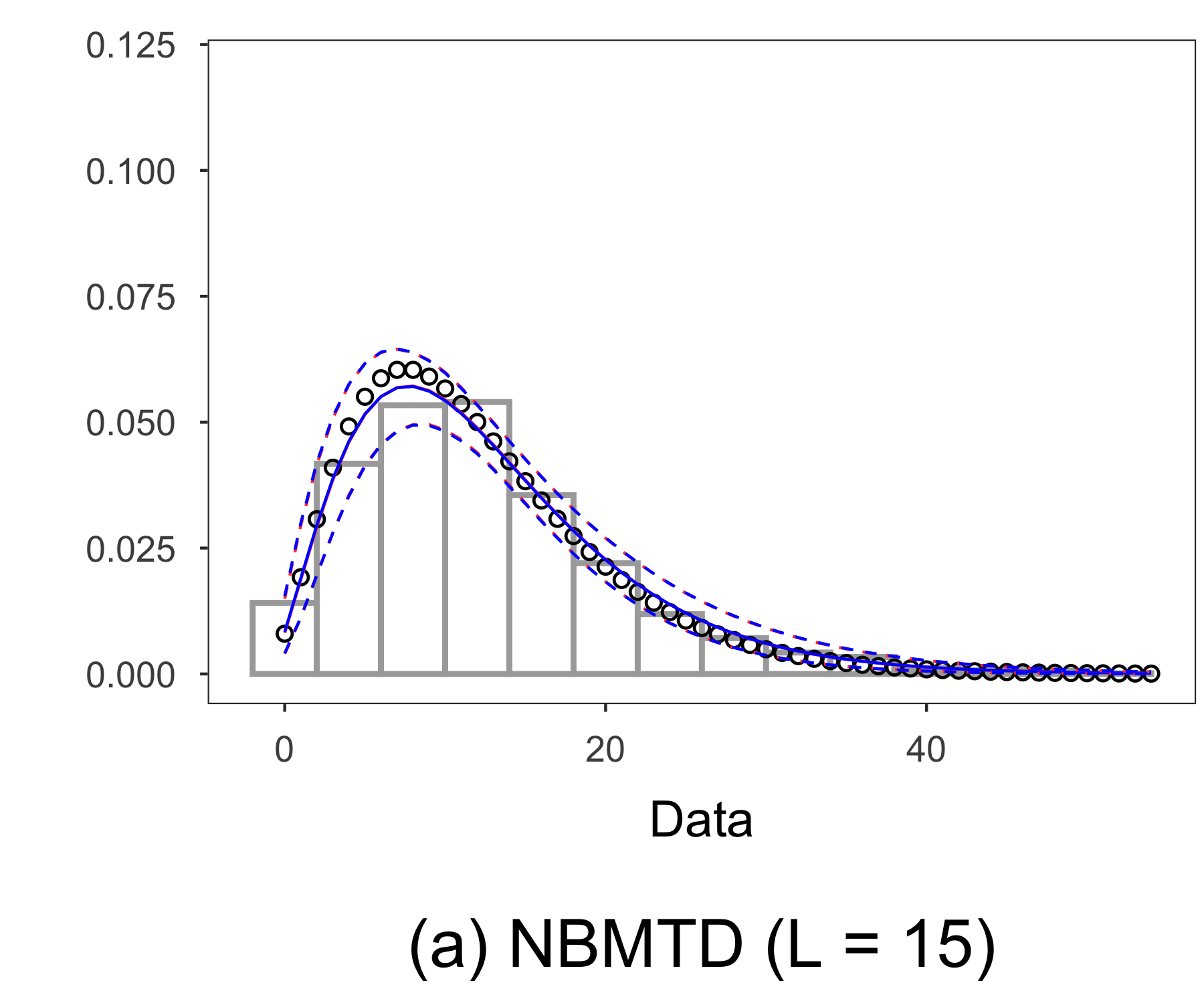}
    \medskip
    \includegraphics[width=.40\textwidth]{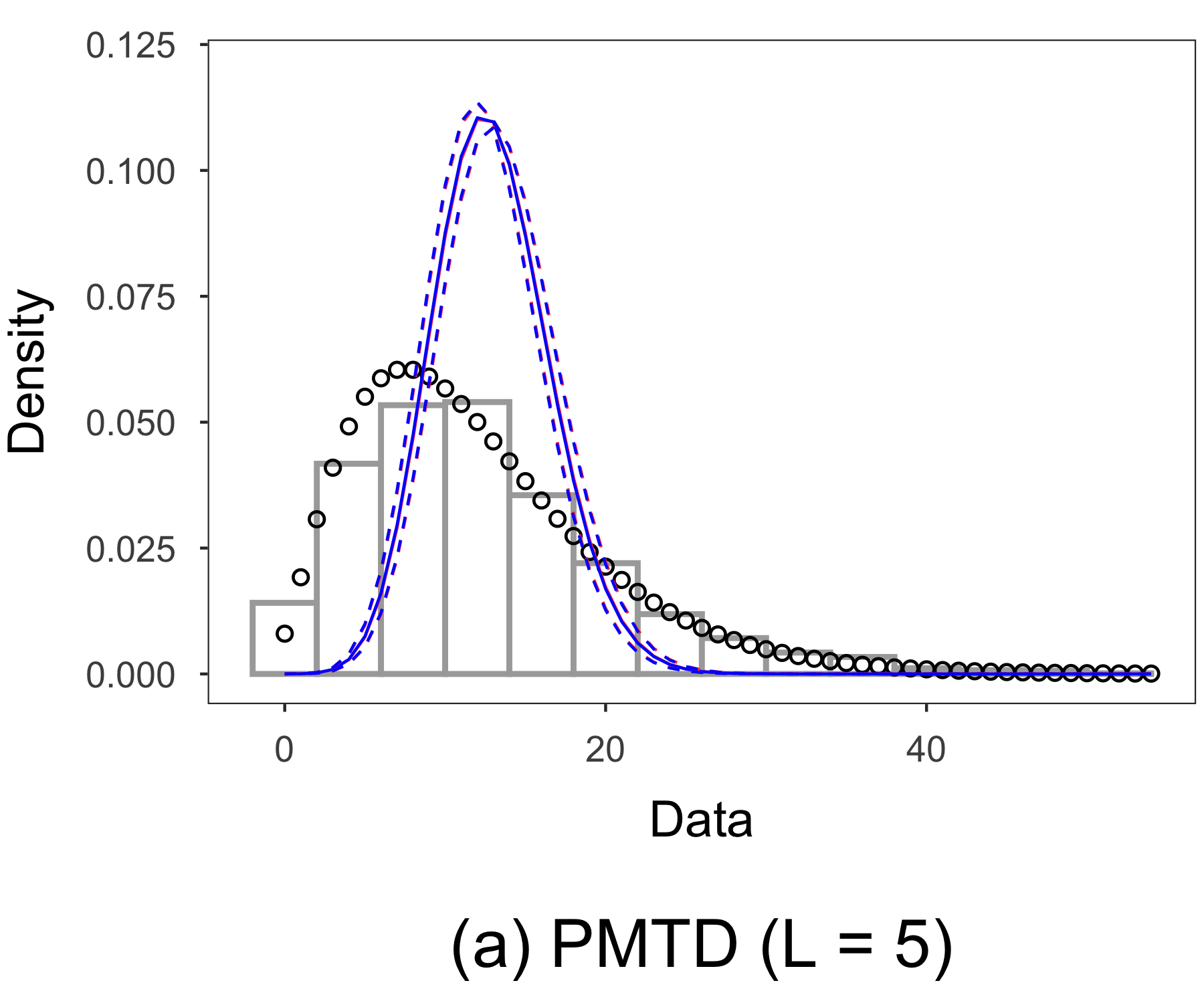}
    \includegraphics[width=.40\textwidth]{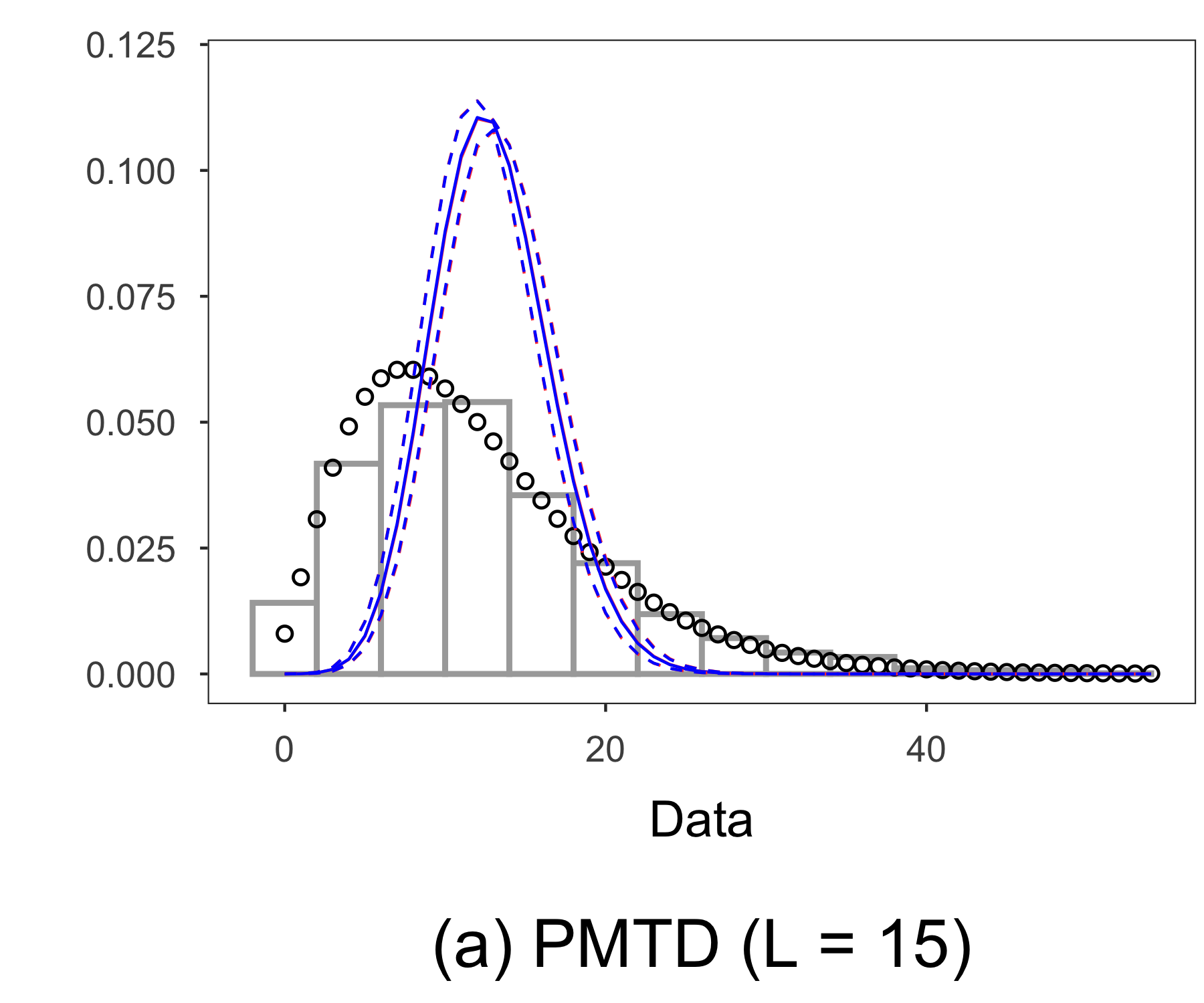}
    \caption{Simulation study 2: stationary marginal estimates. 
    White bars are histogram of the data. Circles are probability mass from
    the true marginal distribution $\mathrm{NB}(3, 0.2)$ evaluated at the effective support. 
    Red (blue) solid lines are posterior means from the fitted model with SB (CDP) prior. 
    Red (blue) dashed lines are 95\% credible intervals from the fitted model with 
    SB (CDP) prior.
    }
    \label{fig:nbmtd_marg}
\end{figure*}

For each model, we chose two different orders,
with one correctly specified, $L = 5$, and the other one over-specified, 
$L = 15$, based on the autocorrelation and partial autocorrelation functions.
The priors for $\theta$ and $\psi$ were
elicited based on priors for $(\lambda,\gamma,\eta)$. We took $\mathrm{Ga}(2,1)$ for each of 
$(\lambda,\gamma,\eta)$, which implies that both $\theta$ and $\psi$ follow beta distributions
$\mathrm{Beta}(\theta\,|\,2,2)$ and $\mathrm{Beta}(\psi\,|\,6,2)$, respectively. We also 
assigned $\mathrm{Ga}(2,1)$ to $\kappa$. 
For the weights,  we considered both the truncated stick-breaking and cdf-based priors.
For the former prior, we took $\alpha_s = 1,2$ corresponding to the orders, and 
for the latter one, we chose $\alpha_0 = 1$, and $b_0 = 3, 6$ respectively for the orders.
To obtain the estimates, in each case, we ran a Gibbs sampler for 85000 iterations, 
discarding the first 5000 samples as burn-in, and collected samples every 10 iterations.

\begin{figure*}[t!]
    \centering
    \includegraphics[width=.90\textwidth]{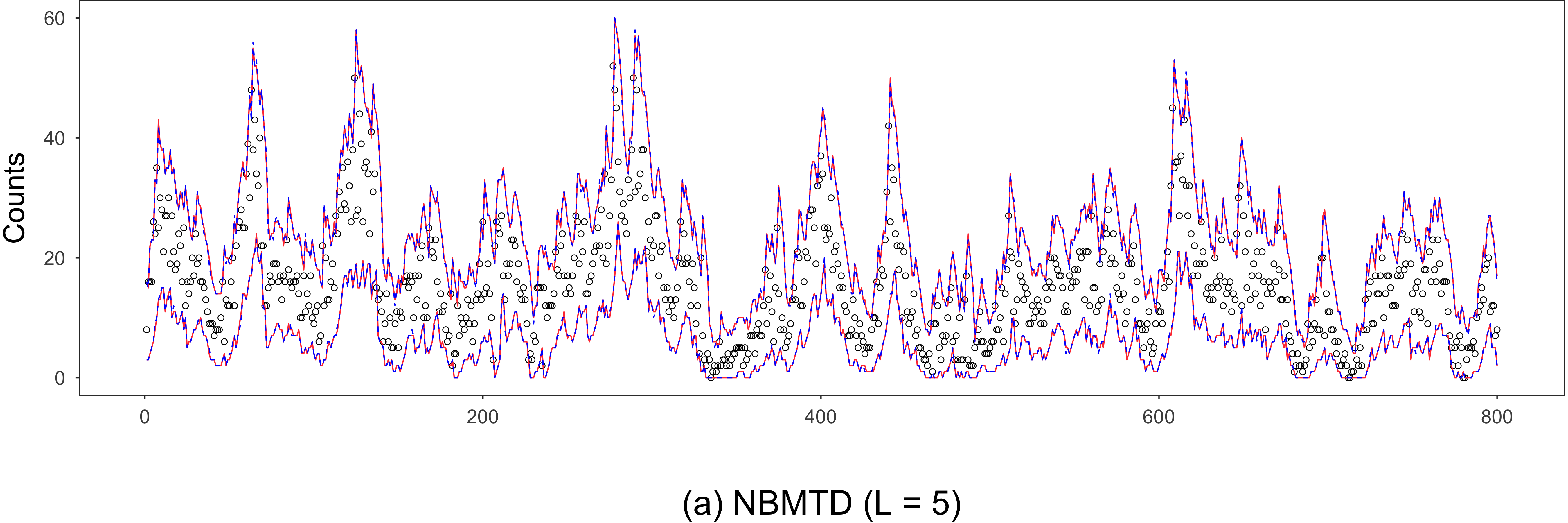}
    \medskip
    \vspace{3pt}
    \includegraphics[width=.90\textwidth]{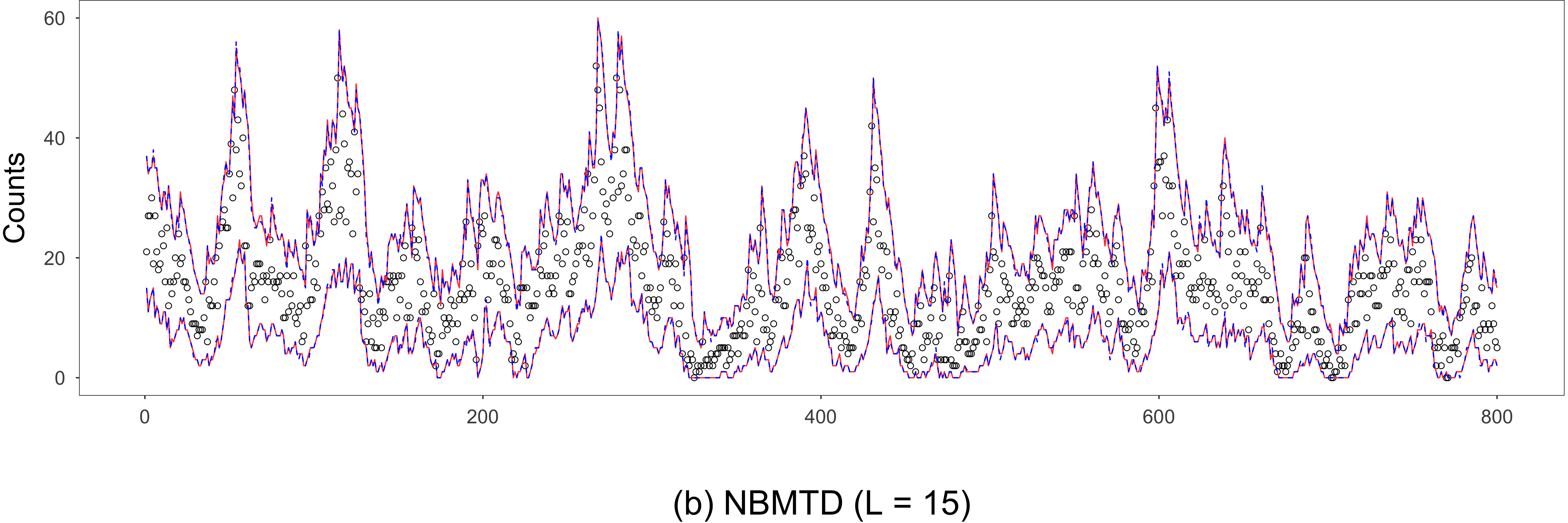}
    \medskip
    \vspace{3pt}
    \includegraphics[width=.90\textwidth]{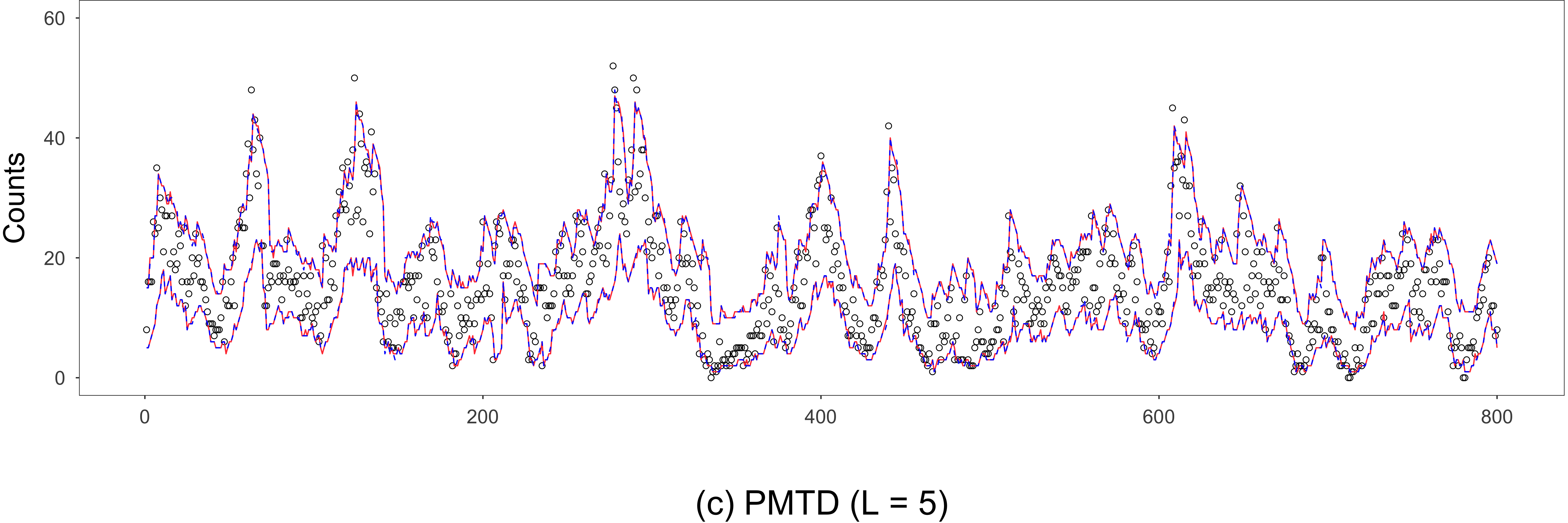}
    \medskip
    \vspace{3pt}
    \includegraphics[width=.90\textwidth]{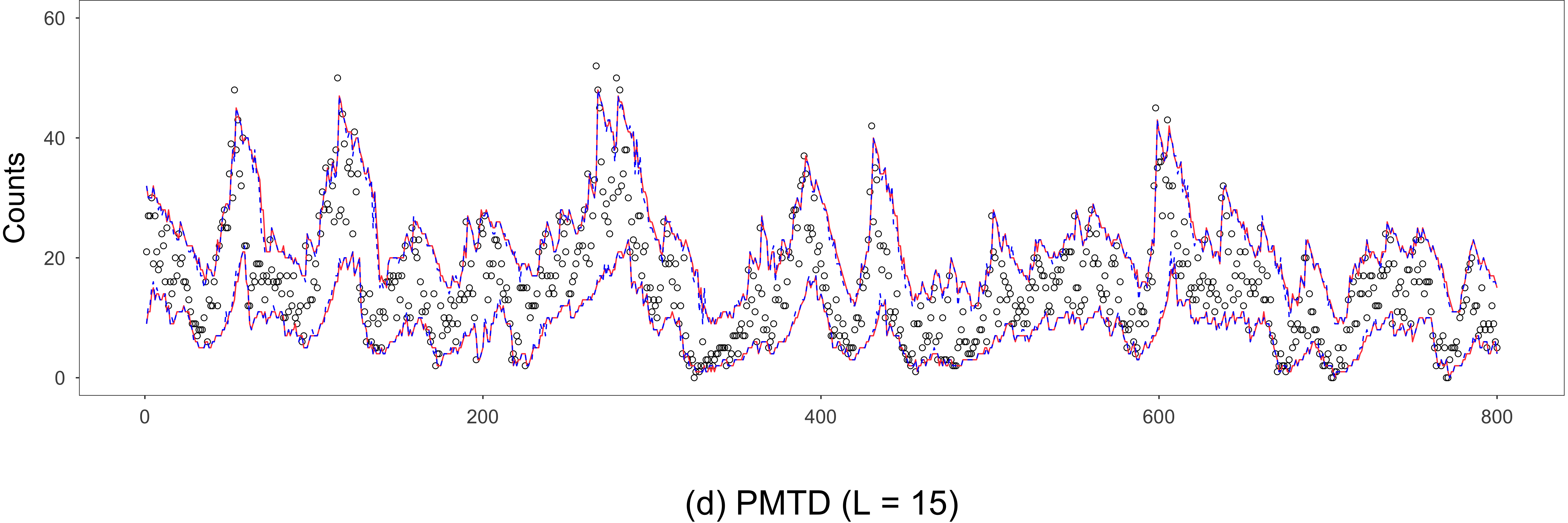}
    \caption{
    Simulation study 2: 95\% one-step ahead posterior predictive intervals. 
    Red (blue) dashed lines are predictive intervals from the fitted model with SB (CDP) prior.
    }
    \label{fig:nbmtd_ppd}
\end{figure*}

We focus on the results in estimating the weights and the stationary marginal distributions.
Figure \ref{fig:nbmtd_weight} shows that the NBMTD model was able to capture the weights
in all cases, while the PMTD model systematically missed the first weight,
in terms of the 95\% credible interval estimates. Moreover, 
even when $L$ is correctly specified, the PMTD model missed the last three weights.
Figure \ref{fig:nbmtd_marg} illustrates the stationary marginal 
estimated by the two models. In each case, the same model with the two proposed priors 
provided estimates that were almost identical. As expected,
the PMTD model was not capable of recovering the marginal, 
while the NBMTD model provided an accurate estimate.

\begin{table}[t!]
    \caption{Simulation study 2: empirical coverage of the $95\%$ predictive intervals.}
    \centering
    \begin{threeparttable}
    \begin{tabular*}{\hsize}{@{\extracolsep{\fill}}lcccc}
    \\[-5pt]
\hline
  & NBMTD-SB & NBMTD-CDP & PMTD-SB & PMTD-CDP\\
\hline
L = 5 & 0.956 & 0.954 & 0.858 & 0.858\\
\hline
L = 15 & 0.958 & 0.957 & 0.877 & 0.873\\
\hline
    \end{tabular*}
    \begin{tablenotes}[para,flushleft]
    \end{tablenotes}
    \end{threeparttable}
    \label{tbl:cover}
\end{table}

Turning to the predictive performance of the two models, 
Figure \ref{fig:nbmtd_ppd} shows the one-step ahead 95\% posterior predictive intervals
for the data. Under a visual examination, we can observe that the predictive intervals estimated by 
the NBMTD model were able to cover most of the small or large values, 
while the estimated predictive interval by the PMTD model missed many such values.
Table \ref{tbl:cover} presents the empirical coverage of the 95\% predictive intervals.
We see that the NBMTD model provided an accurate estimate, while the PMTD model
underestimated the coverage by a large margin.

Overall, we note the NBMTD model's ability to account for over-dispersion.
Moreover, even when $L$ was over-specified, the model provided estimates that 
were very close to the ones under the model with $L$ correctly specified. 

\section{Model checking for the real data examples}

Here, we provide model checking results for the 
real data examples presented in Sections 5.2 and 5.3 of the paper.
Figure \ref{fig:mc} consists of quantile-quantile
plot, histogram and autocorrelation for the residuals. 
If the model is correctly specified, the residuals will be independently and 
identically distributed as a standard Gaussian distribution. 
The results indicate a good fit of all models applied to both data sets.

\begin{figure*}[htbp]
    \centering
    \includegraphics[width=.85\textwidth]{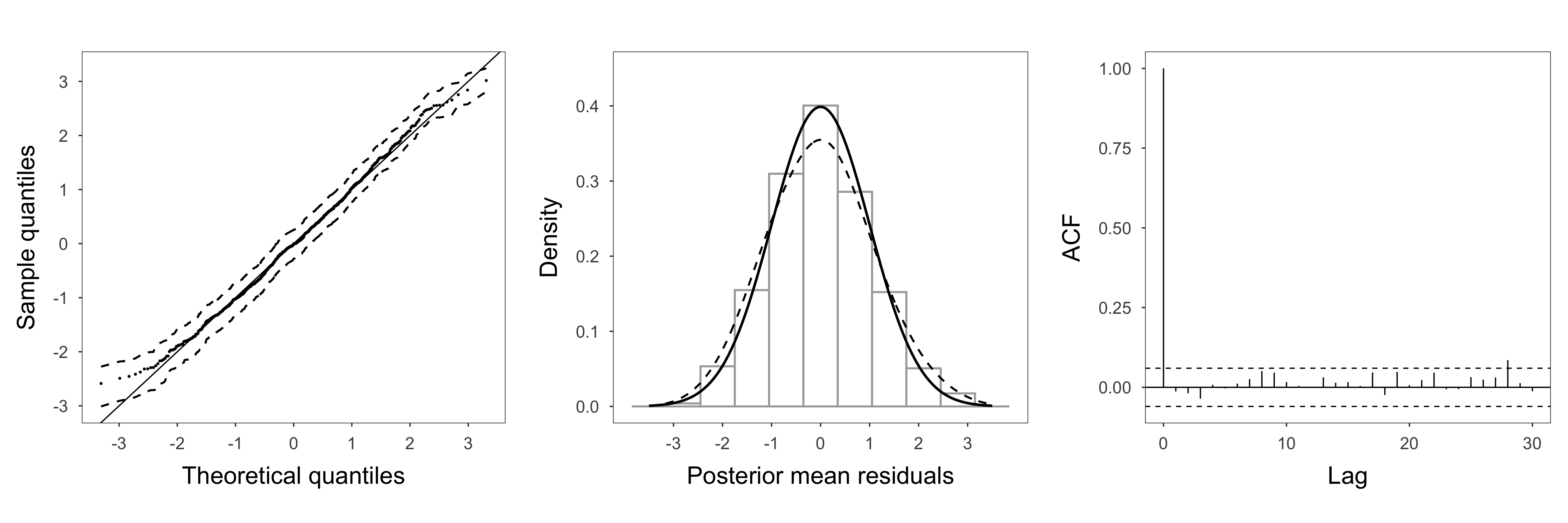}
    \medskip
    \includegraphics[width=.85\textwidth]{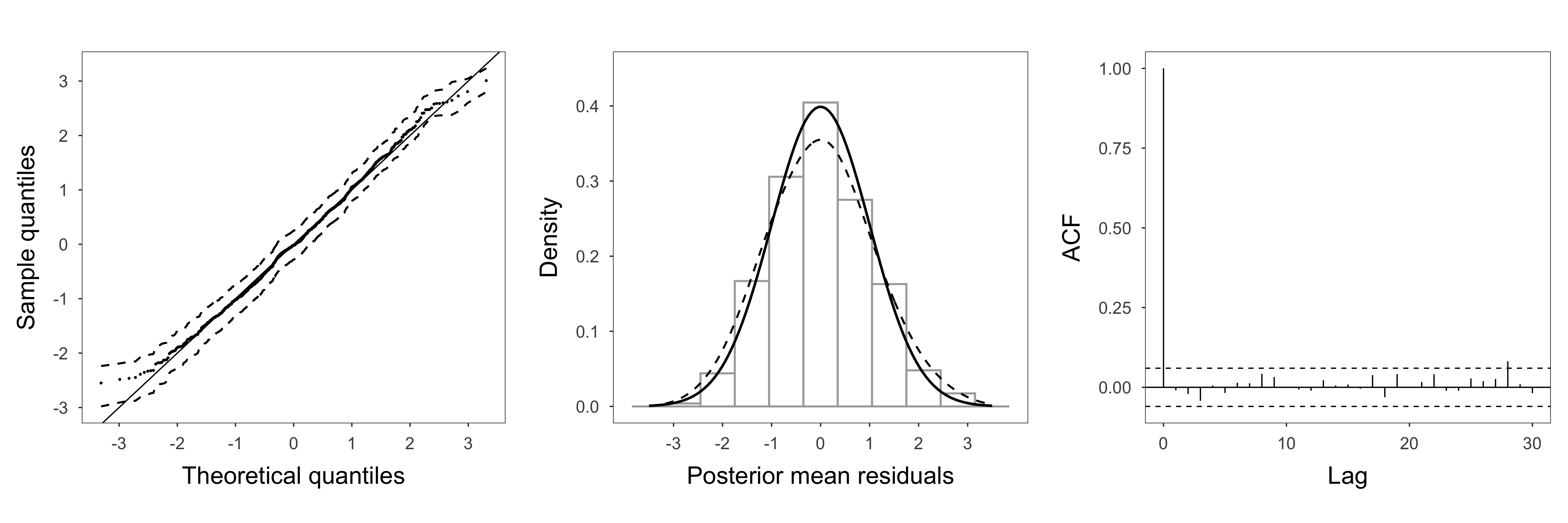}
    \medskip
    \includegraphics[width=.85\textwidth]{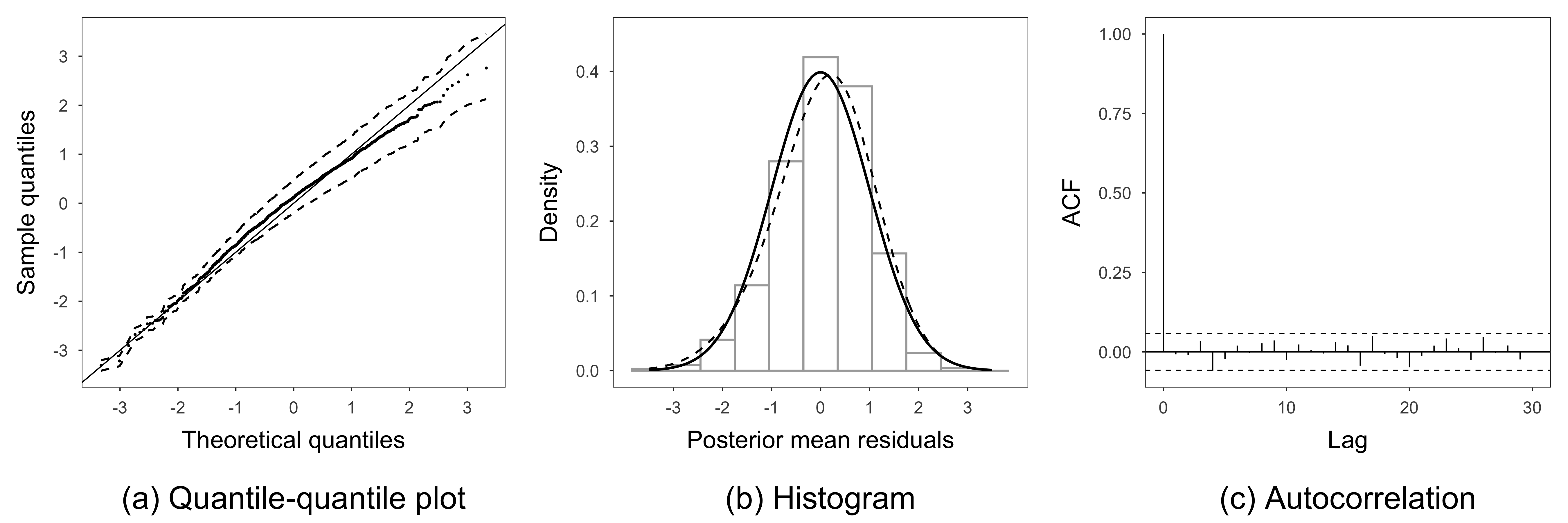}
    \medskip
    \includegraphics[width=.85\textwidth]{Precip/mc_sb.png}
    \caption{
    Randomized quantile residual analysis for the crime data models (first two rows)
    and for the precipitation data models (last two rows).
    In each example, the first and second row corresponds to the fitted model with 
    SB and CDP priors, respectively. In the left column, the
    circles and dashed lines correspond to the posterior mean and 95\% credible 
    interval for the quantile-quantile plot. 
    In the middle column, the solid and dashed line are the standard Gaussian density
    and the kernel density estimate of the posterior means of the residuals. 
    The right column is based on the posterior means of the residuals.}    
    \label{fig:mc}
\end{figure*}

\newpage
\section{Implementation Details}

We provide necessary details of the posterior simulation for
the Gaussian, Poisson, Negative binomial and Lomax MTD models.

We consider for the Gaussian MTD the following prior 
$p(\{\rho_l\}_{l=1}^L,\mu,\sigma^2) = 
\prod_{l=1}^L\mathrm{Unif}(\rho_l\,|\, -1,1)N(\mu\,|\,\mu_0,\sigma_0^2)
\mathrm{IG}(\sigma^2\,|\, u_0,v_0)$. The posterior full conditional 
distribution of $\mu$ is $N(\mu\,|\,\mu_1, \sigma_1^2)$ where 
$\mu_1 = \sigma_1^2\left(\mu_0/\sigma_0^2+c/\sigma^2\right)$ and 
$\sigma_1^2 = \left(1/\sigma_0^{2} + b/\sigma^2\right)^{-1}$ with 
$b = \sum_{t=L+1}^n(1-\rho_{z_t})^2/(1-\rho_{z_t}^2)$ and 
$c = \sum_{t=L+1}^n(1-\rho_{z_t})(x_t-\rho_{z_t}x_{t-z_t})/(1-\rho_{z_t}^2)$. 
The inverse gamma prior for $\sigma^2$ yields a conjugate full conditional 
distribution $\mathrm{IG}(\sigma^2\,|\, u_1,v_1)$ where $u_1 = u_0+(n-L)/2$ and 
$v_1 = v_0 + \sum_{t=L+1}^L(x_t-\rho_{z_t}x_{t-z_t}-(1-\rho_{z_t})\mu)^2/(2(1-\rho_{z_t}^2))$.
Finally, we update each $\rho_l$ using a slice sampler with target density
$\mathrm{Unif}(\rho_l\,|\,-1,1)\prod_{t:z_t=l}N(x_t\mid(1-\rho_l)\mu+
\rho_lx_{t-l},\sigma^2(1-\rho_l^2))$, for $l = 1,\dots, L$.
For each time $t,\;t = L+1,\dots,n$, the posterior probability of $z_t = l$
is proportional to $w_lN(x_t\,|\,(1-\rho_l)\mu+\rho_lx_{t-l},(1-\rho_l^2)\sigma^2)$.

For the Poisson MTD, we reparameterize the model such that
the $l$th component transition density of the model is sampled through
$$
x_t\,|\,q_t,x_{t-l},\theta \sim\mathrm{Bin}(x_t-q_t\,|\,x_{t-l},\theta),\;\;
q_t\,|\,\lambda \sim \mathrm{Pois}(q_t\,|\,\lambda).
$$
We consider conjugate prior $p(\lambda,\theta) = 
\mathrm{Ga}(\lambda\mid u_{\lambda},v_{\lambda})\mathrm{Beta}(\theta\mid u_{\theta},v_{\theta})$. 
The posterior full conditional distribution of $\lambda$ is gamma distribution
with shape parameter $u_{\lambda}+\sum_{t=L+1}^nq_t$ and rate parameter $v_{\lambda}+n-L$.
The posterior full conditional distribution of $\theta$ is a beta distribution
$\mathrm{Beta}(\theta\,|\,u_{\theta}+\sum_{t=L+1}^n(x_t-q_t), v_{\theta}+\sum_{t=L+1}^n(x_{t-z_t}-x_t+q_t))$.
We update $q_t$ with an independent Metropolis step with target density
$\mathrm{Bin}(x_t-q_t\,|\, x_{t-z_t},\theta)\mathrm{Pois}(q_t\,|\,\lambda)$
and proposal distribution being a discrete uniform distribution over the interval 
$[0\vee(x_t-x_{t-z_t}), x_t]$, for $t = L+1,\dots, n$.
For each time $t,\;t = L+1,\dots,n$, the posterior probability of $z_t = l$
is proportional to $w_l\mathrm{Bin}(x_t-q_t\,|\,x_{t-l}, \theta)$.

Similar to the Poisson model, we reparameterize the negative binomial MTD to facilitate
posterior simulation. In particular, the $l$th component transition density of the model 
is sampled through
$$
x_t\,|\,q_t,x_{t-l},\theta \sim\mathrm{Bin}(x_t-q_t\,|\,x_{t-l},\theta),\;\;
q_t\,|\,x_{t-l},\kappa,\psi \sim \mathrm{NB}(q_t\,|\,\kappa+x_{t-l}, \psi),
$$
where $p(\theta,\psi,\kappa) = \mathrm{Beta}(\theta\,|\,u_{\theta},v_{\theta})
\mathrm{Beta}(\psi\,|\,u_{\psi},v_{\psi})\mathrm{Ga}(\kappa\,|\,u_{\kappa},v_{\kappa})$.
The beta priors for $\theta$ and $\psi$ yield conjugate posterior full conditional distributions.
They are $\mathrm{Beta}(\theta\,|\,u_{\theta}+\sum_{t=L+1}^n(x_t-q_t), v_{\theta}+\sum_{t=L+1}^n(x_{t-z_t}-x_t+q_t))$
and $\mathrm{Beta}(\psi\,|\,u_{\psi}+(n-L)\kappa+\sum_{t=L+1}^nx_{t-z_t}, v_{\psi}+\sum_{t=L+1}^nq_t)$.
We update $\kappa$ using a random-walk Metropolis step with target density
$\mathrm{Ga}(\kappa\,|\,u_{\kappa},v_{\kappa})\prod_{t=L+1}^n\mathrm{NB}(q_t\,|\,\kappa+x_{t-z_t},\psi)$.
We update $q_t$ with an independent Metropolis step with target density
$\mathrm{Bin}(x_t-q_t\,|\,x_{t-z_t},\theta)\mathrm{NB}(q_t\,|\,\kappa+x_{t-z_t},\psi)$
and proposal distribution being a discrete uniform distribution over the interval 
$[0\vee(x_t-x_{t-z_t}), x_t]$, for $t = L+1,\dots, n$.
For each time $t,\;t = L+1,\dots,n$, the posterior probability of $z_t = l$
is proportional to $w_l\mathrm{Bin}(x_t-q_t\,|\,x_{t-l}, \theta)\mathrm{NB}(q_t\,|\,\kappa+x_{t-l},\psi)$.

For the Lomax MTD, we consider prior $p(\alpha,\phi,\beta) \propto
\mathrm{Ga}(\alpha\,|\, u_{\alpha},v_{\alpha})\mathrm{IG}(\phi\,|\, u_{\phi},v_{\phi})$.
The posterior full conditional distribution of $\alpha$ is 
$\mathrm{Ga}\left(\alpha\,|\, u_{\alpha}+n-L, v_{\alpha}'\right)$,
where the rate parameter $v_{\alpha}' = v_{\alpha} + \sum_{t=L+1}^n
\log\left(1+y_t\exp(-x_t^{\mathrm{T}}\beta)/(\phi+y_{t-z_t}\exp(-x_{t-z_t}^{\mathrm{T}}\beta))\right)$.
To improve mixing, we integrated out $\alpha$ from the posterior full conditional of $\phi$ and 
that of $\beta$. Then we use random walk Metropolis steps to update $\phi$ and $\beta$ with target densities
$\mathrm{IG}(\phi\,|\, u_{\phi},v_{\phi})g(\{y_t\}_{t=L+1}^n,\phi,\beta)$ and 
$\prod_{t=L+1}^n\exp(-x_t^{\mathrm{T}}\beta)g(\{y_t\}_{t=L+1}^n,\phi,\beta)$, respectively, 
where
$$
\begin{aligned}
g(\{y_t\}_{t=L+1}^n,\phi,\beta) & = 
\left\{\prod_{t=L+1}^n\left(\phi+y_{t-z_t}\exp(-x_{t-z_t}^{\mathrm{T}}\beta)+y_t\exp(-x_t^{\mathrm{T}}\beta)\right)^{-1}\right\}\\
&\;\;\;\;\left(v_{\alpha} + \sum_{t=L+1}^n\log(1+y_t\exp(-x_t^{\mathrm{T}}\beta)/(\phi+y_{t-z_t}\exp(-x_{t-z_t}^{\mathrm{T}}\beta)))\right)^{-(u_{\alpha}+n-L)}.
\end{aligned}
$$

\vspace{10pt}

\section*{Additional References}

Brockwell, P. J. and Davis, R. A. (1991), $\,$
\emph{Time Series: Theory and Methods}, $\,$ Springer.

\end{document}